\documentclass[aps,prb,twocolumn,showkeys]{revtex4-1}
\usepackage{graphics,graphicx,epsfig,color,hyperref,latexsym,float,amsmath}
\usepackage{times}
\input pdfcolor.tex

\begin{document}

\title{A Langevin approach to lattice dynamics in a charge ordered 
polaronic system 
}
\author{Sauri Bhattacharyya, Sankha Subhra Bakshi, Samrat Kadge 
and Pinaki Majumdar}
\affiliation{Harish-Chandra Research Institute, HBNI, 
Chhatnag Road, Jhusi, Allahabad 211019}
\date{\today}

\begin{abstract}
We use a Langevin approach to treat the finite temperature dynamics of 
displacement variables in the half-filled spinless Holstein model. Working
in the adiabatic regime we exploit the smallness of the adiabatic parameter 
to simplify the memory effects and estimate displacement costs from an 
``instantaneous'' electronic Hamiltonian. We use a phenomenological damping 
rate, 
and uncorrelated thermal noise.  The low temperature state has checkerboard 
charge order (CO) and the Langevin scheme generates equilibrium thermodynamic 
properties that accurately match Monte Carlo results. It additionally yields 
the dynamical structure factor, $D({\bf q}, \omega)$, from the displacement 
field $x({\bf r}, t)$.  We observe four regimes with increasing temperature, 
$T$, classified in relation to the charge ordering temperature, $T_c$, and 
the `polaron formation' temperature $T_P$, with $ T_c \ll T_P$. For $T \ll 
T_c$ the oscillations are harmonic, leading to dispersive phonons, with 
increasing $T$ bringing in anharmonic, momentum dependent, corrections. For 
$T \sim T_c$, thermal tunneling events of the $x({\bf r})$ 
field occur, with a propagating `domain' pattern at wavevector ${\bf q} 
\sim (\pi, \pi)$ and low energy weight in $D({\bf q}, \omega)$. When $T_c 
< T < T_P$, the disordered polaron regime, domain structures vanish, the 
dispersion narrows,  and low energy weight is lost. For $T \gtrsim T_P$ we 
essentially have uncorrelated local oscillations.  We propose simple models 
to analyse this rich dynamics.  
\end{abstract}

\maketitle

\section{Introduction}

Apart from its  ubiquitous effect on electronic resistivity
\cite{ziman}, 
electron-phonon (EP) interaction leads to collective 
states like superconductivity\cite{bcs} and charge
or orbital order \cite{gruner}. 
While the physics at weak EP 
coupling is perturbative, strong EP 
coupling leads to the formation of 
an electron-phonon bound state - a small 
polaron \cite{emin}.  
Residual interaction among the polarons
can lead to long range order, but signatures of
a polaronic state survive  
well above any ordering temperature.
Experiments on several materials\cite{tokura1,tokura2} 
over the last couple of decades
have established this.

Experiments probe strong coupling EP physics at various 
levels. The static structural properties that
result from EP coupling, including charge ordering, 
have been characterised in detail \cite{tokura3,pol2,millis1}.
The dynamical structure factor associated with lattice
fluctuations can be directly probed via
inelastic neutron scattering (INS)\cite{ins}.
The dynamics of the large amplitude lattice
displacements feed back on the electron system
leading to unusual spectral signatures observable through
angle resolved photoemission spectroscopy (ARPES).
Such data is already available in the manganites\cite{mannella}.
Beyond equilibrium, several studies have probed
the response of EP systems to intense radiation,
via `pump-probe' experiments
\cite{kemper}, exploring the exchange of
energy between the electron and phonon subsystems and the
approach to equilibrium.
While static structural properties are well understood,
dynamical properties and the physics out of equilibrium 
remain sparsely explored. Our focus in this paper is on
the dynamics at equilibrium.

At weak EP coupling the lattice
dynamics is affected via the electronic polarizability 
- modifying the dispersion and causing damping - and
the band susceptibility adequately describes
phonon properties \cite{phonon} over a reasonable temperature range. 
At strong EP coupling, however, when the electronic state itself
is strongly renormalised and temperature
dependent, one needs a fully self-consistent treatment
of the coupled electron-phonon problem. 
Amongst the non perturbative tools available, 
quantum Monte Carlo (QMC)\cite{QMC}
is numerically exact but subject to sign 
problems, large computation cost, and difficulty in
extracting real frequency information. 
Dynamical mean field theory (DMFT)\cite{DMFT} avoids the 
size dependence by exploring a self-consistent impurity
problem, but ignores spatial correlations which are
important near the thermal transition.

There are several puzzling issues that remain only partially
understood in phonon dynamics. These include:
(i)~the relation between the anomalous softening and 
broadening of phonons and 
short range charge-orbital order \cite{weber1,weber2,weber3},
(ii)~spatio-temporal fluctuations near an ordering transition, 
and (iii)~the relaxation from a `non-equilibrium' initial state,
created, for instance, by intense radiation,
to a thermal distribution.

Theoretical studies till now focus either on one 
dimension - addressing physics near the Peierls transition 
\cite{creff,hohen} and at dilute filling \cite{fehske}, 
or within DMFT \cite{bulla,millis2}.
To approach issues (i)-(iii) above we need a method
that  (a)~handles strong EP coupling, spatial correlations,
and thermal fluctuations, (b)~yields real time dynamics, and,
hopefully, (c)~handles non equilibrium situations!

A systematic approach to this problem requires the 
Keldysh framework\cite{neq}. We will show 
in the paper how a tractable
scheme can be derived from the Keldysh starting point by
assuming smallness of the `adiabatic parameter',
{\it i.e}, the ratio of bare phonon and electron energy scales,
and a high temperature approximation for the `noise' that
acts on the phonon variables. With these
assumptions, and a microscopically motivated choice
of phonon damping, $\gamma$,  
a Langevin equation 
\cite{martin,egger,brandbyge} 
can be written for the displacement field.
The equation that emerges has a parallel 
in classical many body 
physics, in particular the study of dynamical
critical phenomena \cite{hohenberg}. The approach
has also seen recent use in the study  
of spin dynamics in the Hubbard model\cite{chern}.

In this paper we use the Langevin dynamics (LD) 
approach
to study the half-filled spinless Holstein model
in two dimensions and intermediate coupling. 
Our focus is on the dynamical signatures 
as the temperature is increased through the charge ordering 
transition at $T_c$ into the `polaron liquid' phase. 
The charge order (CO) 
is at wavevector ${\bf Q} = (\pi,\pi)$.
We benchmarked the LD based 
charge ordering results against Monte Carlo (MC)
and found excellent agreement.
We focus on the dynamical structure factors,
$D_{nn}({\bf q}, \omega)$ and $D_{xx}({\bf q}, \omega)$, which 
are respectively the Fourier transforms of the
correlation function $\langle n_{\bf r}(t)n_{\bf r'}(t^{\prime})\rangle$
and $\langle x_{\bf r}(t)x_{\bf r'}(t^{\prime})\rangle$, $n_{\bf r}$ 
being the electron density and $x_{\bf r}$ the phonon
displacement. 
Most of our results show 
$D_{xx}({\bf q}, \omega) $ which we directly compute as 
$\vert X({\bf q},\omega)\vert^2$, where $X({\bf q},\omega)$ 
is the Fourier transform of the displacement field $x({\bf r}, t)$.
We measure time in units of $\tau_0$, the bare oscillation 
period for the local phonons.
Our key results are the following:

\begin{enumerate}
\item
For $T \ll T_c$ we observe dispersive phonons with 
frequency $\omega_{\bf q}$ that can be understood within a 
harmonic nearest neighbour model. The damping $\Gamma_{\bf q}$ is 
dictated by $\gamma$ and is momentum independent. 
Increasing $T$ leads to anharmonic signatures -
${\bf q}$ dependent softening of $\omega_{\bf q}$ and
increase of $\Gamma_{\bf q}$. Both changes, $\delta \omega_{\bf q}$
and $\delta \Gamma_{\bf q}$, are proportional to $T$ to leading
order.
\item
As $T$ approaches $T_c$
we observe three effects - (a)~occasional large
displacements at some site, with a quick reversal
to the original state, 
(b)~near $T_c$, a spatially correlated oscillatory 
pattern of large displacements mimicking `domain growth', 
and (c)~oscillations with large period, $\gtrsim 100 \tau_0$, 
generate
huge low energy weight in $D({\bf q} \sim {\bf Q}, \omega)$,
leading to a dramatic softening of the dispersion and reduction
of the damping. 
\item
In the polaron liquid phase, $T_c < T < T_P$, 
short range correlated polaronic distortions 
persist
(over a window upto $\sim 10T_c$ in our case).
As regards spectral
quantities, the low energy feature for ${\bf q} 
\sim {\bf Q}$ is gradually lost and the dispersion tightens.
We also see a quick reduction in phonon linewidths.
The dispersion is undetectable for $T \gtrsim 10 T_c \sim T_P$.
\end{enumerate}

\section{Model and method}

\subsection{Hamiltonian and parameter space}

We study the single band, spinless, Holstein model on a 
2D square lattice:
\begin{equation}
H=\sum_{ij}(t_{ij} - \mu \delta_{ij}) c^{\dagger}_{i}c_{j} 
+\sum_{i}(\frac{p^2_{i}}{2M} + \frac{1}{2}Kx^2_{i})-g\sum_{i}n_{i}x_{i}
\end{equation}

Here, $t_{ij}$'s are the hopping amplitudes. We study a 
nearest neighbour model 
with $t=1$ at $n=0.5$ (half-filling). 
$K$ and $M$ are the stiffness constant and mass, respectively,
of the optical phonons, 
and $g$ is the electron-phonon coupling constant. 
We set $K=1$. In this paper, we 
report studies for $\Omega = \sqrt{K/M} = 0.1$, 
which is a reasonable value for real materials.
We focus on a fixed, intermediate coupling value 
$g=2.0$.  
The chemical potential $\mu$ is set so that 
$n=0.5$.

\subsection{Keldysh to Langevin}

The Holstein problem can be set up in the Keldysh language in terms of
coherent state fields corresponding to $x_{i}$ and $c_{i}$ operators, 
with their full space-time 
dependence retained. We will indicate how 
a Langevin-like equation of motion can be obtained from the Keldysh action 
in the adiabatic limit. Physically, taking this limit corresponds to a 
``small $\dot{x}$'' approximation- namely the velocity of this field is 
assumed to be much smaller than Fermi velocity. 
We outline this below.

The partition function for the $x_i$ `oscillators' can be written as-
\begin{equation}
Z_{osc}=\int Dx_{i,f}Dx_{i,r}e^{ i(S_0+S_1) }
\end{equation}
Here $x_{i,f}$ and $x_{i,r}$ are lattice displacement fields along
forward and return contours respectively. The expressions for $S_{0}$ 
and $S_{1}$ are- 
\begin{eqnarray}
S_{0}&=&\frac{1}{2}\int dt[\sum_{i}(M\ddot{x}_{i,f}+Kx_{i,f})x_{i,f}-
(M\ddot{x}_{i,r}+Kx_{i,r})x_{i,r}] \cr
S_{1}&=&iTr(log([\mathcal{G}^{-1}]_{ij}(t,t^{\prime}))
\end{eqnarray}
where $\mathcal{G}$ is the matrix electron Green's function (with 
$N\times N$ dimension in real space and $2\times2$ in Keldysh space) in a 
time fluctuating $(x_{i,f},
x_{i,r})$ `background'.
To facilitate the derivation, one can transform to new 
`classical' and `quantum' variables 
$$
x_{i,cl}  = \frac{x_{i,f}+x_{i,r}}{2},~~~~~
x_{i,q}   = x_{i,f}-x_{i,r}
$$
The next important step is to assume the characteristic oscillator
frequency $\Omega$ to be much smaller than the electronic
energy scales (nominally the hopping $t$ in our model). In this situation,
one can perturbatively expand $S_{1}$ in powers of $x_{i,q}(t)$
while retaining $x_{i,cl}(t)$ non-perturbatively in the theory. 
The parameter that controls the expansion \cite{martin}
 is $\Omega/t$. 
Physically, the expansion in powers of $x_{i,q}(t)$ means we're adopting a 
semiclassical picture. Expanding up to linear order gives classical 
deterministic phonon dynamics. The quadratic term carries the effect of an 
added noise. This is done following the lines of Ref.26.

The look of the effective action for the oscillators now is-
$$
S_{eff}=S_{0} + [G^{K}_{cl}]_{ii}(t,t)x_{i,q}(t)
+ [\Pi^{K}_{cl}]_{ij}(t,t^{\prime})x_{i,q}(t)x_{j,q}(t^{\prime})
$$
where $[G^{K}_{cl}$] is the Keldysh component of electron Green's function
$\mathcal{G}$ computed setting $x_{i,q}=0$. The quantity 
$[\Pi^{K}_{cl}$] is the Keldysh
component of electronic polarizability for $x_{i,q}=0$, related to the 
Green's functions by the relation-
$$
\Pi^{K}_{ij}(t,t^{\prime})=G^{R}_{ij}(t,t^{\prime})G^{A}_{ji}(t^{\prime},t)
+(R \leftrightarrow A)
+G^{K}_{ij}(t,t^{\prime})G^{K}_{ji}(t^{\prime},t) 
$$
$G^{R}$ and $G^{A}$ being retarded and advanced components of $\mathcal{G}$.

The coefficients of the linear and quadratic terms in $x_{i,q}(t)$ are 
thus determined through computing electronic correlation 
functions in an `arbitrary'
$x_{i,cl}(t)$ background. This calculation can be simplified by expanding
the `trajectories' $x_{i,cl}(t)$ around a reference time $t_{0}$ in powers
of the velocity $\dot{x}_{i,cl}$. The velocity independent term is interpreted
in terms of a force exerted by an instantaneous effective Hamiltonian. The
linear in $\dot{x}_{i.cl}$ term gives rise to `damping' with a frequency 
dependent kernel. 

The next stage of approximation concerns the frequency dependence of
the Keldysh component
of electronic polarizability $\Pi^{K}_{ij}(\omega)$. At equilibrium, the
frequency dependence of this quantity can be factored according to
fluctuation-dissipation theorem\cite{neq} as-
\begin{equation}
 \Pi^{K}_{ij}(\omega)=coth(\frac{\omega}{2k_{B}T})
 (\Pi^{R}_{ij}(\omega) - \Pi^{A}_{ij}(\omega))
\end{equation}
where $\Pi^{R}_{ij}$ and $\Pi^{A}_{ij}$ are the retarded and advanced
components of the polarizability respectively.
These are defined as-
\begin{equation}
\Pi^{R/A}_{ij}(t,t^{\prime})=G^{R/A}_{ij}(t,t^{\prime})G^{K}_{ji}(t^{\prime},t)
+(R/A \leftrightarrow K)
\end{equation}

Next, we make the high temperature ($k_{B}T \gg \omega$) approximation on 
the RHS. The hyperbolic cotangent gives a factor of ($2k_{B}T/\omega$), 
and the low frequency spectral part of $\Pi$ contributes $\gamma\omega$,
where $\gamma=\frac{Im(\Pi^{R}(\omega))}{\omega}$ and we've neglected
the spatial dependence of the polarizability.

If one carefully carries out the evaluation of the linear in velocity
($\dot{x}_{i,cl}$) term, the coefficient comes out to be the the spectral
part of the polarizability $Im(\Pi^{R}_{ij}(\omega))$. Again neglecting
spatial dependences here and taking the low-frequency limit, the term
simplifies to $\gamma\omega$ and becomes the usual non-retarded Langevin
damping coefficient.

Finally, one decouples the quadratic term in $x_{i,q}(t)$ through
a Hubbard-Stratonovich transformation introducing a `noise' field 
$\xi_{i}(t)$ and then integrates over $x_{i,q}(t)$ in the 
partition function to obtain an `equation of motion' \cite{neq} 
for $x_{i,cl}(t)$.
This leads to our dynamical equation, below.
 
\subsection{Effective equation}

The dynamical equation which we solve for the phonon field is the following-
\begin{eqnarray}
M\ddot x_{i}(t) ~~~ & = &~ -\gamma\dot{x}_i(t)  - K x_i(t) - 
{ {\partial {\langle H_{el}\{ x\} \rangle }}   \over {\partial x_i }} 
+ \xi_i(t) \cr \cr
H_{el}~~~~~~ & = &~ \sum_{ij}(t_{ij} - \mu \delta_{ij}) 
c^{\dagger}_{i}c_{j} - g\sum_in_ix_i
\cr \cr
{\partial {\langle H_{el}\{ x\} \rangle }}   \over {\partial x_i }
& = &~ -g {\bar n}_i(t) \cr
{\bar n}_i(t) ~~~~~&  = &~ \sum_{\epsilon_{n}(t)}\vert U_{in}(t)\vert^2
n_f(\epsilon_{n}(t))
\end{eqnarray}
where $U_{in}(t)$ are site amplitudes of the instantaneous eigenvectors 
of $H$ (as in Eq.1) for a given $x_{i}(t)$ configuration and $\epsilon_{n}(t)$
are the corresponding eigenvalues. $n_f(\epsilon_{n}(t))$ denotes Fermi
factors needed to calculate the instantaneous density field.
Note that the spatial correlations in this arise only via
the dependence of the density ${\bar n}_i$ on the field
$\{ x_i \}$.

The first term describes damping, second and third are effective forces 
and the last one is the noise field, which is specified by the conditions-
\begin{eqnarray}
\langle \xi_{i}(t)\rangle~~~~&=&0 \cr
\cr
\langle \xi_{i}(t)\xi_{j}(t^{\prime}) \rangle&=&2\gamma 
k_{B}T\delta_{ij}\delta(t-t^{\prime})
\nonumber
\end{eqnarray}
The unit of time is taken to 
be the inverse of the bare oscillator frequency $\tau_{0}=2\pi/\Omega$.  
For most of our simulations, we chose $\gamma=1.0t$, which sets the
damping timescale to $2M/\gamma= 3\tau_{0}$. The imaginary part of the
retarded polarizability $Im \Pi^{R}(\bf {q},\omega)$ is gapped at low $T$
in the present model. At intermediate temperatures, it picks up a low
energy contribution proportional to $\omega$. The microscopic estimate
of $\gamma$, based on $Im \Pi^{R}$, is smaller and also $T$ dependent.
To minimise parameter variation and ensure reasonably rapid
equilibriation we have set $\gamma=1$.

We integrate the equation numerically using the well-known Euler-Maruyama 
method. The time discretization for most calculations was set to 
$\Delta t= 1.6 \times 10^{-4} \tau_{0}$. We typically ran the simulations for 
$\sim10^7$ steps, ensuring a time span of almost a few hundred times the 
equilibration time. This ensured enough frequency points to 
analyze the power spectrum.   

\subsection{Indicators}

We quantify the equal-time and dynamical properties through several indicators.
We first define some timescales. We set an `equilibriation time' $\tau_{eq}
= 100 \tau_0$ before saving data for the power spectrum. The outer timescale,
$\tau_{max} \sim 10 \tau_{eq}$.  The `measurement time' $\tau_{meas}
= \tau_{max} - \tau_{eq}$, and the number of sites is $N$. We calculate the 
following:
\begin{enumerate}
\item
Dynamical structure factor, $D({\bf q}, \omega) = \vert
X({\bf q}, \omega) \vert^2$, where
$$
X({\bf q}, \omega) = 
\sum_{ij} \int_{\tau_{eq}}^{\tau_{max}} 
dt e^{i {\bf q}. ({\bf r}_i - {\bf r}_j) }
e^{i \omega (t - t')}
x({\bf r}_i, t) x({\bf r}_j, t')
$$
\item
The instantaneous structure factor 
$$
S({\bf q},t)= 
{1 \over N^2} 
\sum_{ij} e^{i {\bf q}. ({\bf r}_i - {\bf r}_j) }
x({\bf r}_i, t) x({\bf r}_j, t) 
$$
The instantaneous `order parameter' is $S({\bf Q}, t)$, and the 
time averaged structure factor is
$$
{\bar S}({\bf q}) = {1 \over \tau_{meas}} 
\int_{\tau_{eq}}^{{\tau}_{max}} dt e^{i \omega t} S({\bf q},t) 
$$
\item
The distribution of distortions:
$$
P(x)= \frac{1}{N \tau_{meas}} \sum_{i} 
\int_{\tau_{eq}}^{\tau_{max}} dt \delta(x - x_{i}(t))
$$
\item
Dispersion $\omega_{\bf q}$ and damping $\Gamma_{\bf q}$: 
\begin{eqnarray}
\omega_{\bf q} &=& \int_{0}^{\omega_{max}} d\omega 
\omega  D({\bf q}, \omega) \cr 
\Gamma_{\bf q}^{2} & =  &
\int_{0}^{\infty} d\omega 
(\omega - \omega_{{\bf q}})^2 D({\bf q}, \omega) 
\nonumber
\end{eqnarray}
While calculating moments, we've normalized by ($(g/K)^{-2}$),
to ensure dimensional consistency.
\item
We show spatial maps for 
the overlap of $x_{i}(t)$ with a perfect CO
with alternating distortions $0$ and $g/K$.
\end{enumerate}
 
\begin{figure}[b]
\centerline{
\includegraphics[width=5.7cm,height=4.6cm]{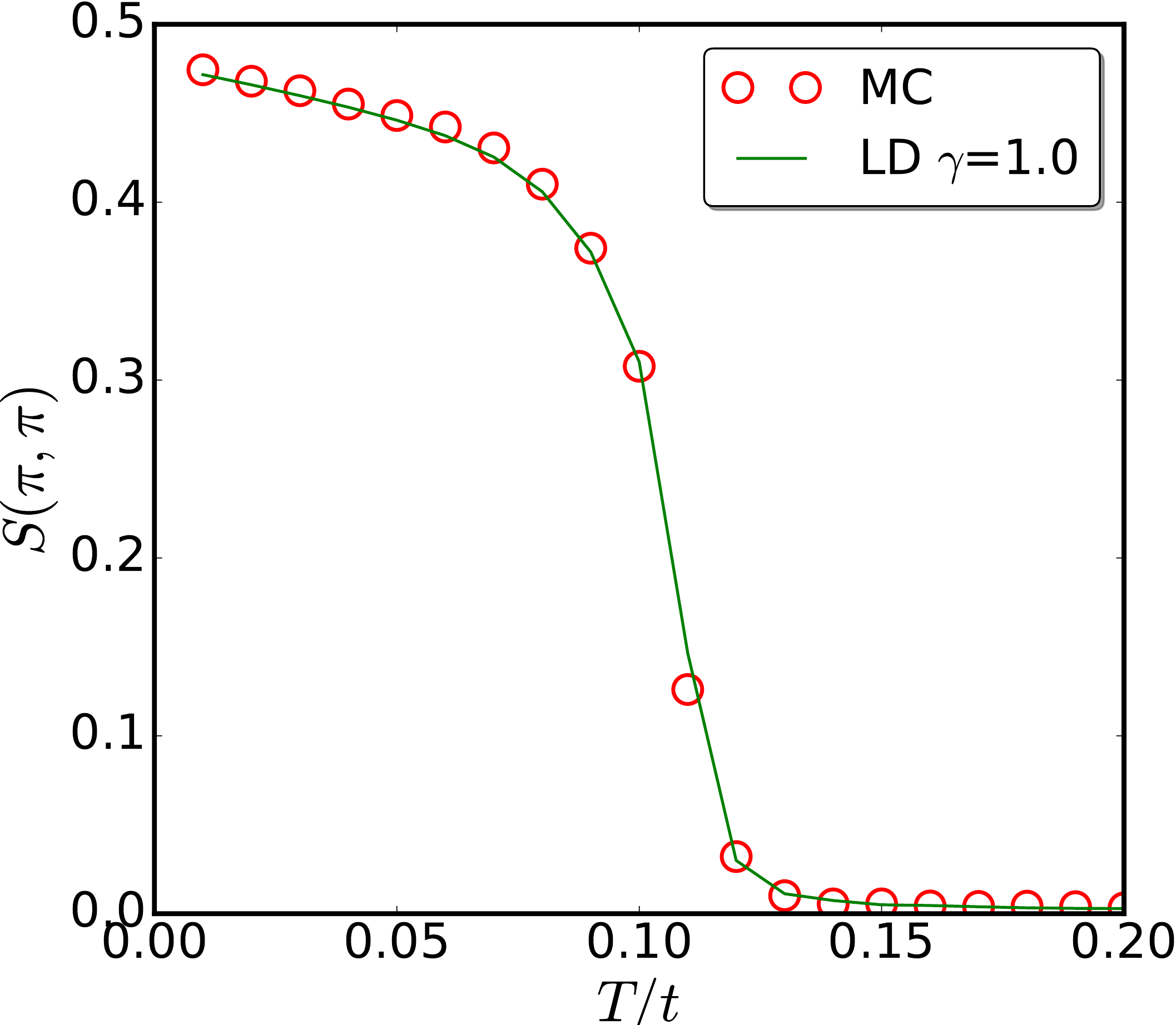}
}
\caption{Comparison of $S(\pi,\pi)$ computed
using Langevin dynamics (green line) and MC annealing
(red open circles).
}
\end{figure}
\begin{figure}[t]
\centerline{
\includegraphics[width=7.5cm,height=5.0cm]{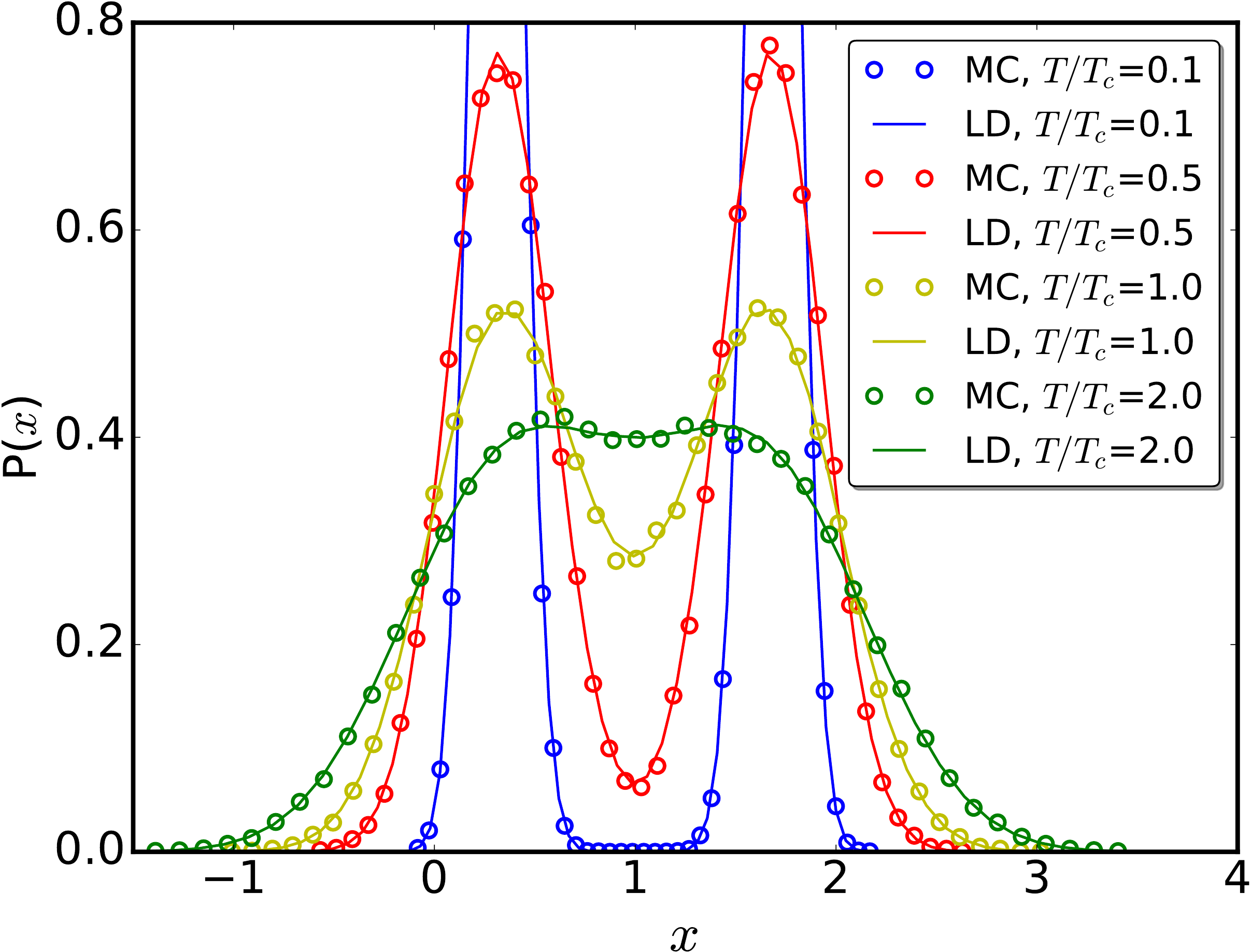}
}
\caption{
Comparison of displacement distribution $P(x)$ obtained
using Langevin dynamics (solid lines) and MC annealing
(open circles) in various temperature regimes.
}
\end{figure}

\section{Benchmarking with~ Monte Carlo}

In this section we compare the CO order parameter and the distribution,
 $P(x)$, of lattice deformations, obtained via Langevin dynamics and
via Monte Carlo simulation.  

\subsection{Order parameter}

Fig.1 compares $S(\pi,\pi)$  
from Langevin dynamics for various $\gamma$ values with that from 
MC annealing (red open circles). The agreement is excellent at
$\gamma =1$, and generally good at the other $\gamma$ as well.
The transition temperature $T_c\sim0.12t$ is inferred from the onset of 
rise in both curves. The transition is in the Ising class, however there's
a quantitative reduction of structure factor in the low $T$ regime due to 
the continuous nature of the $x_i$ variable, 
absent in the Ising model.
The dependence on $\gamma$ is weak, but higher
$\gamma$ generally leads to better correspondence with equilibrium MC results.

\subsection{$P(x)$ distributions}

In Fig.2 we show the $P(x)$ obtained from a Monte Carlo calculation,
compared to results from LD at $\gamma = 1$.
The dynamical method gives histograms quantitatively comparable 
to the MC results. We plot the distributions at four temperatures
-  $0,~0.5T_c,~T_c,~2T_c$.
The solid lines are the Langevin data whereas open circles depict 
results from MC annealing. The agreement suggests that the two
methods should predict the same `equal-time' properties in equilibrium. 

\begin{figure}[b]
\centerline{
\includegraphics[width=4.5cm,height=4.2cm]{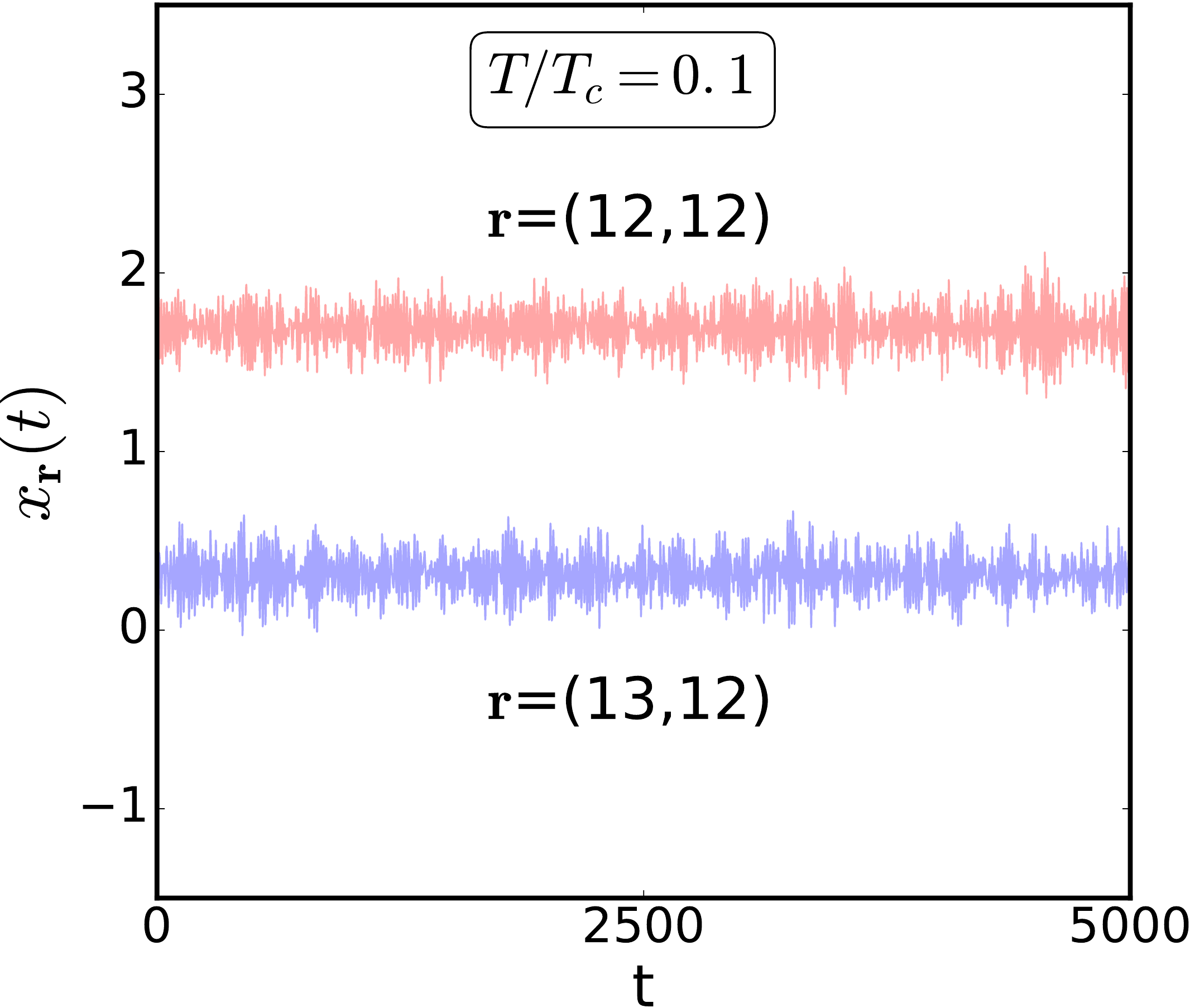}
\includegraphics[width=4.5cm,height=4.2cm]{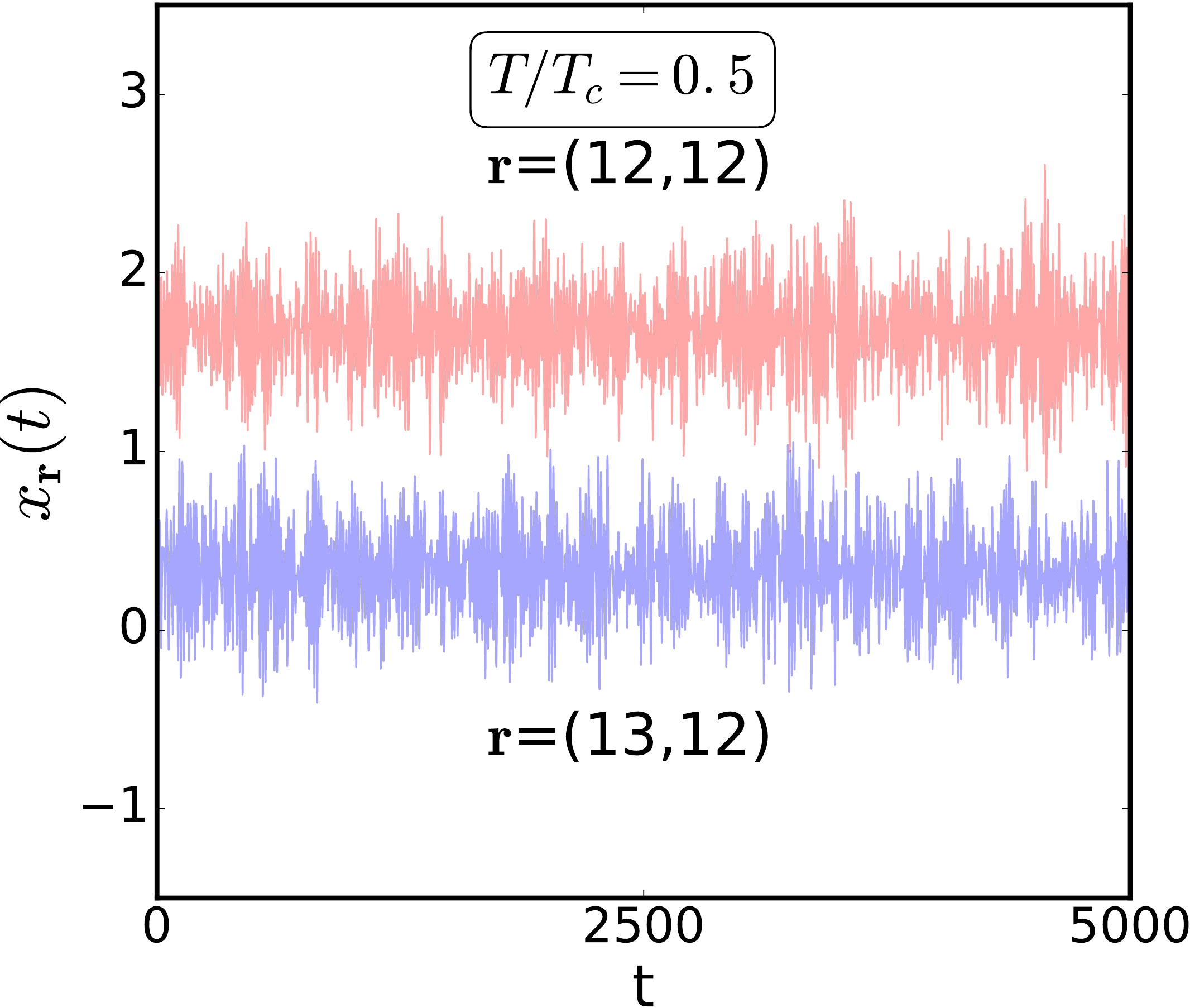}
}
\centerline{
\includegraphics[width=4.5cm,height=4.2cm]{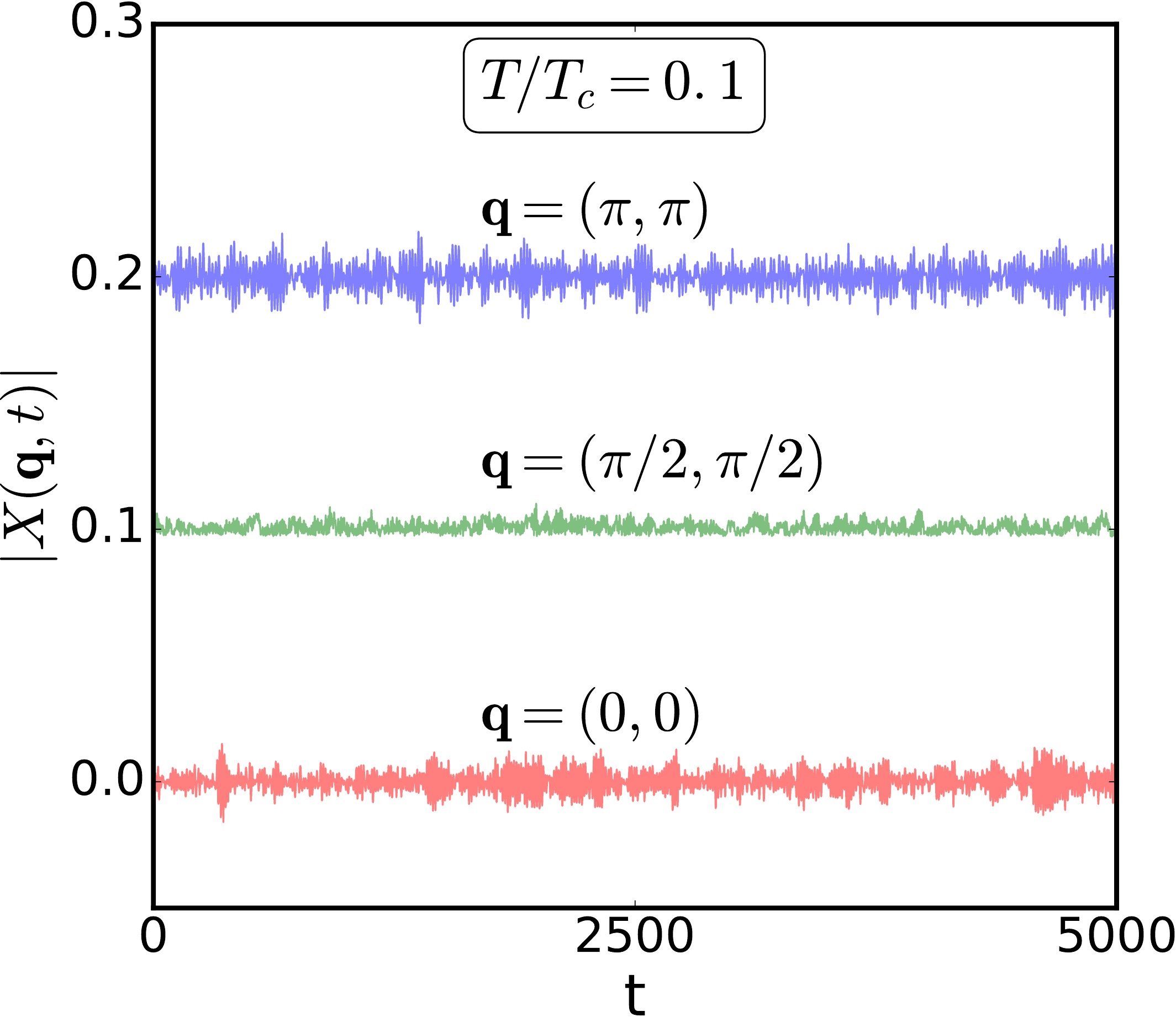}
\includegraphics[width=4.5cm,height=4.2cm]{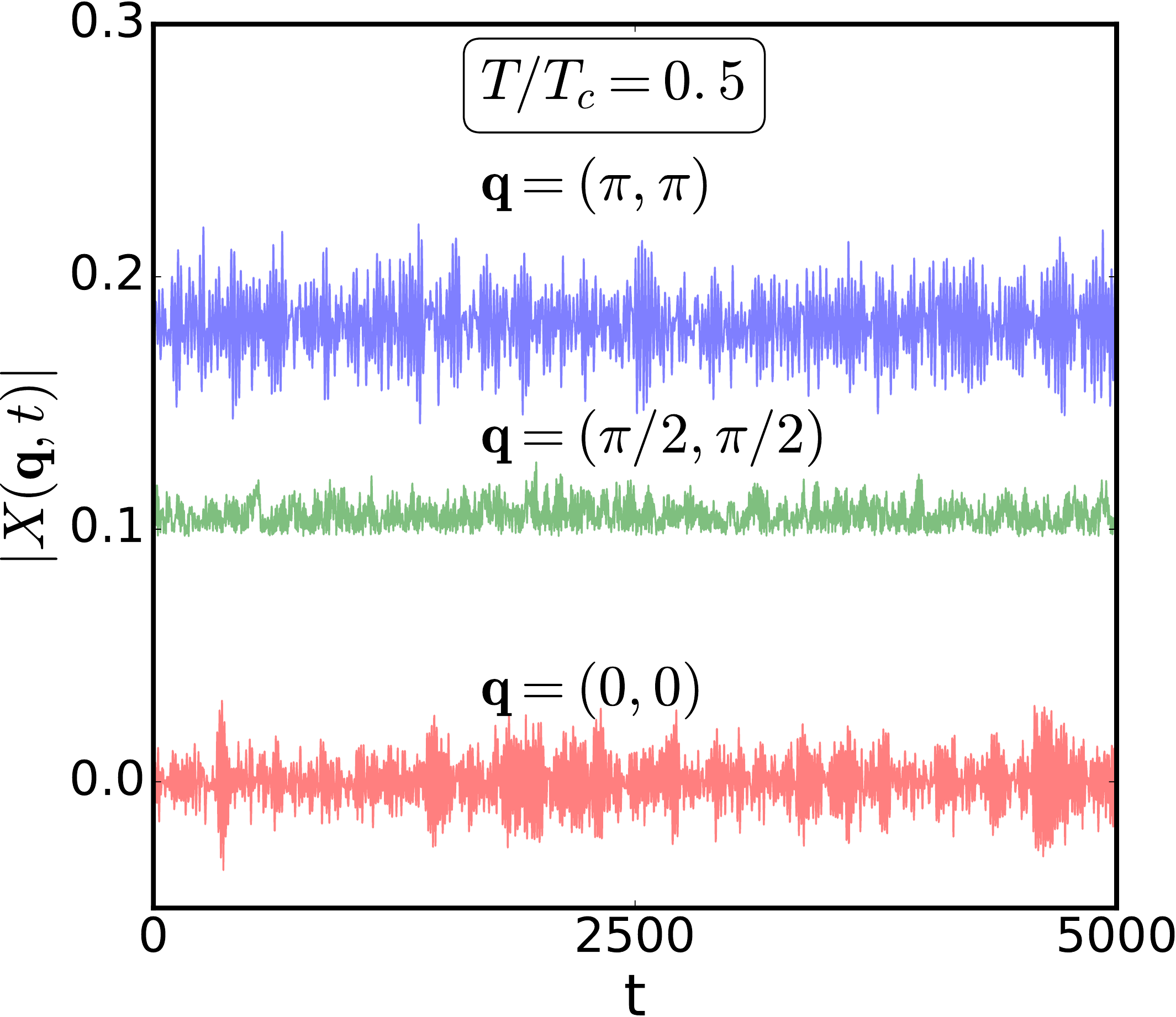}
}
\caption{Top panels: Trajectories of nearest neighbour sites for part
of the full time series. We see harmonic vibrations about equilibrium
positions in the left figure. The fluctuation window increases
considerably on heating ($0.5T_c$), featured in the right figure.
Bottom panels: Corresponding Fourier mode trajectories at the same
temperatures. The means are subtracted and trajectories shifted by a
constant in the left figure ($0.1T_c$). The effect of anharmonicity 
is to reduce the mean $X(\pi,\pi)$ and visibly enhance fluctuations in
both $(0,0)$ and $(\pi,\pi)$ modes.
}
\end{figure}

\begin{figure*}[t]
\centerline{
~~~
\includegraphics[width=17.5cm,height=4cm]{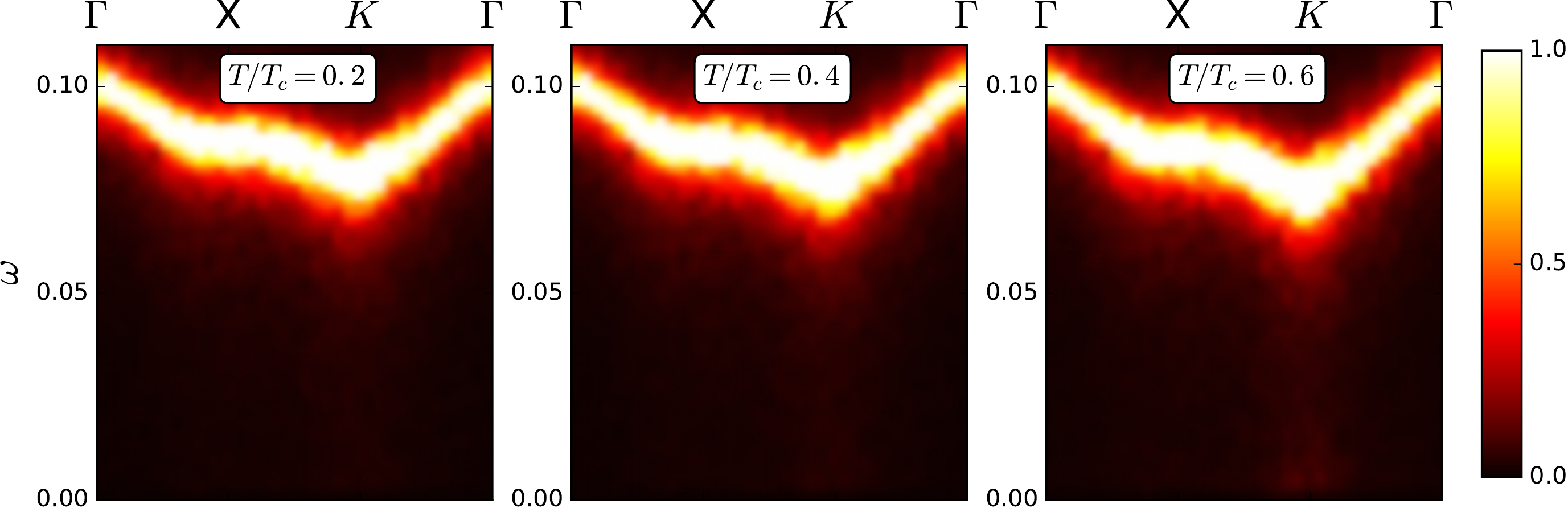}
}
\vspace{.3cm}
\centerline{
~
\includegraphics[width=5cm,height=4cms]{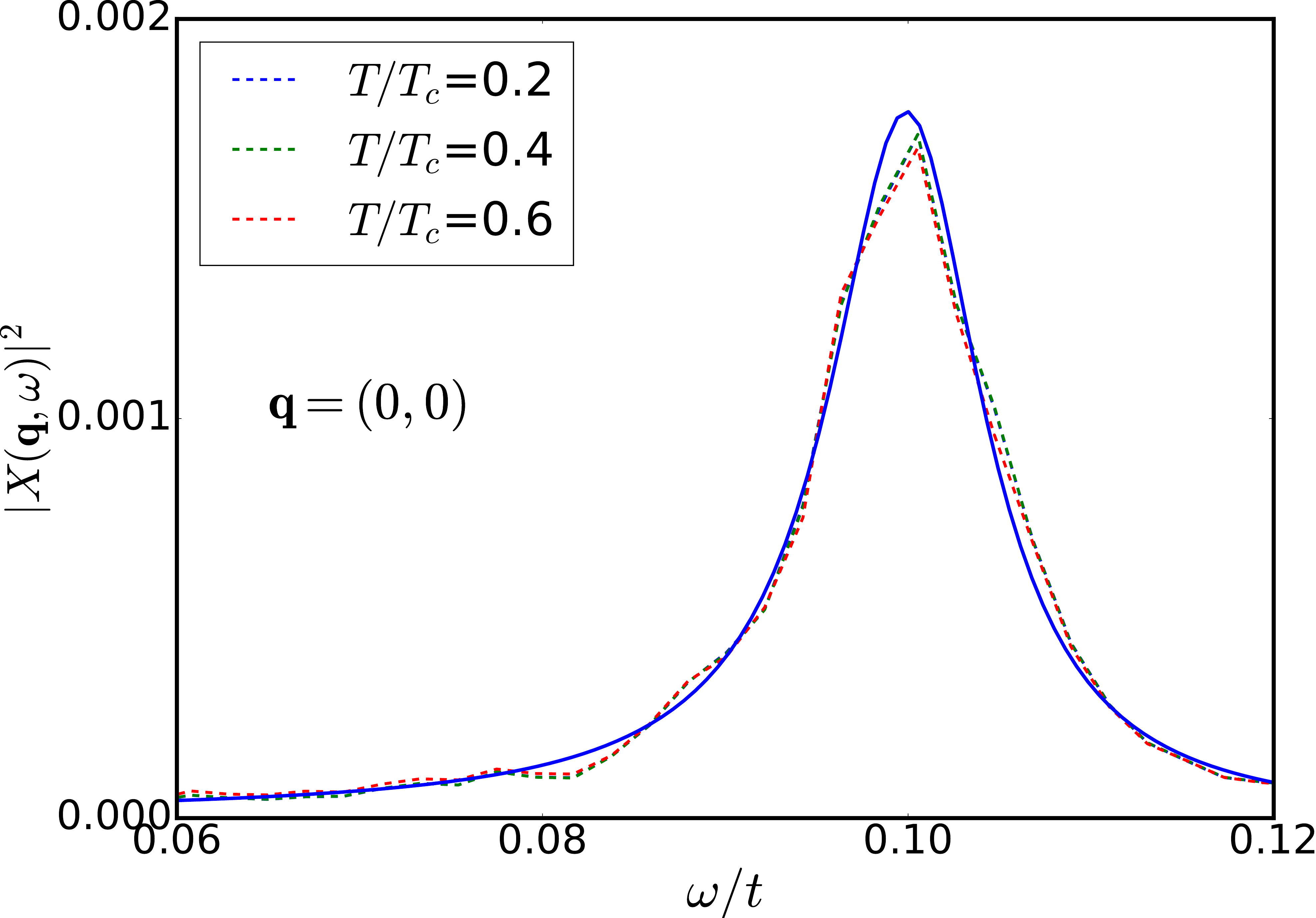}
\includegraphics[width=5cm,height=4cms]{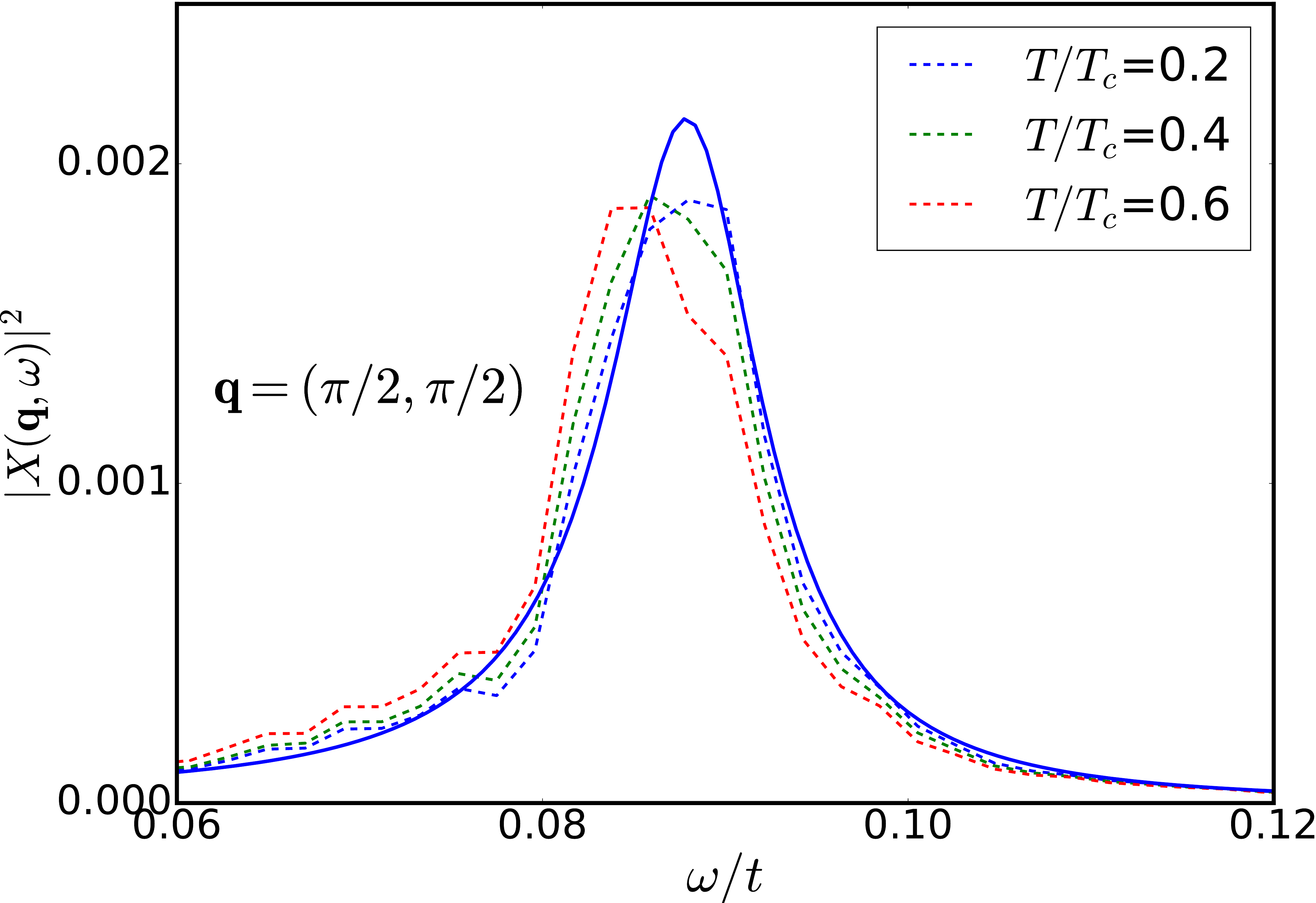}
\includegraphics[width=5cm,height=4cms]{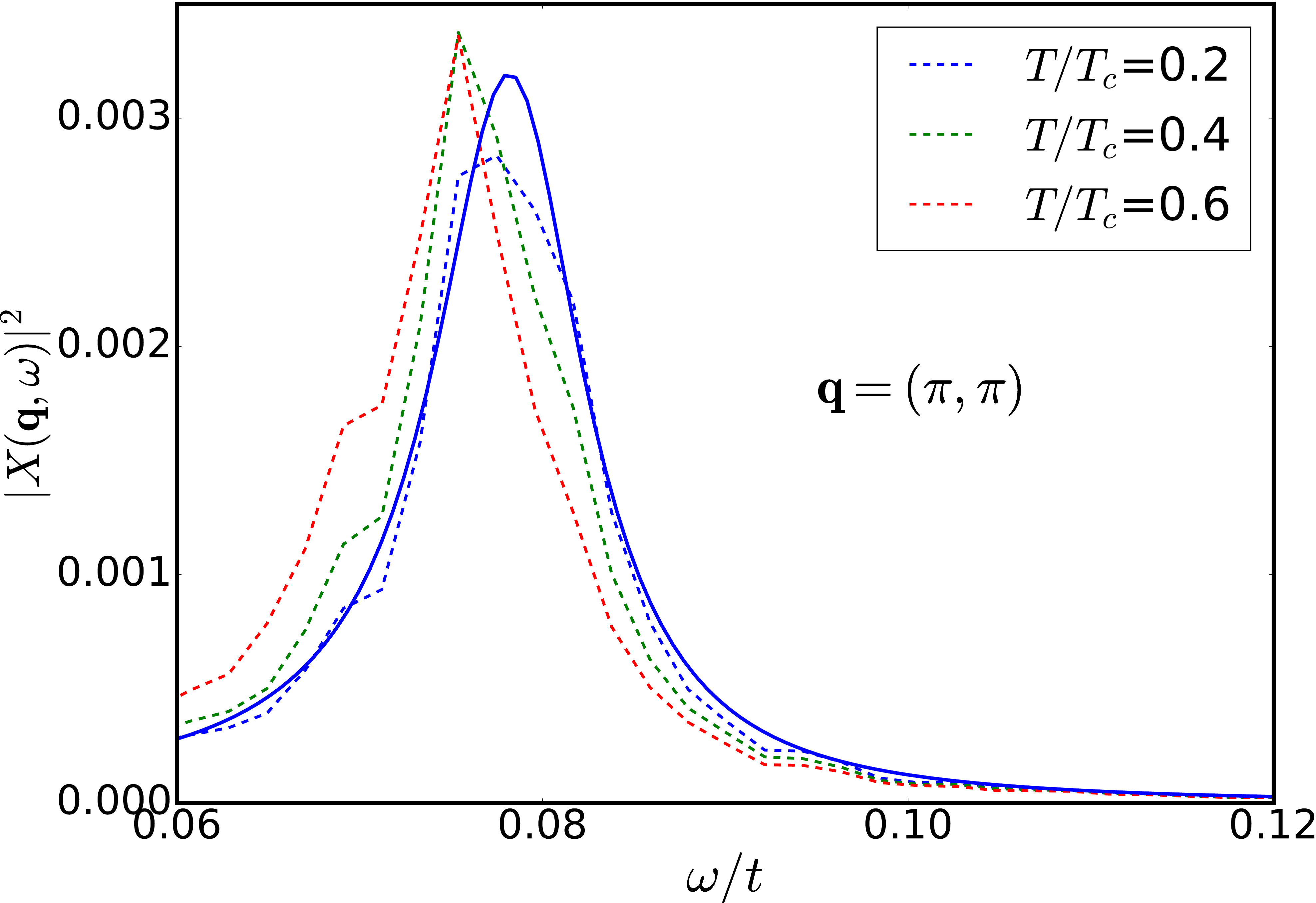}
~~~~~~~~~~
}
\caption{Top row: False color maps of the power
spectrum $\vert X({\bf q},\omega) \vert^2$ in the low temperature 
regime. The spectral intensities are plotted with the momentum
trajectory $(0,0)\rightarrow(\pi,0)
\rightarrow(\pi,\pi)\rightarrow(0,0)$ along the x-axis.
Bottom row: Lineshapes at corresponding temperature points for
three momentum points along the BZ diagonal: $(0,0)$,
$(\pi/2,\pi/2)$ and $(\pi,\pi)$. All power spectra are
normalized by $k_{B}T$.
}
\end{figure*}

\section{Real time dynamics}

We classify our results into four thermal regimes: 
(a)~low temperature, $T \lesssim 0.6T_c$, where the dynamics
is harmonic or mildly anharmonic, (b)~the `critical'
window, $0.6T_c \lesssim  T \lesssim 1.5T_c$, where thermal 
tunneling events dominate, (c)~the `polaron liquid' phase,
$1.5 T_c \lesssim T \lesssim 10T_c \sim T_P$,
where the distortions and density still have a bimodal character
but spatial correlations are only short range, and
(d)~the `polaron dissociated' phase, $T \gtrsim T_P$, where
we have essentially independent local oscillations.
For each of these regimes we typically show some trajectories for
the real space dynamics, time dependence of some Fourier modes,
the power spectrum, and sometimes damping and dispersion scales
extracted from the power spectrum. 

\subsection{Harmonic and weakly anharmonic regime: $T \ll T_c$}

\subsubsection{Real time trajectories}

Fig.3 shows the time dependence of phonon variable
both in real and momentum space 
at  $T = 0.1T_c$ and  $T = 0.5T_c$.
The top left panel shows the trajectories
$x_{\bf r}(t)$ at 
nearest neighbour sites, ${\bf r} = (12,12) \& (13,12)$
at  $T = 0.1T_c$, while the top right panel shows the
same at  $T = 0.5T_c$. 
The $0 < T < 0.6T_c$ window roughly
defines the `low temperature' regime as we discuss below.
The left panel shows small amplitude vibrations about
mean distortions $\sim0$ and $2$ respectively. The right
panel shows a 
qualitative increase in fluctuation amplitude, retaining similar
mean values. 

The bottom panel depicts trajectories $\vert X({\bf q},t) \vert$ 
at three momenta along the Brillouin Zone (BZ) diagonal.
The left panel is at  $T = 0.1T_c$, the right at  $T = 0.5T_c$.
In the left panel we subtracted the mean values and 
shifted the curves by a constant ($0.1$) to aid visualization.
The ${\bf q} = (0,0)$ and $(\pi,\pi)$ modes are seen to 
fluctuate more compared to $(\pi/2,\pi/2)$. In
the right panel the mean $\vert X(\pi,\pi,t) \vert$ reduces. 
Oscillations are most prominent at $(\pi,\pi)$.

\begin{figure}[b]
\centerline{
\includegraphics[width=5.5cm,height=4.5cm]{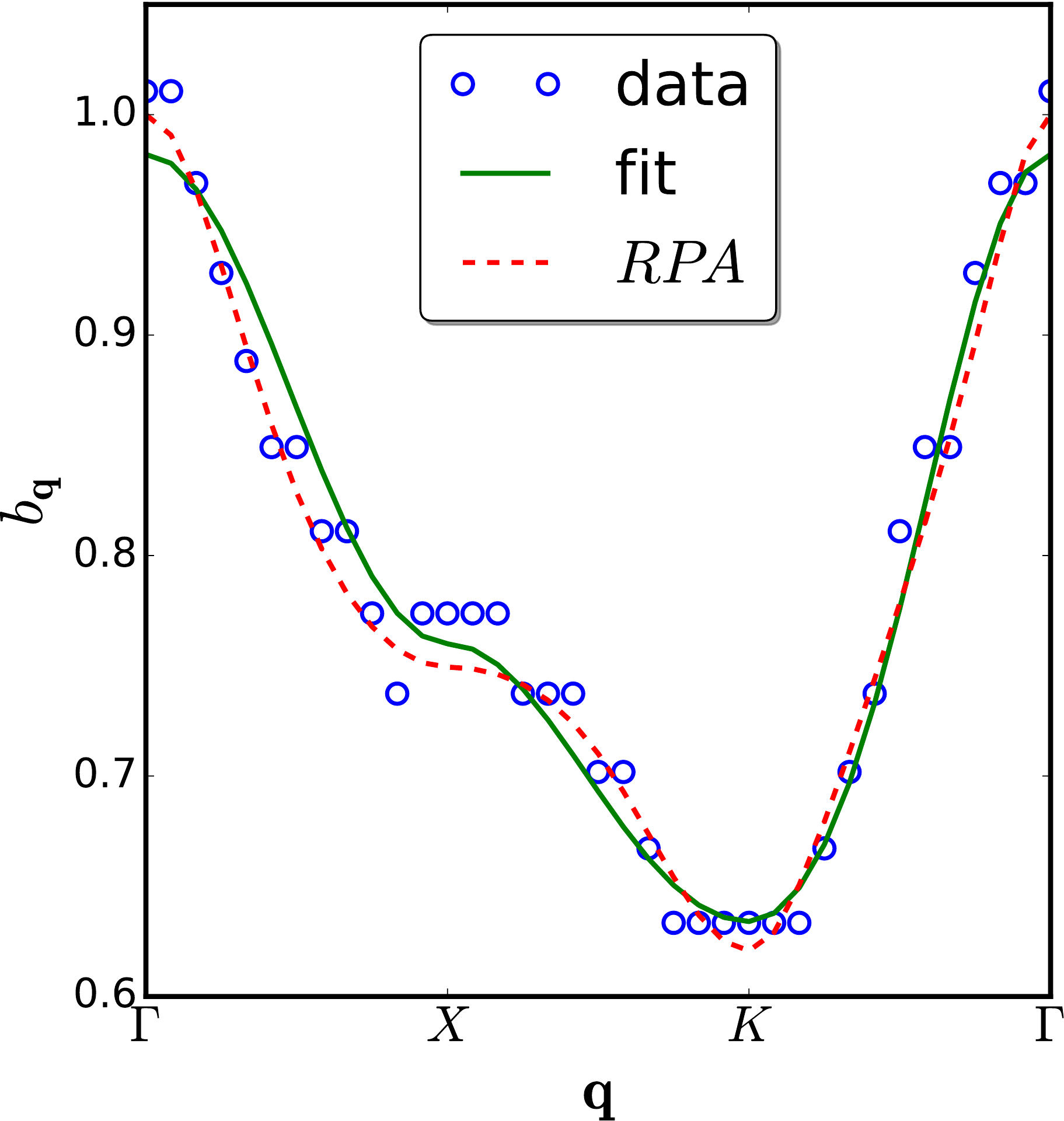}
}
\caption{Fitting the dispersion $b({\bf q})$ using harmonic functions
on the BZ at the lowest $T=0.1T_c$.
We find that a nearest neighbour model is reasonable accurate
in describing the dispersion of mode values. The blue open circles
are extracted from the data, green solid line is the best fit with free
parameters $K_{eff}$ and $J$ and the red dashed curve features RPA result
computed using the perfect ordered state at zero $T$.
The obtained fitting parameters are $K_{eff}=0.78$
and $J=0.09$, which denote the effective local stiffness and
the nearest neighbour intersite coupling respectively.
}
\end{figure}
\begin{figure}[t]
\centerline{
\includegraphics[width=4.5cm,height=4.5cm]{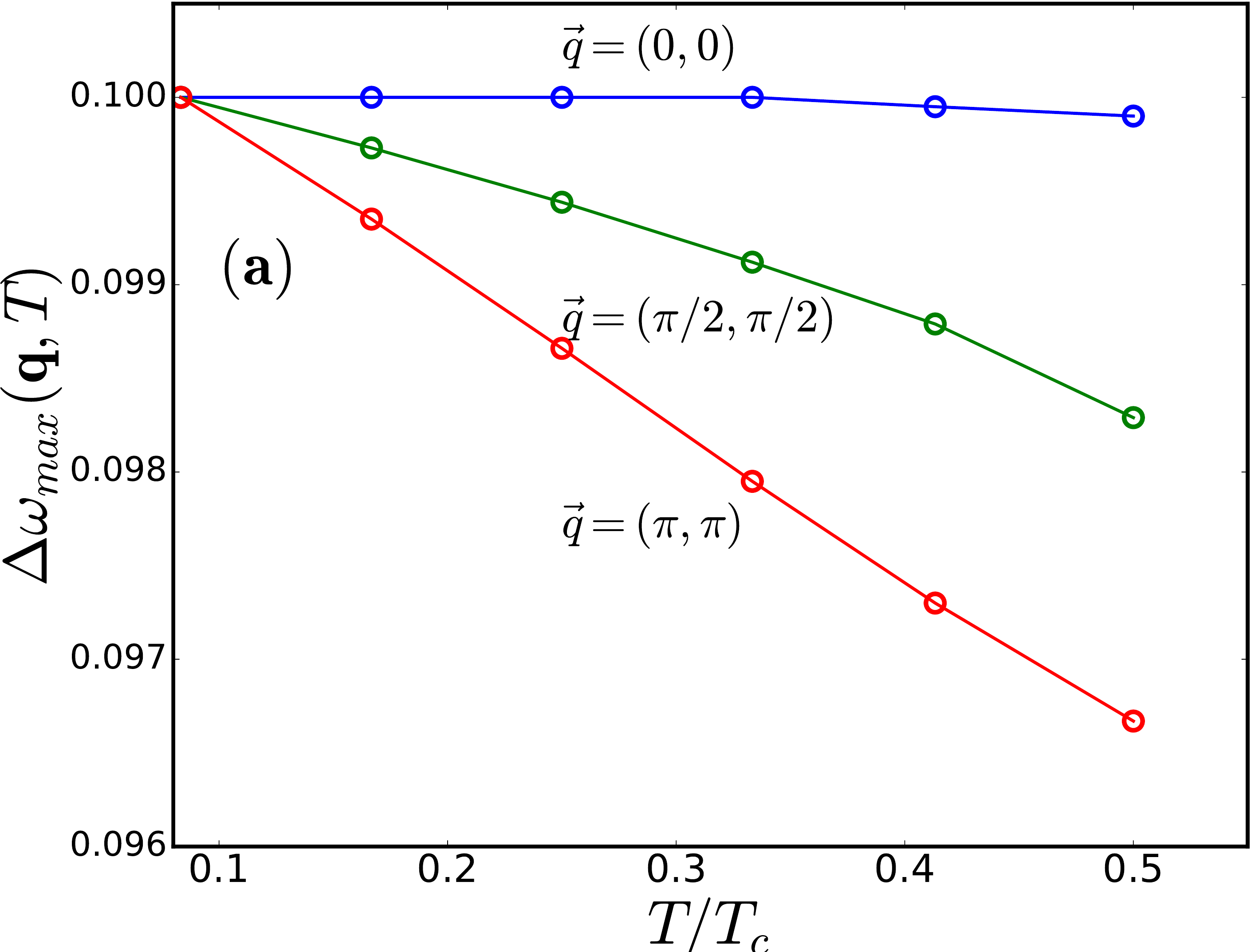}
\includegraphics[width=4.5cm,height=4.5cm]{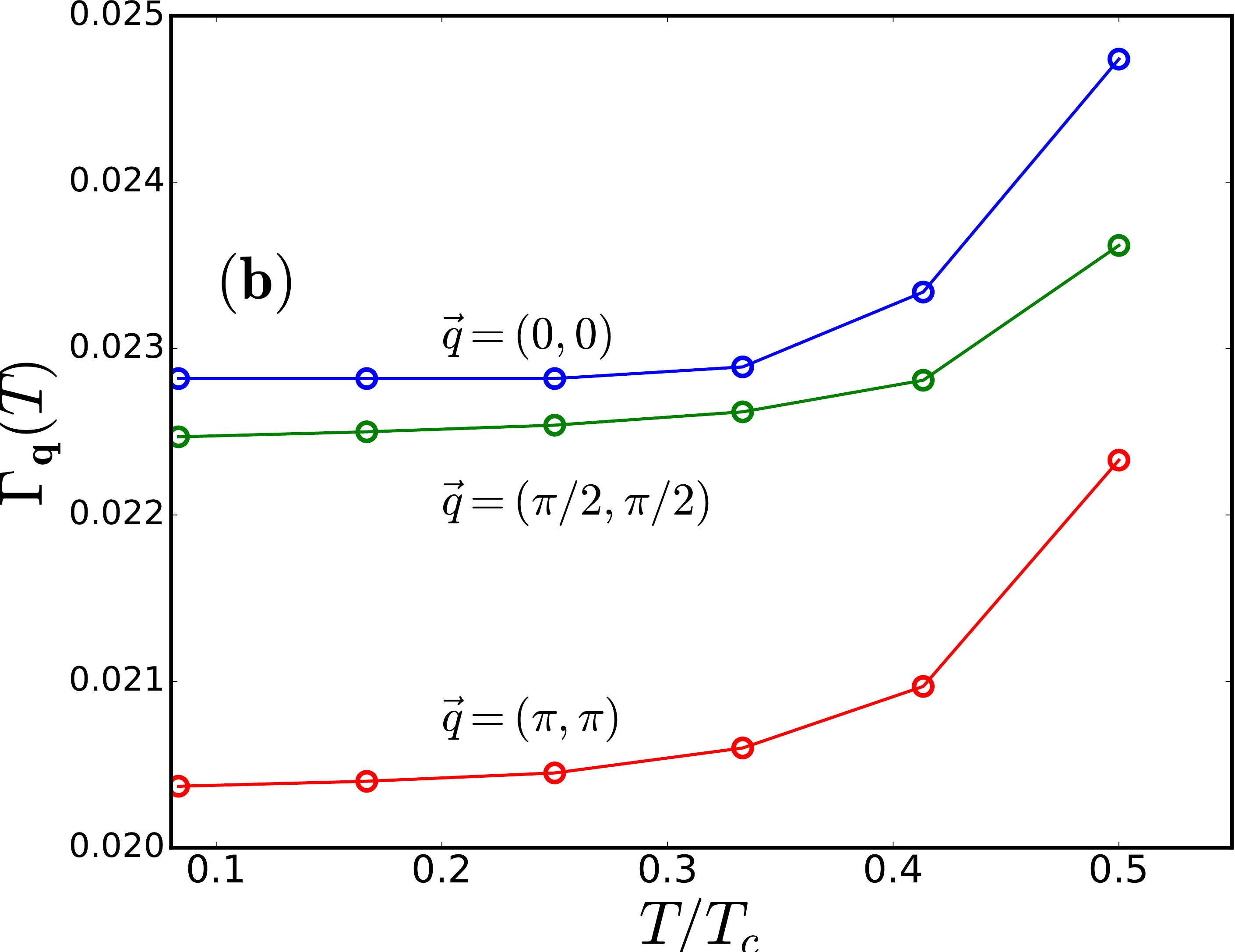}
}
\caption{Quantifying the effect of anharmonicity.
Left: $\Delta\omega_{max}({\bf q},T)$: 
the difference of peak locations from
that at $T=0$, and Right: the linewidth $\Gamma_{\bf q}(T)$.
Results are for three
characteristic momenta- $(0,0)$, $(\pi/2,\pi/2)$ and
$(\pi,\pi)$. We see linear softening
of mode values and an initially linear increase in damping for all
momenta. Non-linear corrections take over in the latter as one
raises the temperature.
}
\end{figure}

\subsubsection{Power spectrum}

We turn now to the description of dynamics in terms of 
$\vert X({\bf q},\omega)\vert^2$.
The top row of Fig.4 features the results in 
the low temperature regime. The maps are color coded in terms of 
varying intensity and the X-axis shows a momentum scan along the
trajectory $(0,0)
\rightarrow (\pi,0)\rightarrow (\pi,\pi)\rightarrow (0,0)$ in the 2D BZ.
We have divided out all  $\vert X({\bf q},\omega)\vert^2$ 
by $k_{B}T$.

In the bottom panel, we've plotted the lineshapes at three 
characteristic momenta ($(0,0),(\pi/2,\pi/2),(\pi,\pi)$ in terms
of their $T$ dependence. Here small changes are observed on 
increasing temperature as one goes from the harmonic to the 
anharmonic regime. The $(\pi,\pi)$ mode is most sensitive to this 
effect, whereas the lineshapes at BZ center don't respond appreciably.

The asymptotically low temperature regime is understandable in terms
of an effective harmonic model. The assumption is that the deviations of 
displacement fields from the checkerboard ordered ground state pattern
are small. The 
most general distortion cost one can write to quadratic order is:
$
V^{(2)}_{eff} \approx \sum_{i,j}b_{ij}\Delta x_i\Delta x_j
$,
where the $\Delta x_i = x_i - x_i^0$, with $x_i^0$ being the distortion
in the $T=0$ CO state.
The resulting Langevin equation is linear in $\Delta x_i$, but spatially
coupled, and can be solved by Fourier transformation. 
The $b_{ij}$ can be obtained from an expansion of the ground state
energy, $b_{\bf q}$ is its Fourier transform.
The power spectrum that emerges has  the form:
\begin{eqnarray}
\vert X ({\bf q}, \omega) \vert^2
& = & \frac{2\gamma k_{B}T}{\vert f({\bf q}, \omega) \vert^2} \cr
\cr
f({\bf q}, \omega) ~~& = & (-M\omega^2 + b_{\bf q} + i\gamma \omega)
\end{eqnarray}
The dispersion, found by plotting the peak locations 
at the lowest $T=0.1t$, has been fitted in Fig.5
to obtain the coefficients of 
nearest and next-nearest neighbour contributions in $b_{\bf q}$.
The fit parameters 
are $K_{eff}=0.78$ and $J=0.09$, respectively 
the local stiffness and the nearest neighbour coupling.
The further neighbour contributions are significantly smaller, owing
to the gap in density of states of the ordered state.
The dispersion compares well with a 
`random phase 
approximation' (RPA) calculation done on the $T=0$ CO background,
using
the static polarizability,  $\Pi({\bf q},\omega =0)$.

\begin{figure}[b]
\centerline{
\includegraphics[width=4.5cm,height=4cm]{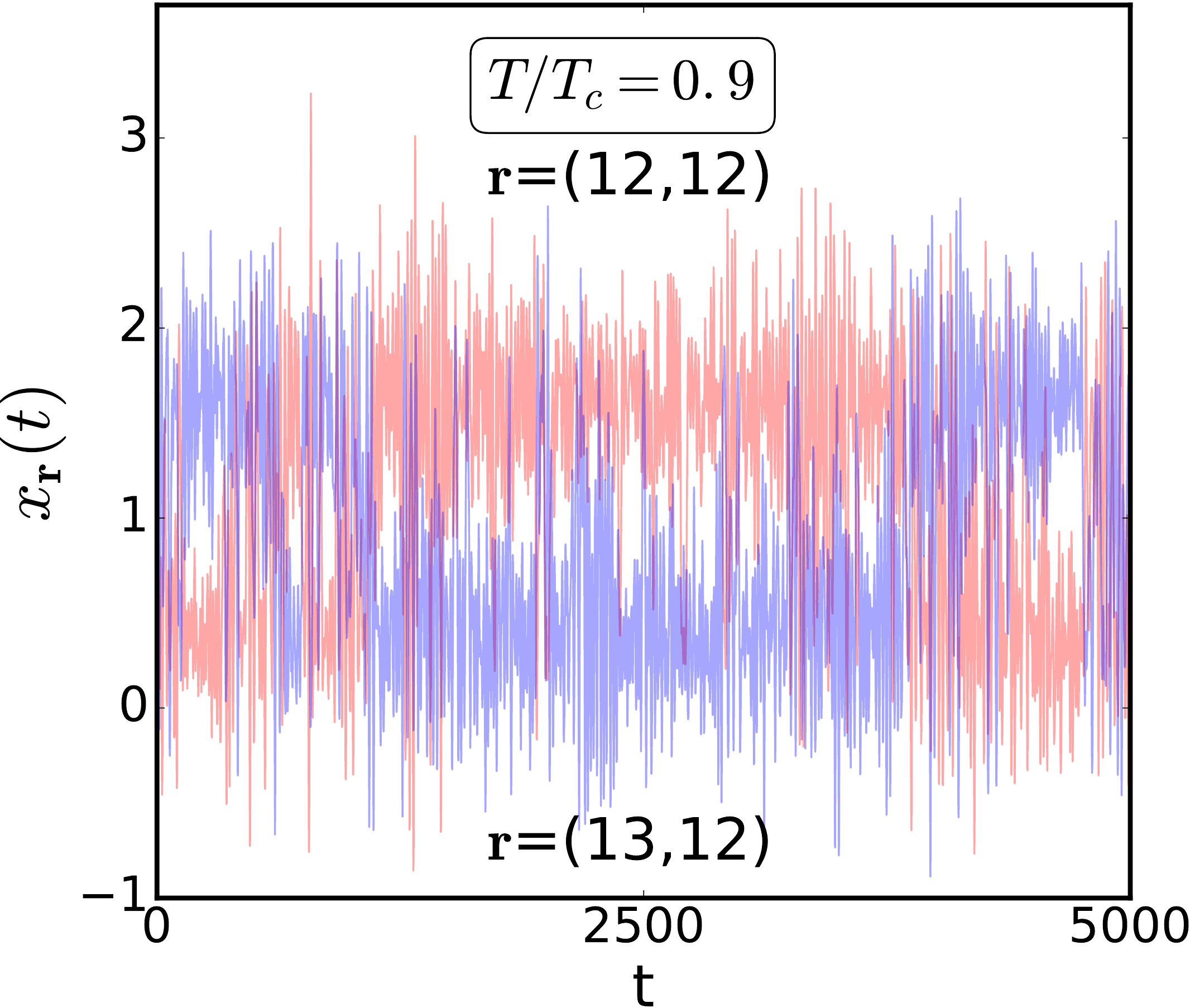}
\includegraphics[width=4.5cm,height=4cm]{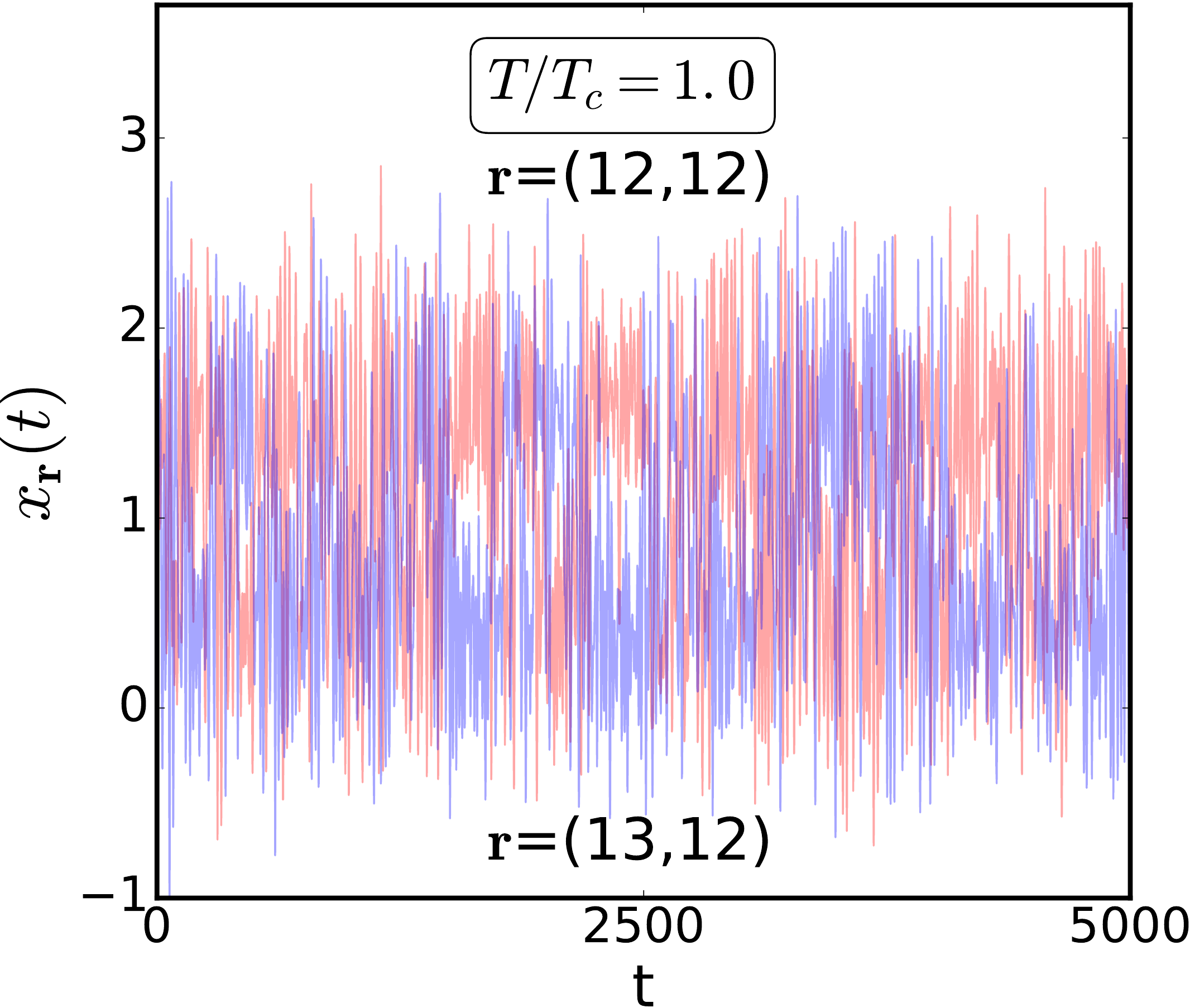}
}
\centerline{
\includegraphics[width=4.5cm,height=4cm]{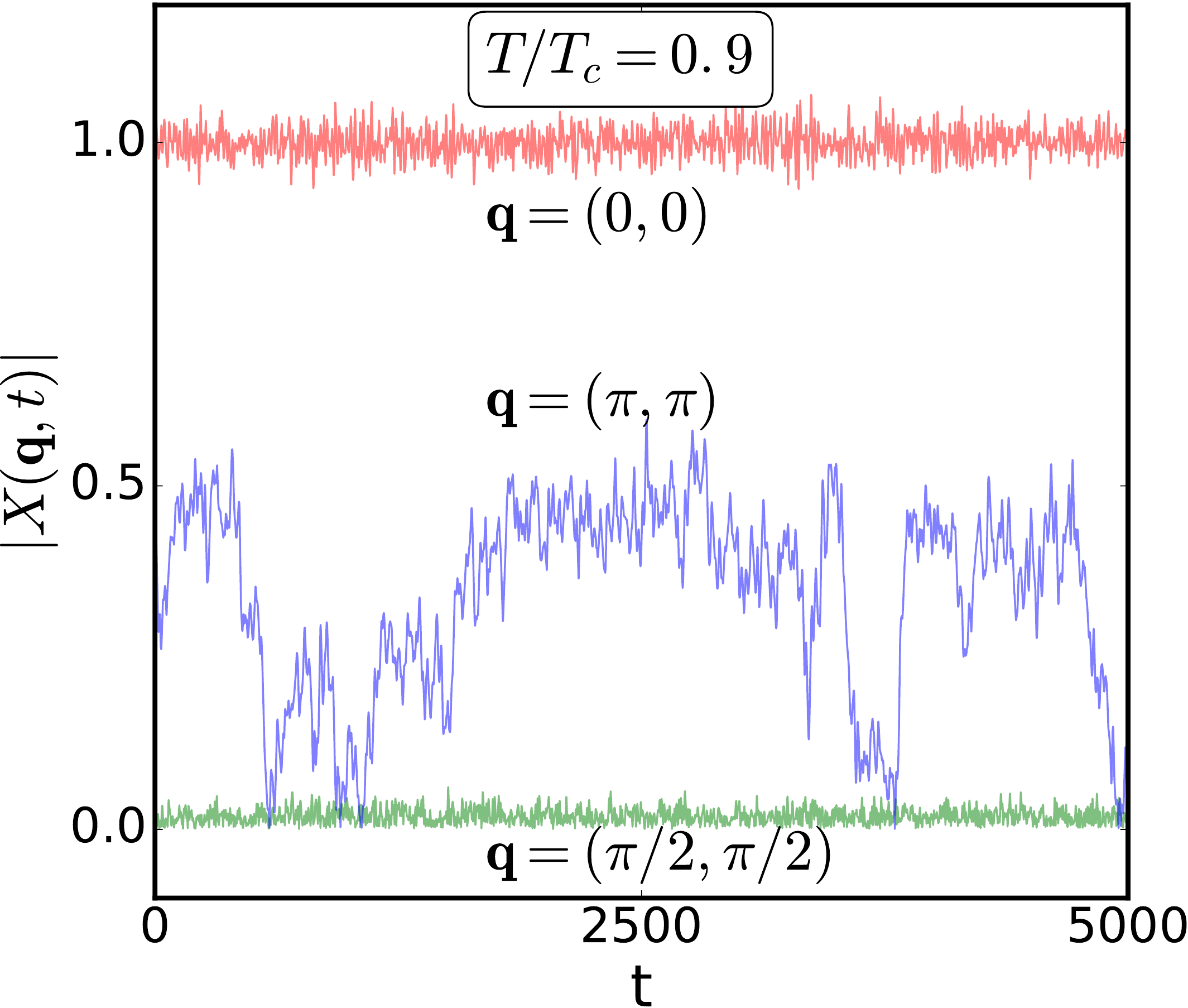}
\includegraphics[width=4.5cm,height=4cm]{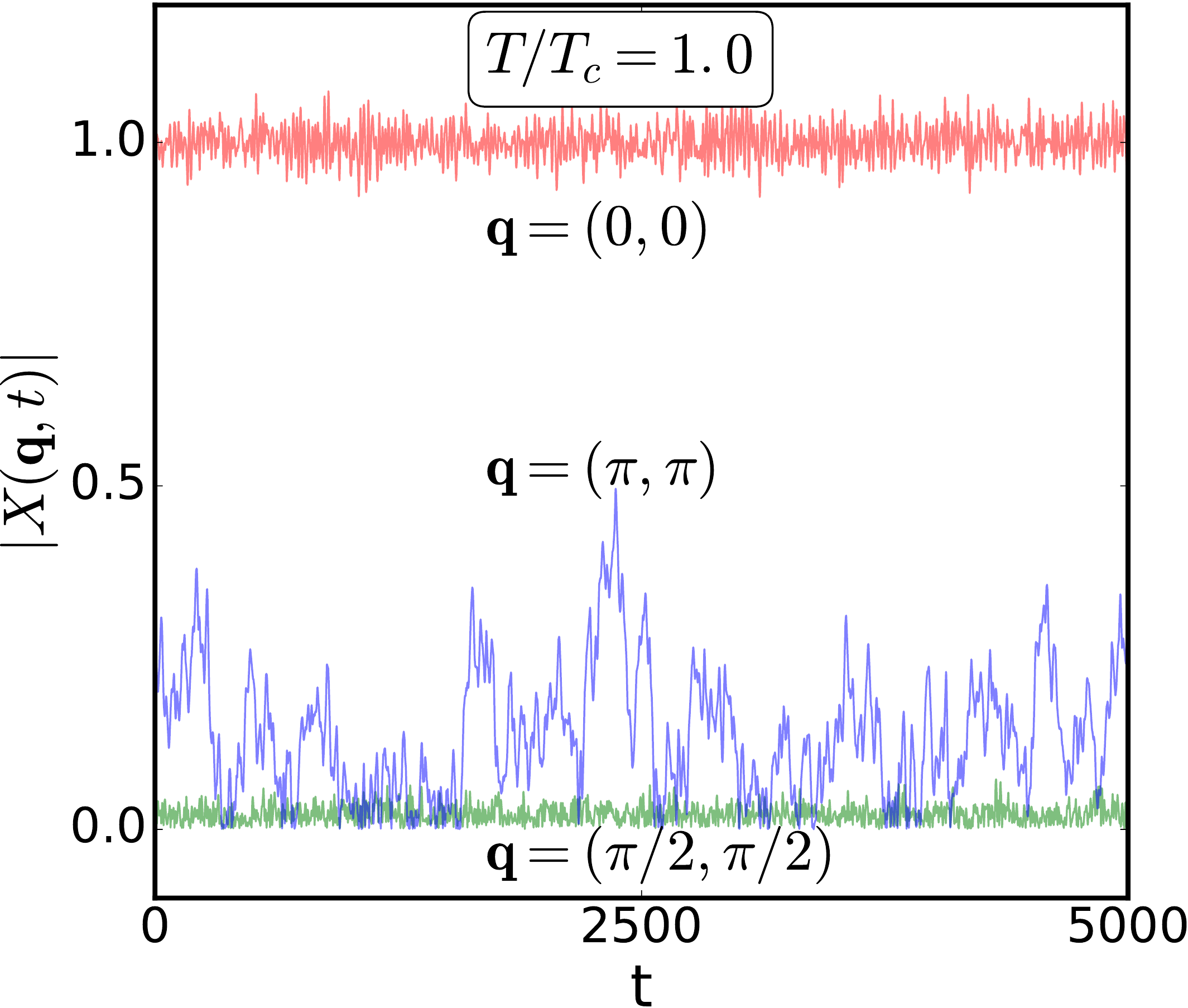}
}
\caption{Top panel: Trajectories of nearest neighbour sites for part
of the full time series. We see `resident flips' first
appearing in the
left figure ($0.9T_c$). The nearest neighbours switch in
mean values
and stay for tens of oscillation timescales. Moving
to the actual $T_c$, featured in the right figure, we see
more frequent exchange moves
that ultimately leads to vanishing of $S(\pi,\pi)$.
Bottom panel: Corresponding Fourier mode trajectories at the
same temperatures. The mean values are retained and actual
trajectories are
shown without shifting, as opposed to the low $T$ figure.
We see the special behaviour of the $(\pi,\pi)$ mode quite
clearly near criticality. Large oscillations feature in the
left figure while a reduction of mean value to near zero
is visible at $T_c$.
}
\end{figure}

\begin{figure*}[t]
\centerline{
~~~
\includegraphics[width=17.5cm,height=4cm]{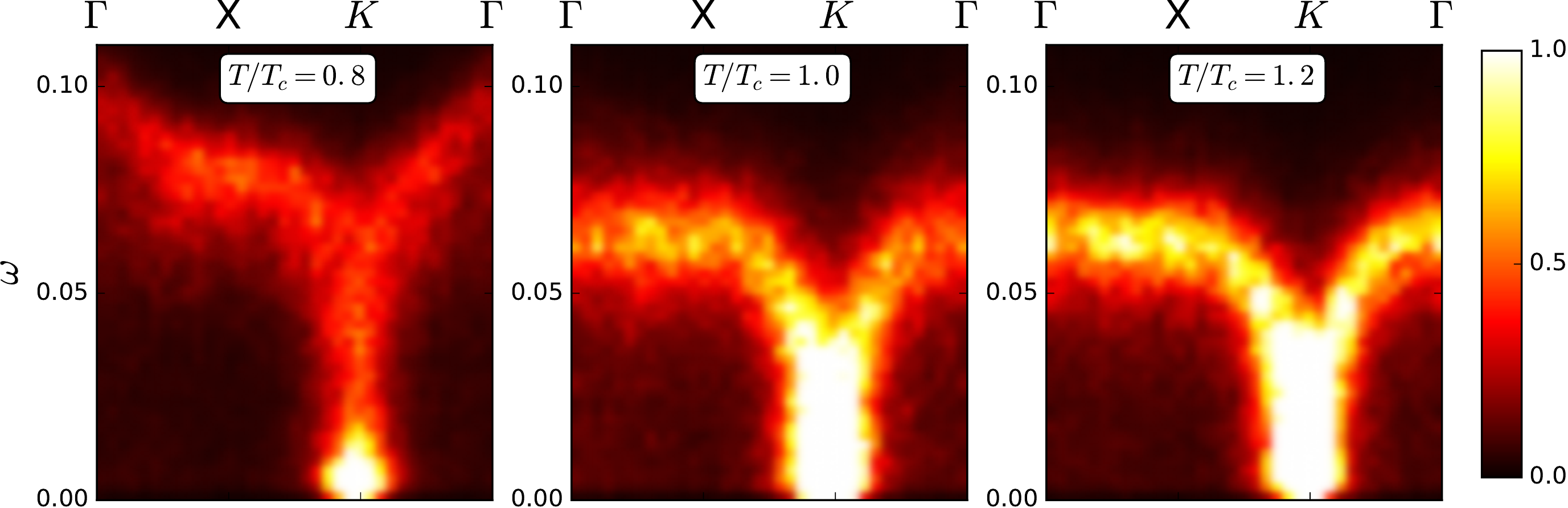}
}
\vspace{.3cm}
\centerline{
~~
\includegraphics[width=5.1cm,height=4.0cms]{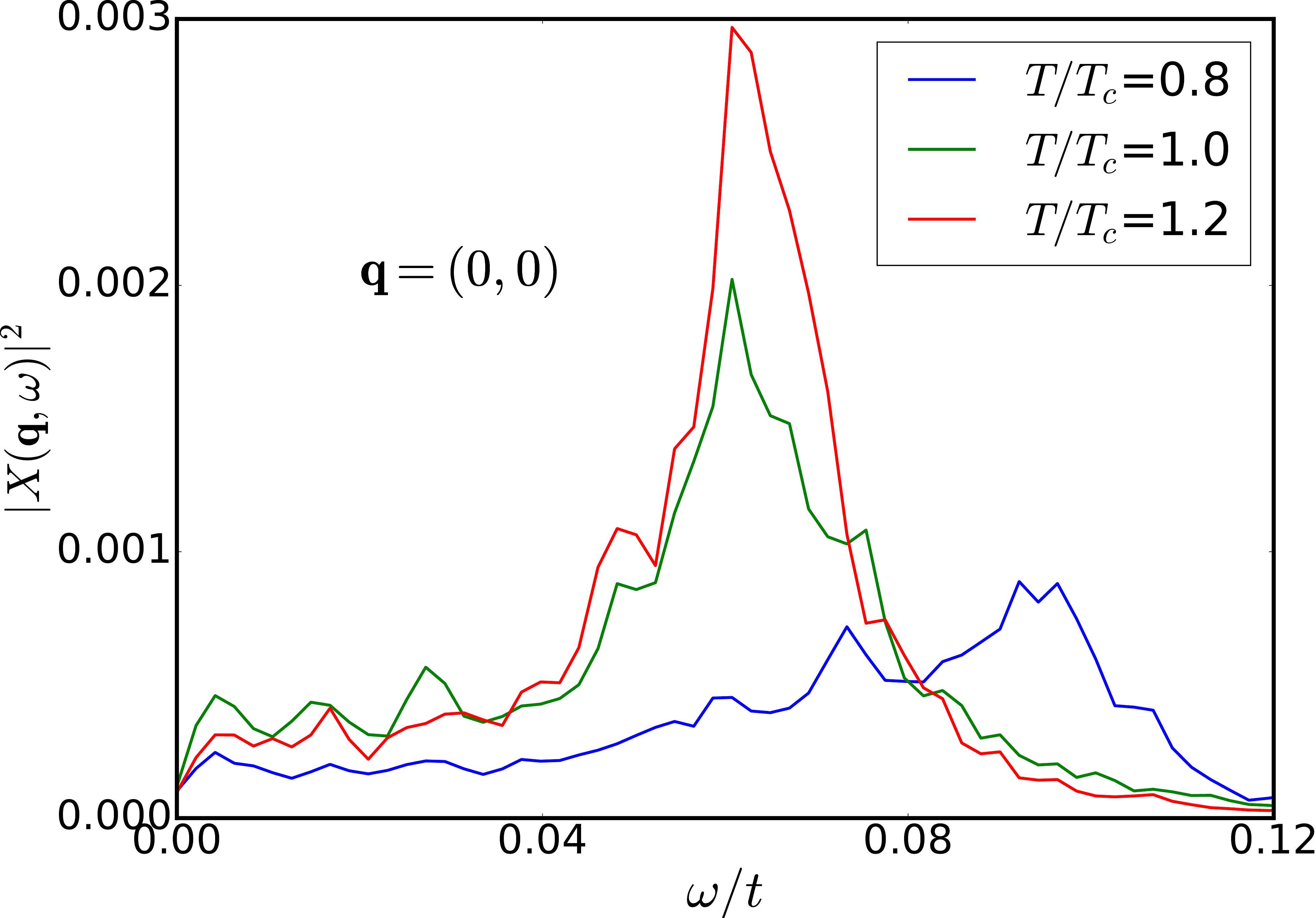}
\includegraphics[width=5.1cm,height=4.0cms]{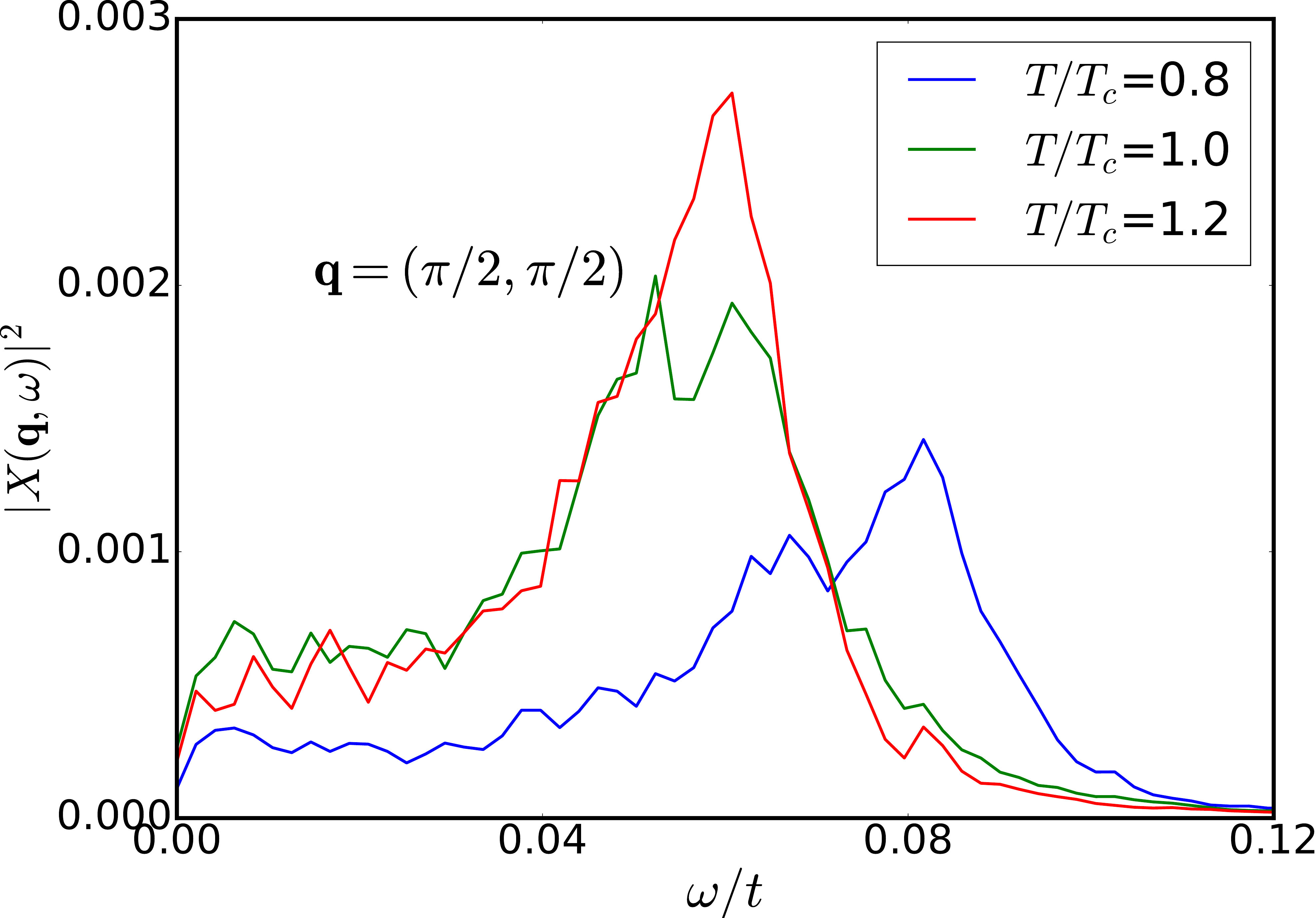}
\includegraphics[width=5.1cm,height=4.0cms]{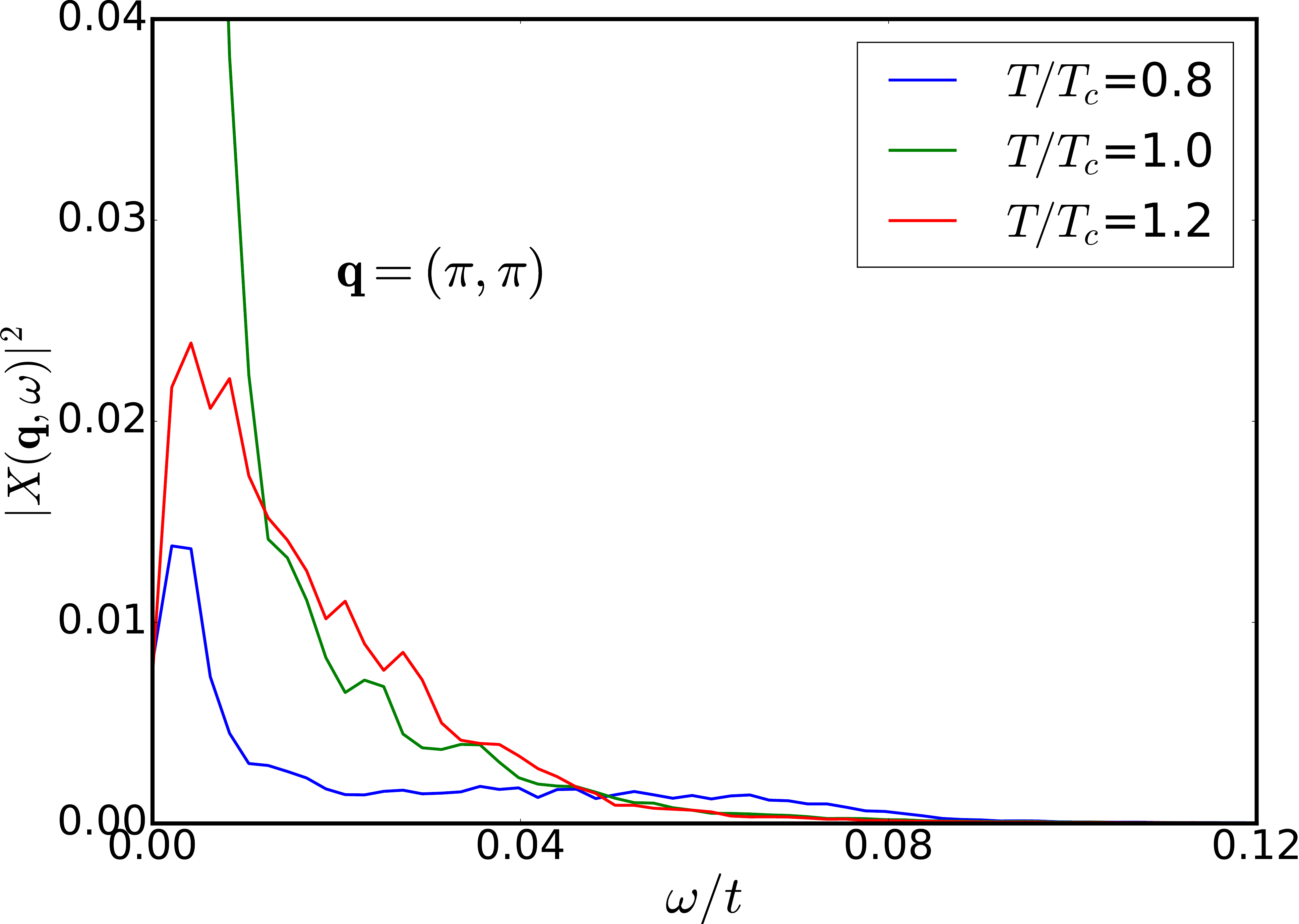}
~~~~~~~~~~
}
\caption{Top row: Maps of the power spectrum $\vert X({\bf q}, 
\omega)\vert^2$ in the vicinity of $T_c$.  The spectral 
intensities are plotted with the momentum trajectory $(0,0)
\rightarrow(\pi,0) \rightarrow(\pi,\pi)\rightarrow(0,0)$ along 
the x-axis.  Bottom row: Lineshapes at corresponding temperature 
points for three momentum points along the BZ diagonal:
$(0,0)$, $(\pi/2,\pi/2)$ and $(\pi,\pi)$. All power spectra
are normalized by $k_{B}T$.
}
\end{figure*}

The departure from harmonic behaviour with
increasing $T$ can be motivated by using a cubic term
in the energy, below. The energy 
cost with cubic terms has a form:
$
V^{(3)}_{eff} = \sum_{ij}b_{ij}\Delta x_{i} \Delta x_{j} + 
\sum_{ijk}c_{ijk} \Delta x_{i} \Delta x_{j} \Delta x_{k} 
$
Fourier transforming the Langevin equation leads to
$$
f({\bf q}, \omega)X({\bf q},\omega)  = 
\xi({\bf q},\omega) - \alpha({\bf q}, \omega) $$
where 
$ \alpha({\bf q}, \omega)  = 
\sum_{{\bf q}^{\prime},\omega^{\prime}}
C_{{\bf q},{\bf q^{\prime}}}
X({\bf q}-{\bf q}^{\prime},\omega-\omega^{\prime})
X({\bf q}^{\prime},\omega^{\prime})
$.
This can be dealt with
perturbatively by substituting the harmonic solution in
the nonlinear term.  
One formulates the perturbation expansion in terms of the response
function $\langle X({\bf q},\omega)\xi({\bf -q},-\omega) \rangle$. 
The Fourier
transformed variable $X({\bf q},\omega)$ has an expansion in powers
of the anharmonic coefficient. This has the first non-vanishing correction 
for the response
function at second order, as odd order correlators of the noise vanish
by symmetry.

Before averaging over the noise variable, one has all kinds of diagrams
for the response function with cubic interaction vertices and
`dangling legs' of the noise. After taking 
the noise average, these legs connect up and one
family of diagrams may be identified as the `RPA series', which has repeated
`bubbles' arising as corrections to the free propagator.
One can then selectively resum 
these contributions. 
The effective self-energy 
$\Sigma({\bf q},\omega)$ that emerges 
has the form:
$$
\Sigma({\bf q},\omega)=
\langle \sum_{{\bf q}^{\prime},\omega^{\prime}}
\frac{\xi({\bf q}- {\bf q}^{\prime},\omega-\omega^{\prime})
\xi({\bf q}^{\prime},\omega^{\prime})}
{f({\bf q}- {\bf q}^{\prime},\omega-\omega^{\prime})
f({\bf q}^{\prime},\omega^{\prime})}
\rangle
$$
where angular brackets denote averaging over the noise variable.
This is
an $O(T)$ quantity at this level. The real and imaginary parts
of this corrects the pole location and damping respectively, 
with the corrections varying linearly with $T$ for a fixed $\bf {q}$. 

Fig.6
indicates that the frequency shift and increase in broadening
indeed have a leading linear $T$ behaviour at low temperature.
The damping features non-linear corrections in $T$ 
as one heats up, which
arise from `stray flips' (SF) which are large, isolated exchange of 
displacements on the lattice. The slope of the linear part is of course
${\bf q}$ dependent and the nature of this dependence is monotonic
along the BZ diaginal. 

\begin{figure}[b]
\centerline{
\includegraphics[width=4.5cm,height=4.5cm]{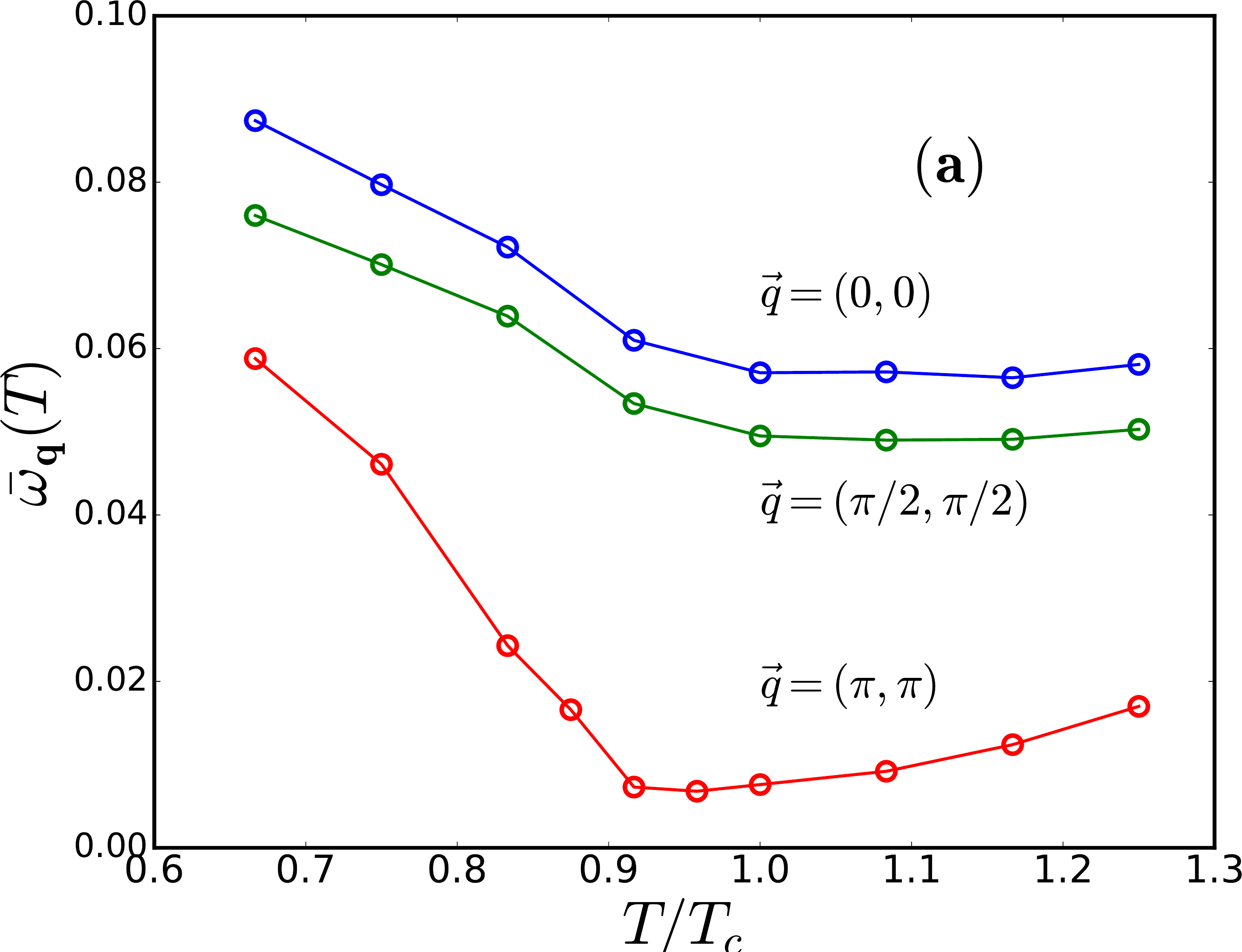}
\includegraphics[width=4.5cm,height=4.5cm]{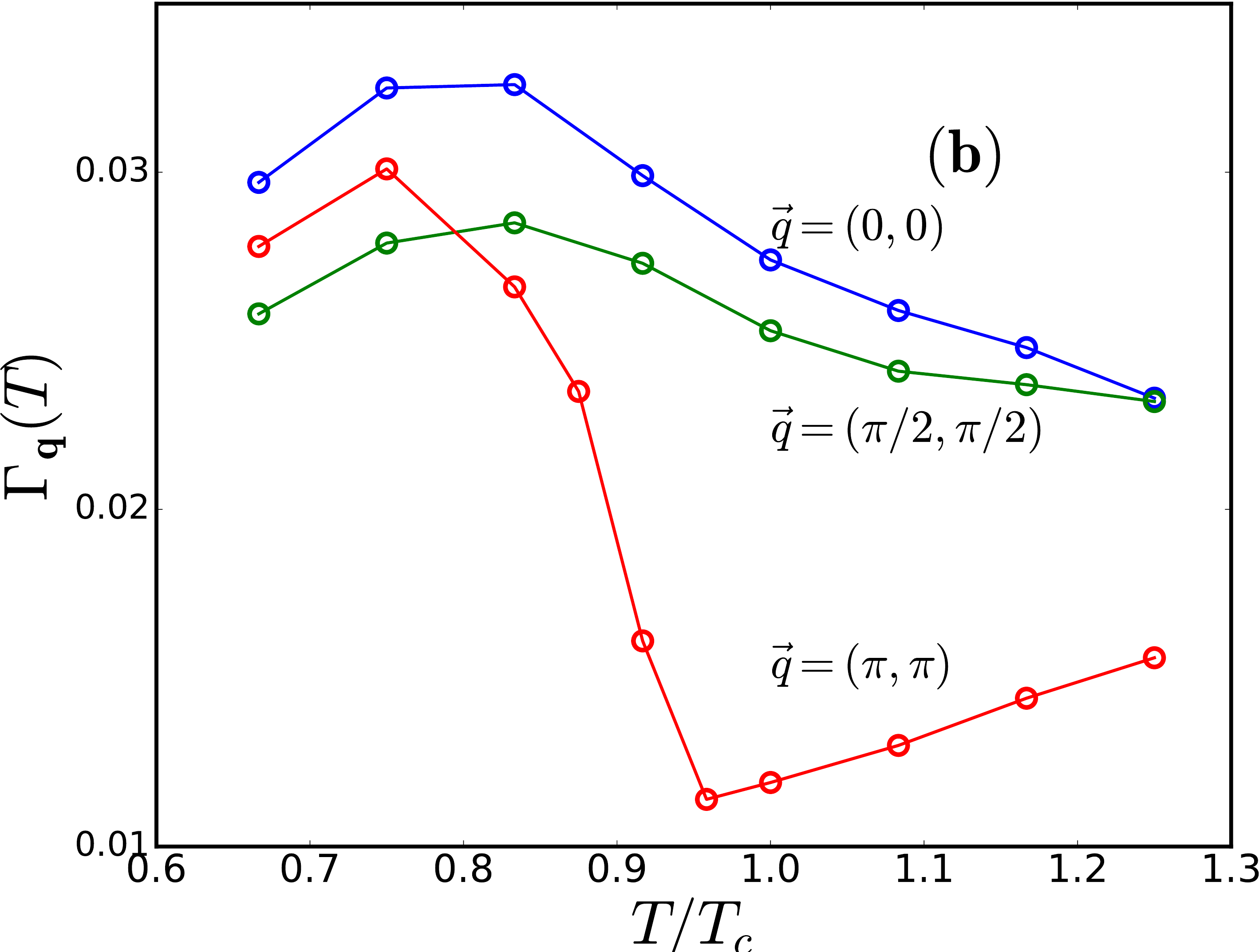}
}
\caption{
Fits to extract the mean frequencies ($\bar{\omega}$) and
standard deviations
($\Delta \omega$) as functions of temperature.
The left panel features mean curves, showing an overall
softening trend that's most prominent at $(\pi,\pi)$.
The right panel features damping rates, where a
non-monotonicity is seen in the thermal behaviour. At
$T_c$, the $(\pi,\pi)$ mode is again sharp.
}
\end{figure}
\begin{figure*}[t]
\centerline{
~~~~~
\includegraphics[width=17.5cm,height=3.4cm]{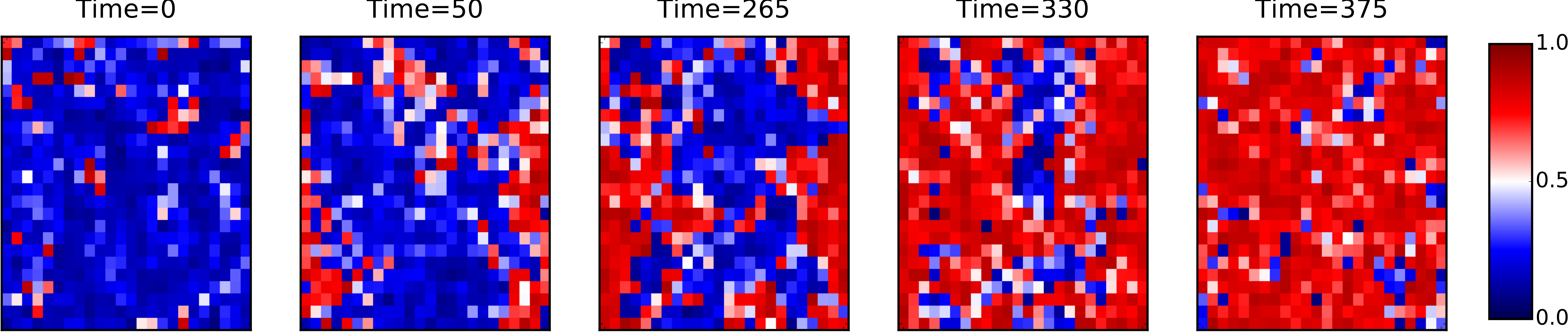}
}
\vspace{.3cm}
\centerline{
\includegraphics[width=17.2cm,height=3.0cm]{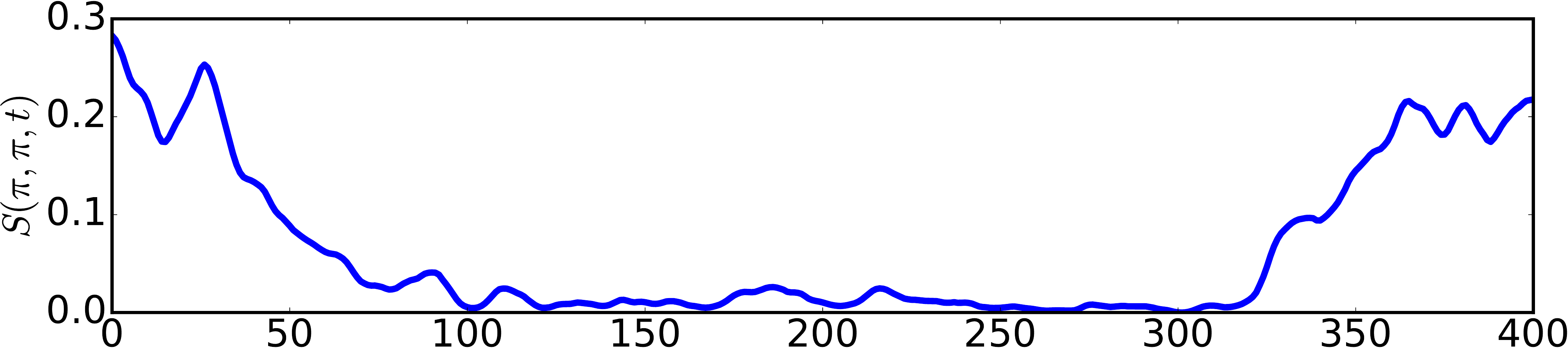}
~~~~~~~~~~~~~
}
\vspace{.1cm}
\caption{Top panel: Growth of a large domain is shown through taking
the actual configurations and subtracting out the prevalent checkerboard
pattern $C$. Bluish regions indicate perfect order alla $C$ and red regions
denote opposite pattern of checkerboard ordering $C^{\prime}$. The white
areas are indicative of boundaries between these two patterns. Time is
measured in units of $\tau_{0}$, the bare oscillation timescale.
Bottom panel: The instantaneous structure factor $S(\pi,\pi,t)$ shown
across the time interval of domain growth. Dominant order alla $C$ or
$C^{\prime}$ result in increased value, while in the middle panel we observe
a near zero value due to almost equal size of the two domains.
        }
\end{figure*}

\subsection{Critical fluctuations: $T \sim T_c$}

\subsubsection{Real time trajectories}

The trajectories in both real and momentum space for this regime are
shown in Fig.7. The top row depicts nearest neighbour displacement
time series. New events show up compared to low $T$, in the form of
`exchange moves' between large and small distortions. The left figure
($T/T_c=0.9$) shows a clean version of what may be called
a `resident flip' (RF), where the exchanged distortions don't flip
back within a short ($\sim5-10\tau_{0}$) timescale.
Moving closer to the actual $T_c$, their frequency increases. These
events have associated spatial correlations as well, which trigger
domain growth and shrinkage, as discussed later. Ultimately, these
lead to destruction of order as a whole.

The bottom row of Fig.7 shows trajectories of different momentum
modes. This time the mean values are shown without shifts. As before,
we observe the benign nature of $(0,0)$ and $(\pi/2,\pi/2)$ modes. The
$(\pi,\pi)$ mode, however, features large oscillations just before $T_c$, 
whose mean value drops rapidly on approaching criticality.
The dynamical event responsible for this is domain growths,
taking place over the full lattice. We will discuss this later.

\subsubsection{Power spectrum}

Moving to the frequency dependent indicators, far richer behaviour 
is seen here compared to low temperatures. The top row of Fig.8 depicts
spectral maps of $\vert X({\bf q},\omega)\vert^{2}$. 
In the leftmost figure
($T=0.7T_c$), a faint tail is seen emerging near $(\pi,\pi)$. 
A more complete weight transfer is observed in the middle panel 
($T=0.8T_c$). This originates from `exchange moves' discussed 
before in the context of trajectories. They have an 
associated timescale that is much larger
($\sim100$ times) than the bare oscillation period. 

Low energy spectral weight arises from these moves.
If one tracks the trajectories long enough, these should be 
present at even lower $T$. However, there these events are 
too rare to have any appreciable impact on the frequency response 
of the system as a whole.
Stray flips (SF) start occuring around $0.6T_c$, whereas 
RF's discussed before only show up around $0.9T_c$. The spectrum
at criticality (right panel) has a different dispersive character
compared to lower $T$ and is considerably more broad around
$(\pi,\pi)$. 

The lineshapes at three characteristic momenta $(0,0)$, 
$(\pi/2,\pi/2)$ and $(\pi,\pi)$ features in the bottom row of Fig.8. A
gradual spectral weight transfer to lower frequencies is observed in the
first two on heating close to $T_c$, whereas a dramatic near-zero frequency
weight develops at $(\pi,\pi)$. The intensity also, even after being
divided out by a factor of $k_{B}T$, is $\sim 10$ times here at $T_c$.

In Fig.9, we've shown mean frequencies (left panel) and standard deviations
(right panel) across the chosen trajectory in BZ in this regime. The softening
at the BZ boundary is quite prominent on approaching $T_c$. The overall 
branch also changes its character compared to low $T$. In the dampings, 
a non-monotonic trend is seen, also most prominent at $(\pi,\pi)$. The
width of the spectrum at this momentum is actually resolution limited, 
rather than $\gamma$ limited. 
Hence, this is an universal feature, irrespective
of microscopic details.

\subsubsection{Domain dynamics near $T_c$}

The ground state of the present problem is $(\pi,\pi)$ ordered, which
corresponds to a checkerboard pattern. 
However, there exists two energetically degenerate patterns, related to 
each other by a $Z_{2}$ transformation.
We call them C and C'. 
During the time evolution there are two kinds of forces 
acting on the each lattice site. 
One is the systematic part of the force ($g\bar{n}_{i}-Kx_i$). This depends 
on the $\{ x_i \}$ configuration at the given instant. 
The other is the thermal noise. 
At zero temperature, if we create a small defect in the lattice 
(a small domain of C in C' or vice versa), it exterts a force on its 
neighbours, trying to convert them into C and hence growing the domain. 
These events are rare 
at low temperature. 
\begin{figure}[b]
\centerline{
\includegraphics[width=4.5cm,height=4cm]{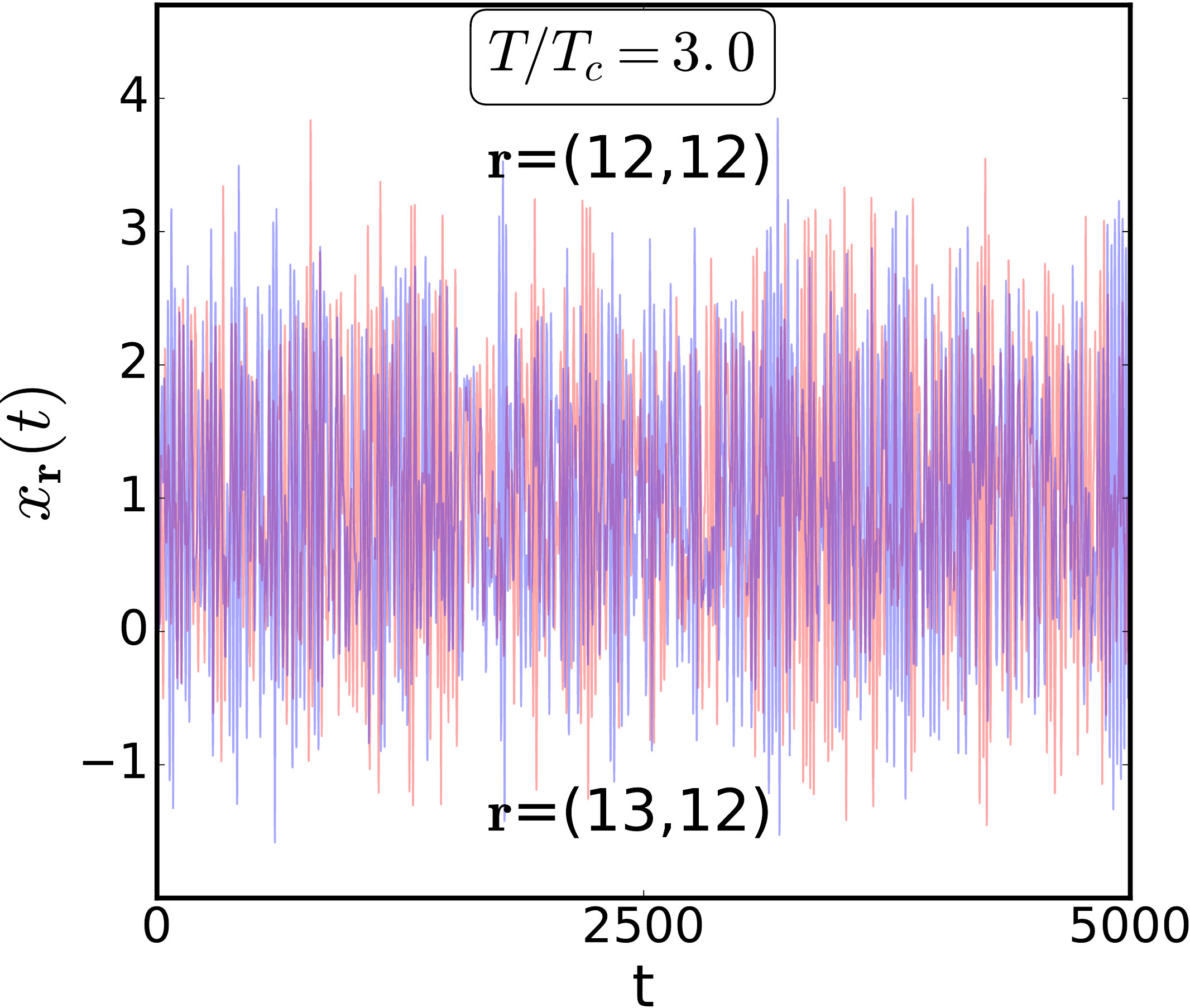}
\includegraphics[width=4.5cm,height=4cm]{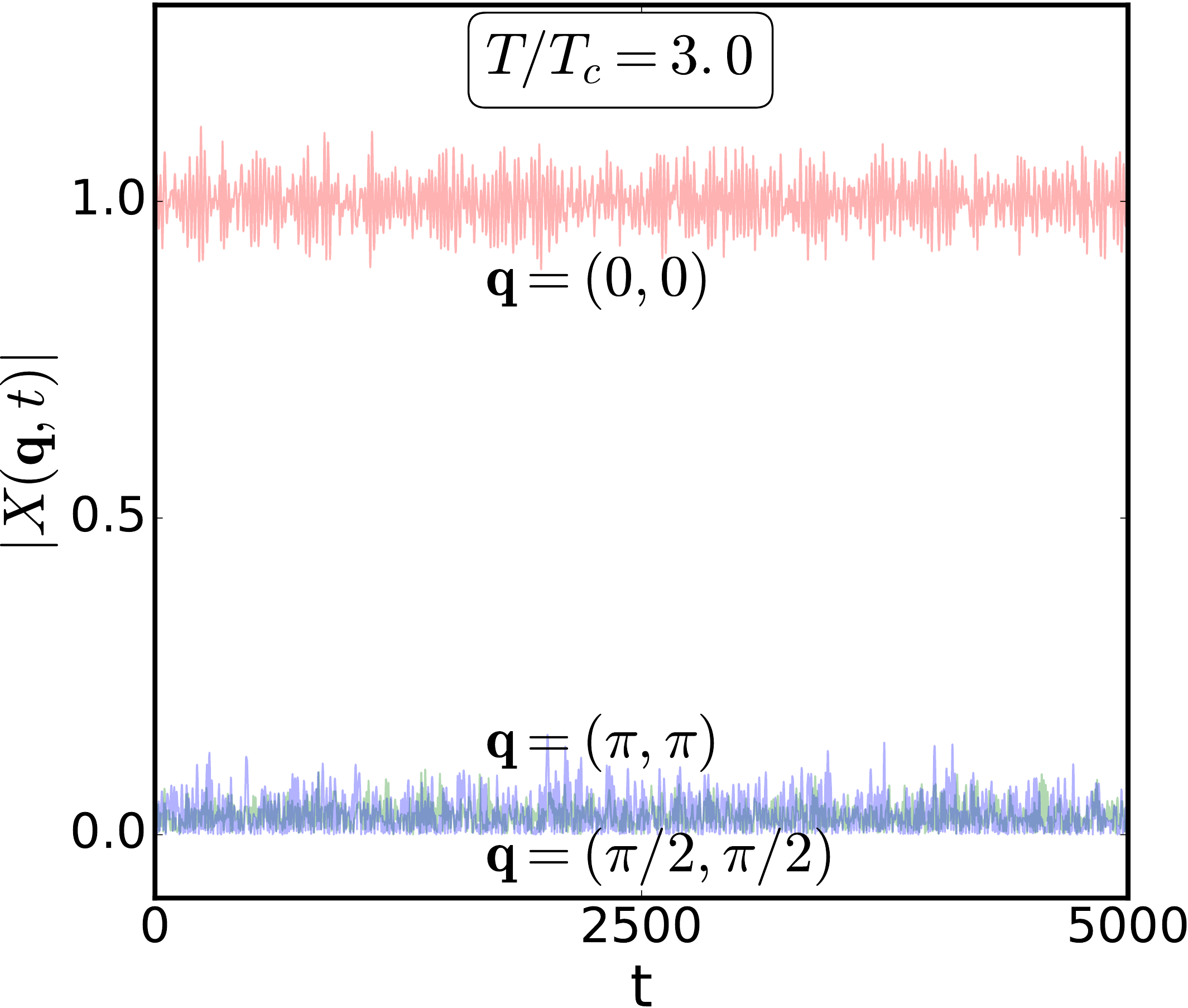}
}
\caption{Left: Trajectories of nearest neighbour sites for part
of the full time series. We see flip moves merging with large
amplitude oscillations. This leads to broad but unimodal
power spectra in this regime.
Right: Corresponding Fourier mode trajectories at the same
temperatures. The mean values are retained and actual trajectories are
shown without shifting, as opposed to the low $T$ figure. Both the
$(\pi/2,\pi/2)$ and $(\pi,\pi)$ trajectories oscillate above zero,
indicating loss of spatial correlations in the dynamics.
}
\end{figure}
\begin{figure*}[t]
\centerline{ ~~~
\includegraphics[width=17.5cm,height=4cm]{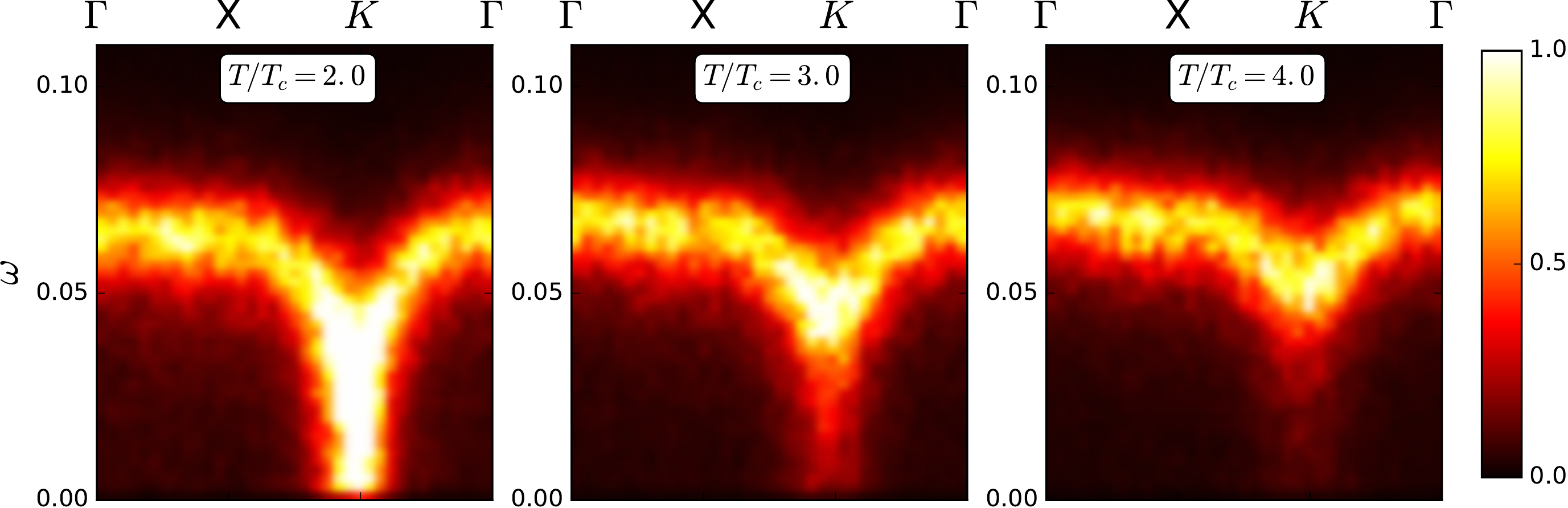}
}
\vspace{.3cm}
\centerline{
\includegraphics[width=5.1cm,height=4cms]{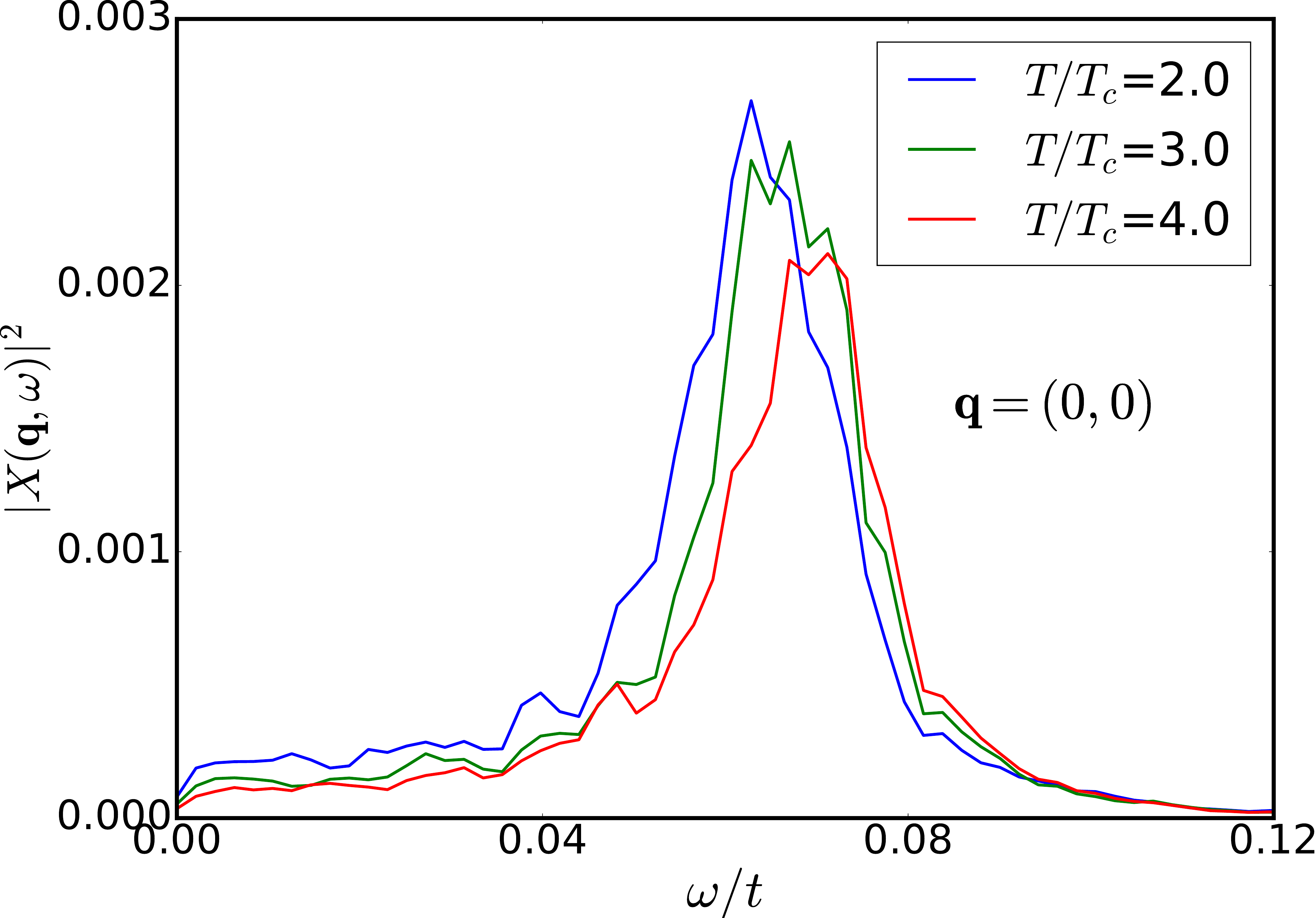}
\includegraphics[width=5.1cm,height=4cms]{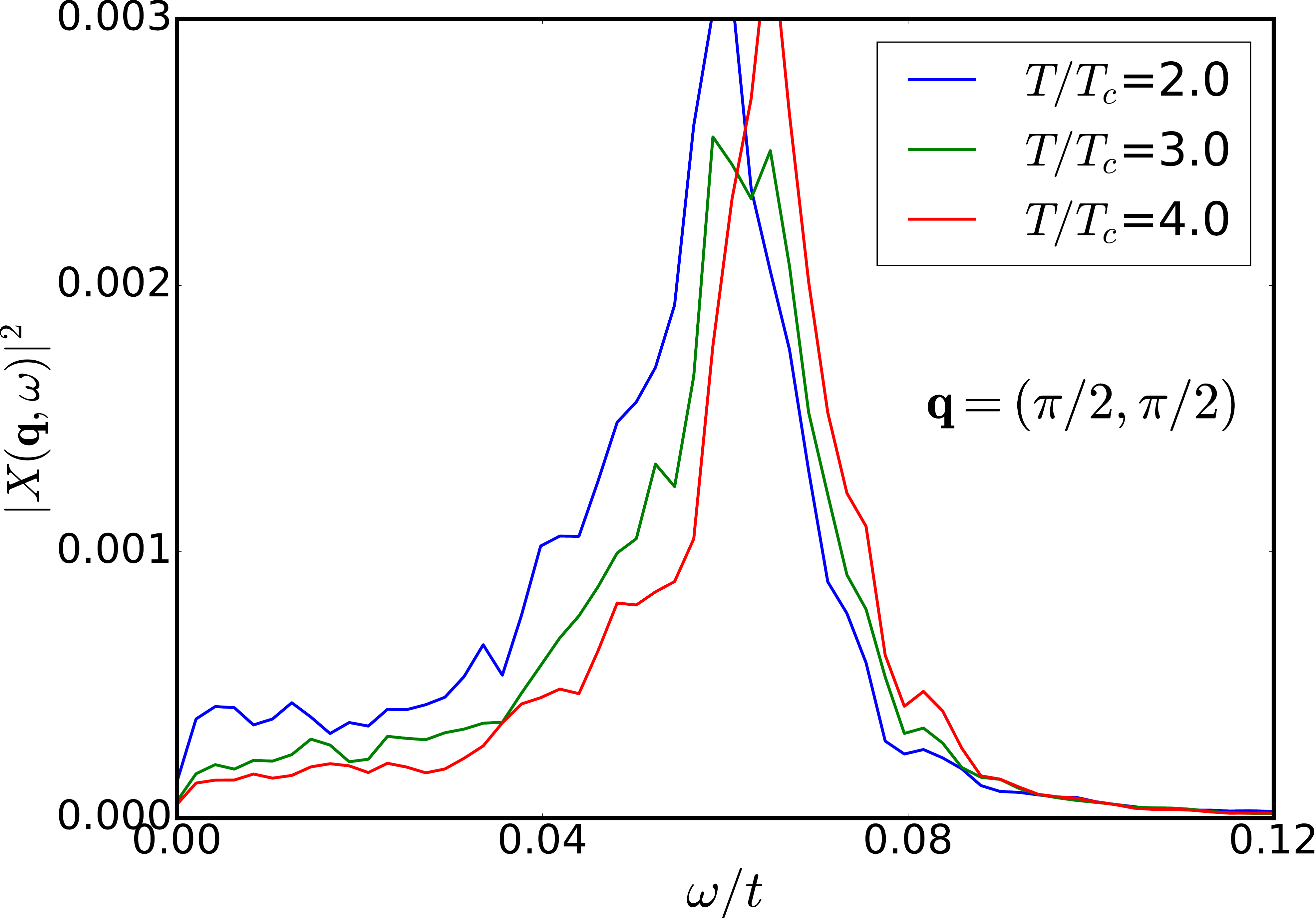}
\includegraphics[width=5.1cm,height=4cms]{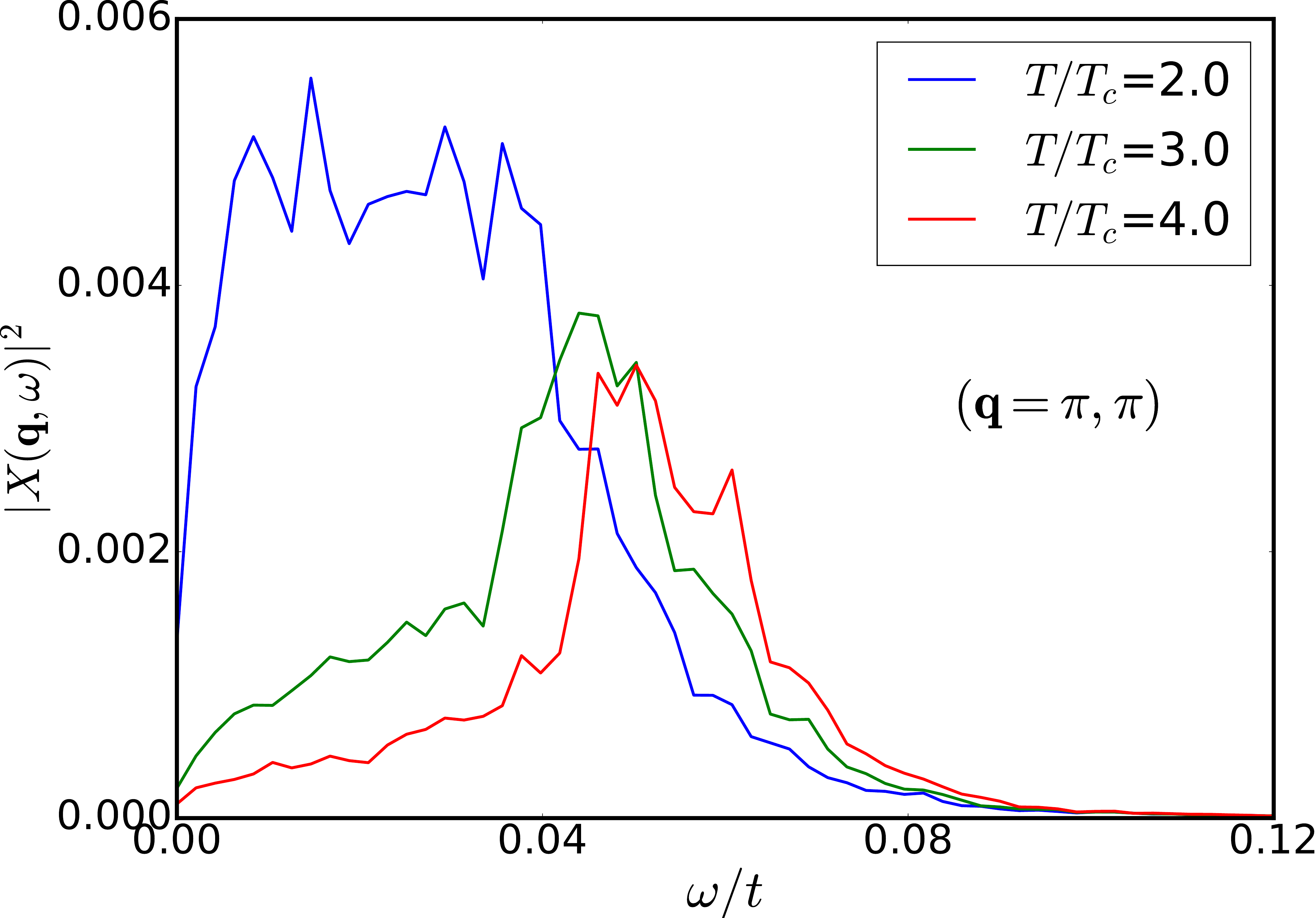}
~~~~~ }
\caption{Top row: False color maps of the power
spectrum $\vert X({\bf q},\omega) \vert^2$ in the disordered polaron
liquid regime. The spectral intensities are plotted with the momentum
trajectory $(0,0)\rightarrow(\pi,0)
\rightarrow(\pi,\pi)\rightarrow(0,0)$ along the x-axis.
Bottom row: Lineshapes at corresponding temperature points for three
momentum points along the BZ diagonal: $(0,0)$, $(\pi/2,\pi/2)$ and
$(\pi,\pi)$. All power spectra are normalized by $k_{B}T$.  }
\end{figure*}

At temperatures $\sim0.5T_c$
there are considerable amplitude 
fluctuations about mean values at different sites. Also,
large kicks
cause `exchange moves' or `flips'.
However, only stray flips (SF) could be found 
(the defect stays only for $\sim\tau_{0}$ time).
This happens because  at low $T$ 
the effective potential 
for two sites in an otherwise frozen background 
has one deep minimum and one very shallow minimum (separated by $g/K$). 
Hence the difference in distortions can't settle on a `flipped' value
and resident flips (RF, flips that stay for $\sim 100\tau_{0}$) cannot 
happen. As temperature increases ($\sim0.8T_c$) the amplitude fluctuations
of $x_{i}$'s 
increase and SF's become more frequent. When an SF creates a defect and 
its neighborhood has a 
small difference in $x_{i}$'s, the force that the defect exerts on the 
neighbourhood might be enough to create spatially
correlated flips, causing a growth of the domain. 
\begin{figure}[b]
\centerline{
\includegraphics[width=6cm,height=4cm]{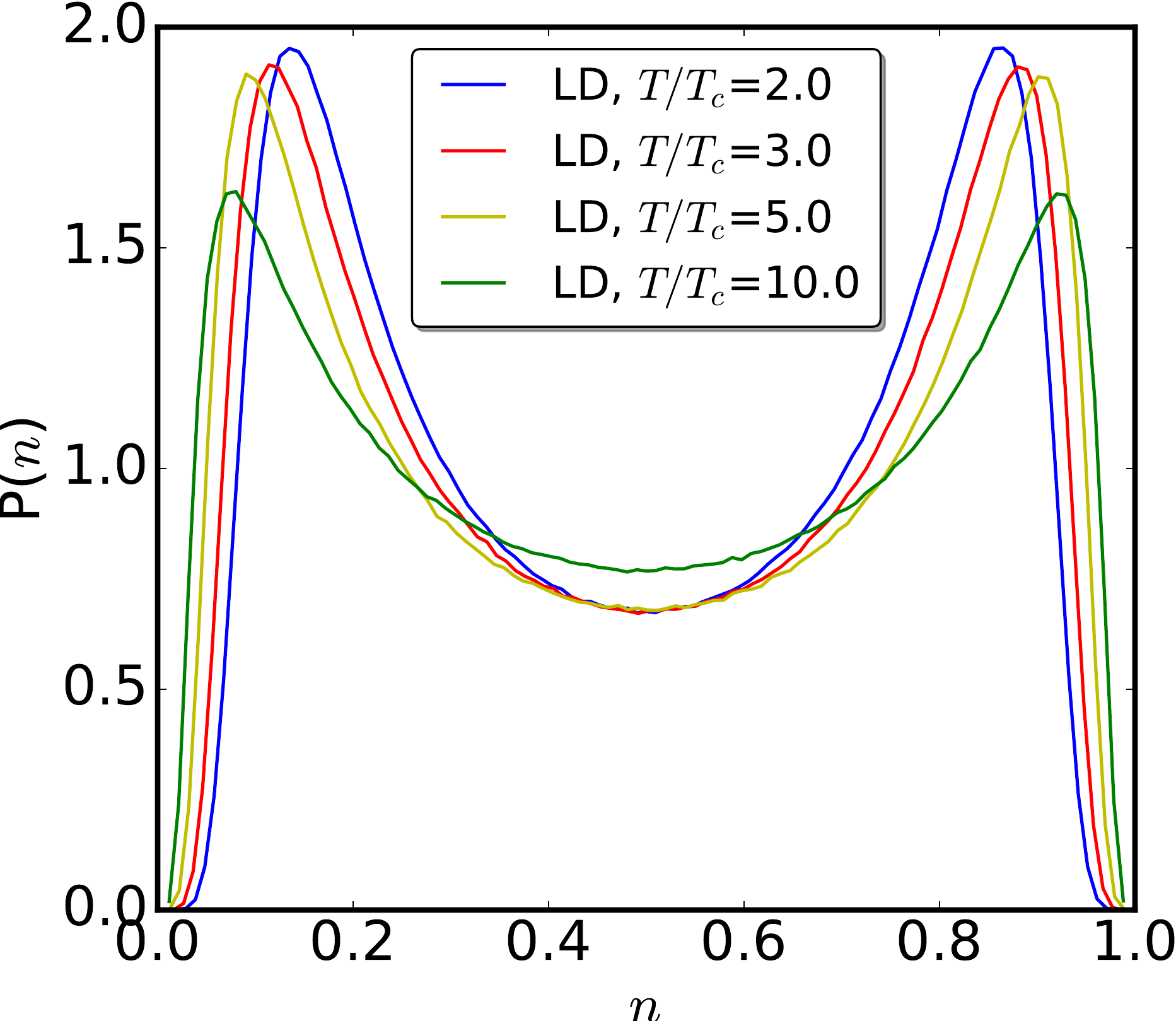}
~~~~~
}
\vspace{.2cm}
\centerline{
\includegraphics[width=8.5cm]{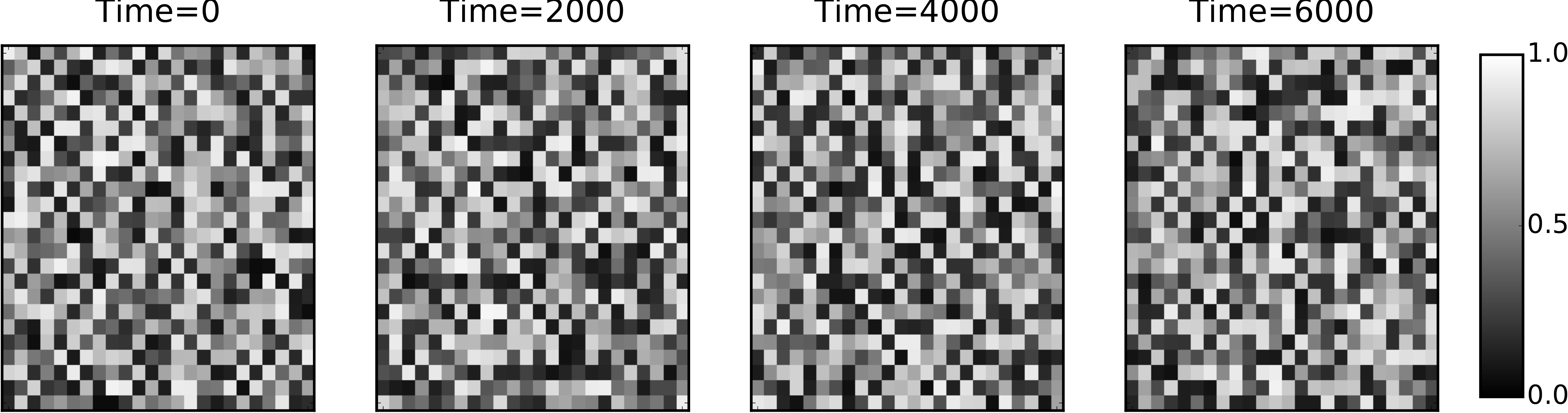}
}
\caption{Top panel: the distribution of local density $P(n)$
showing distinct bimodal features till $\sim10T_c$. This means
there's a large region in
temperature where polarons are present in the lattice, but there's no
global ordering. Bottom panel: Snapshots from time evolution at $T=3.0T_c$.
One sees dynamic patterns of short-range correlated polarons, hinting a
non-trivial spatial structure.
}
\end{figure}

In Fig.10, we depict a domain growth event as a function of time from our data
at $T/T_c=0.9$. The bluish regions depict `C' kind of order, which dominates
initially. However, with increasing time, the opposite pattern C' tries to
grow from within and the structure factor $S(\pi,\pi,t)$ decreases as a 
consequence. In the middle, we see a low structure factor where two kinds
of domains are almost equally present. Later, the C' order encompasses the
lattice and $S(\pi,\pi,t) $ grows again. These `domain oscillations' typically
take place over a large ($\sim500\tau_{0}$) timescale.

\subsection{Polaron liquid:  $T_c \ll  T < T_P $}

In this regime long-range order is lost but 
there are short-range correlations
still prevalent amongst large distortions. There is no obvious small 
parameter, or a universal phenomenology as in the critical regime.

\begin{figure*}[t]
\includegraphics[width=17.5cm,height=4cm]{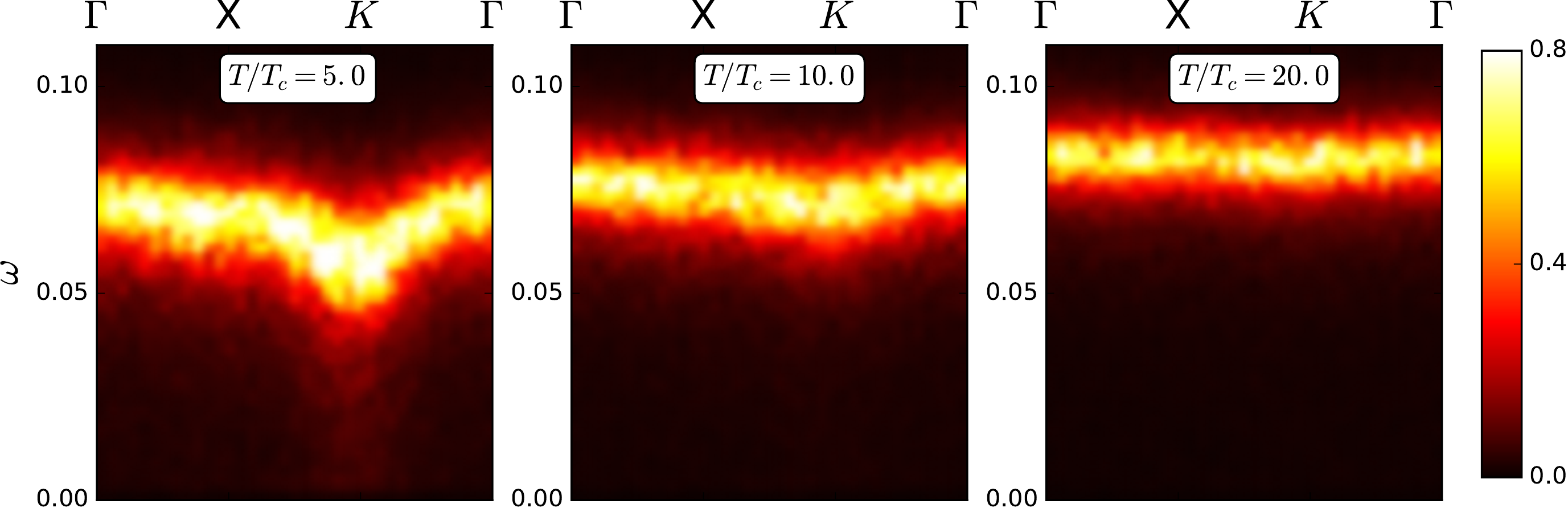}
\caption{False color maps of the power
spectrum $\vert X({\bf q},\omega) \vert^2$ in the high $T$ regime.
The spectral intensities are plotted with the momentum trajectory
$(0,0)\rightarrow(\pi,0)
\rightarrow(\pi,\pi)\rightarrow(0,0)$ along the x-axis.
}
\end{figure*}

\subsubsection{Real time trajectories}

The left panel of Fig.11 features real space trajectories on nearest neighbours
at $T=3.0T_c$, where one sees a merger of large oscillations and flip moves.
The corresponding momentum trajectories at this temperature are all featureless,
as observed in the right panel.

\subsubsection{Power spectrum}

The power spectra (top panel in Fig.12) display a mildly dispersive band with
pronounced softening and broadening near $(\pi,\pi)$. The dip gradually reduces
on heating up, signifying loss of intersite correlations. There's also a mild
stiffening of the branch as a whole with increasing $T$. On heating up even
further, one sees the branch lose dispersion at $\sim10T_c$. The broadening
also becomes $\gamma$ limited eventually. This signifies that the effective
Hamiltonian for the oscillators first reduces to a local one and then the
anharmonicities vanish at asymptotically high $T$.

The lineshapes at specific momenta corroborate the conclusions drawn above.
At lower $T$, the damping is highest at the BZ corner. The mode values shift
on heating up.
There's a quantitative reduction in damping on heating for all momenta, the
most noticeable being $(\pi,\pi)$.

The corresponding density distributions and snaphots are highlighted in 
Fig.13. In the top panel, we do see prominent bimodality for an extended 
range of temperatures ($2T_c-10T_c$). This is indicative of the fact 
that polarons and their short-range correlations (shown in the bottom panel) 
dictate the physics in this regime.

\subsection{Polaron dissociated phase: $T \gtrsim T_P$}

In the high $T$ regime, the power spectrum gradually loses dispersive 
features and is broadened compared to its low $T$ counterpart. 
Intuitively, these features can 
be explained using an effective Hamiltonian with a nonlinear local term 
and a nearest neighbour harmonic part for the $x_{i}$ field. 
The local term may be derived by tracing out electrons from a single 
site Holstein problem. The nearest neighbour coefficient is 
of order $J\sim\frac{t^2}{E_p}$. The dimensionless form of the full
Langevin equation reads-
\begin{equation}
\frac{d^2X_{i}}{d\tau^2}=-\gamma^{\prime}\frac{dX_{i}}{d\tau}-X_{i}+
\langle n_{i} \rangle 
+\sqrt{\frac{k_{B}T}{E_{p}}}\eta_{i}
\end{equation}
where $\gamma^{\prime}=\gamma/M\omega_{0}$, $\tau=\tau_{0}t$ and 
$X_{i}=(g/K)x_{i}$. The noise correlator is given by-
\begin{equation}
\langle \eta_{i}(\tau)\eta_{j}(\tau^{\prime}) \rangle =
\gamma^{\prime}\delta_{ij}
\delta(\tau-\tau^{\prime})
\end{equation}
\begin{figure}[b]
\includegraphics[width=6cm]{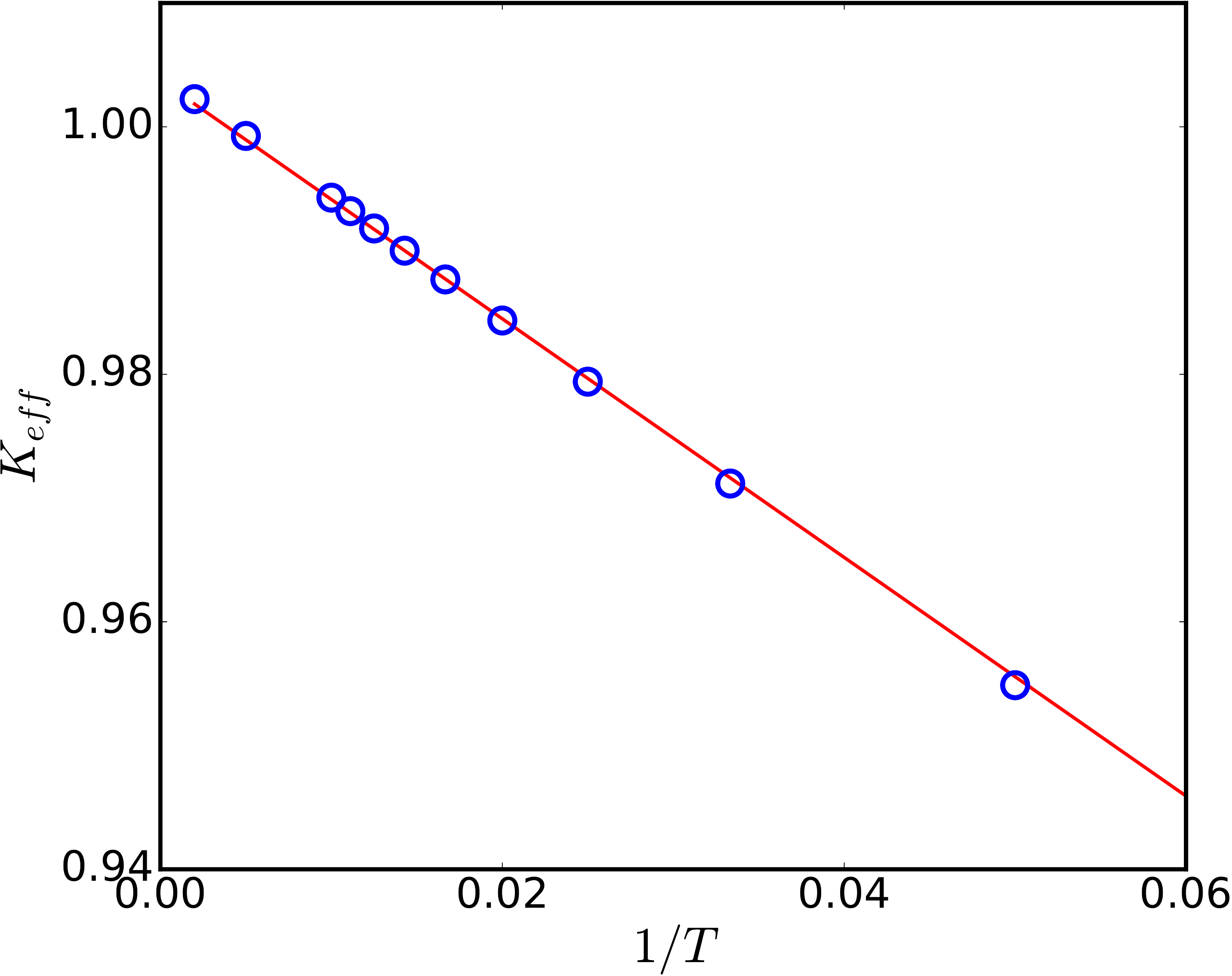}
\caption{Effective stiffness ($K_{eff}$) plotted against $1/T$ in the
asymptotically high $T$ regime. A linear fit (consistent with a high $T$
expansion) is in fairly good agreement with the data. Ultimately, for small
$1/T$, one gets back to the bare stiffness value $K=1$.
}
\end{figure}
\begin{figure}[t]
\centerline{
\includegraphics[height=4cm,width=6cm]{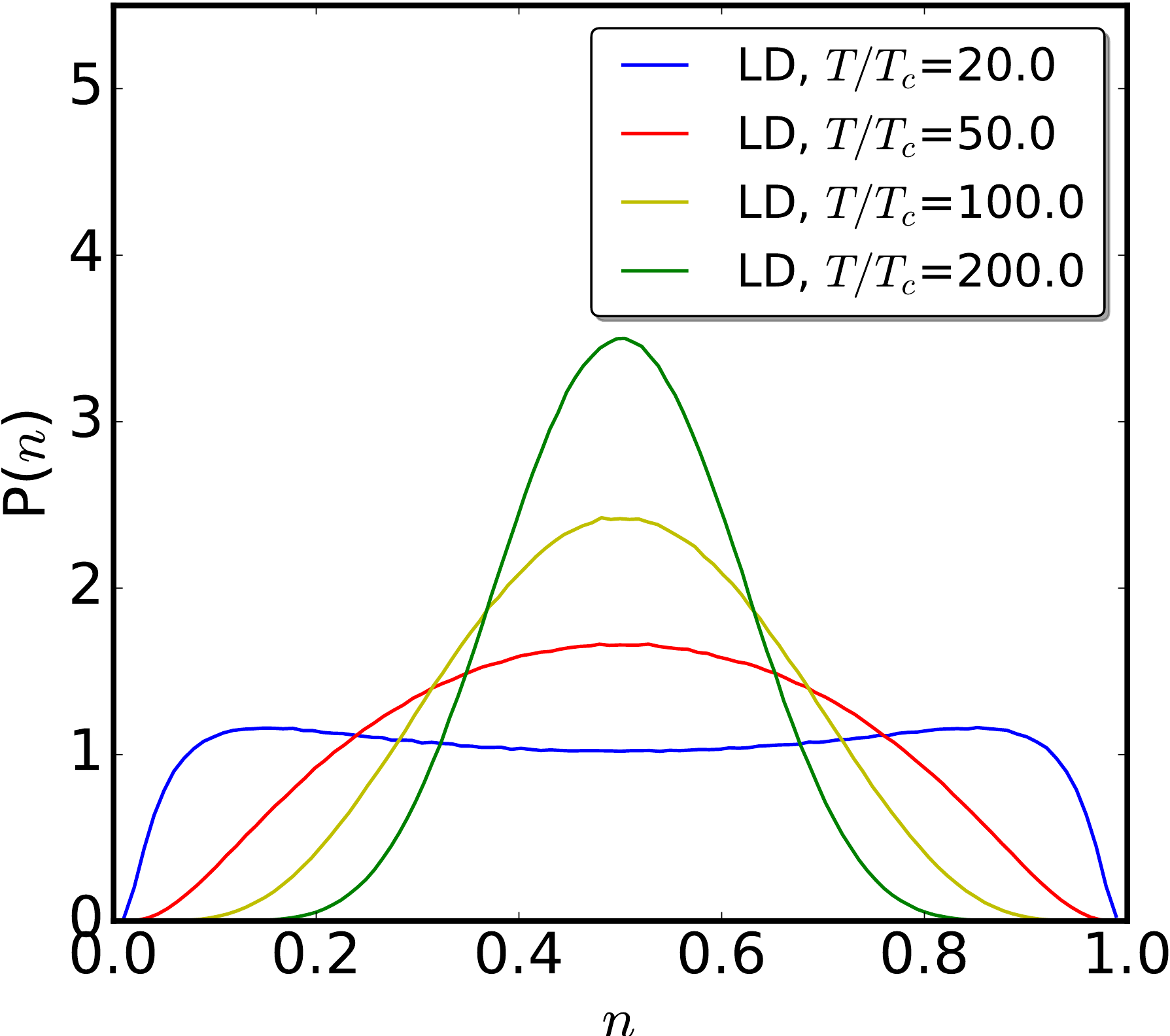}
~~~~~}
\vspace{.2cm}
\centerline{
\includegraphics[width=8.5cm]{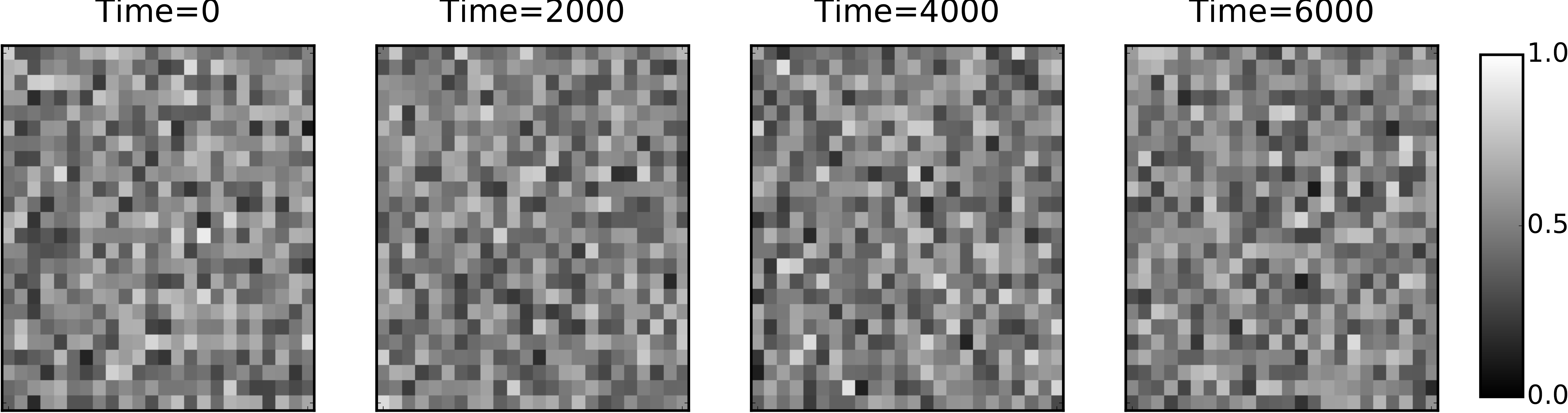}
}
\caption{$P(n)$ and spatial character in the `polaron dissociated'
phase. The top panel depicts a broad distribution at the lowest temperature
shown ($T/T_{c}=20.0$) gradually sharpening to a nearly gaussian distribution
with mean $n=0.5$ at very high temperature ($T/T_{c}=200.0$). The spatial
character shows `melted' polarons, which don't have any interesting dynamics.
}
\end{figure}

At asymptotically high temperature, $T \gg E_p$, the density becomes a local, 
linear function of $X_{i}$
and the corresponding power spectrum becomes identical to that of a single
harmonic oscillator with a renormalized frequency. 
This frequency approaches
the bare value at asymptotically high temperature. 
On lowering $T$, higher order corrections in $\beta E_p$ feature in the 
$<n_{i}>$. We have a linear ($O(\beta E_p$)) correction to the bare stiffness
featuring first, and then a higher order ($O(\beta^{3}t^{2}E_p)$) intersite
correction which gives rise to `dispersion' at lower $T$.
In Fig.14, the power spectra
in this regime are featured, which ceases to show dispersive features and
only exhibits gradual band tightening, with reduction in damping as one heats
up.

In Fig.15, we fit the effective stiffness, calculated from the $P(x)$ 
distributions against $1/T$. One expects a linear variation from an 
analytic high $T$ expansion, which is borne out by the actual data.

The density distribution and snapshots from dynamics in this regime are
shown in the top and bottom panels of Fig.16. respectively. The distributions
converge to a gaussian at large enough $T$, with a mean $n=0.5$ and width
proportional to $\sqrt{T}$. The snapshots exhibit `melted' polarons, which
are represented as greyish regions.

\begin{figure}[b]
\centerline{
\includegraphics[width=3.0cm,height=4cm]{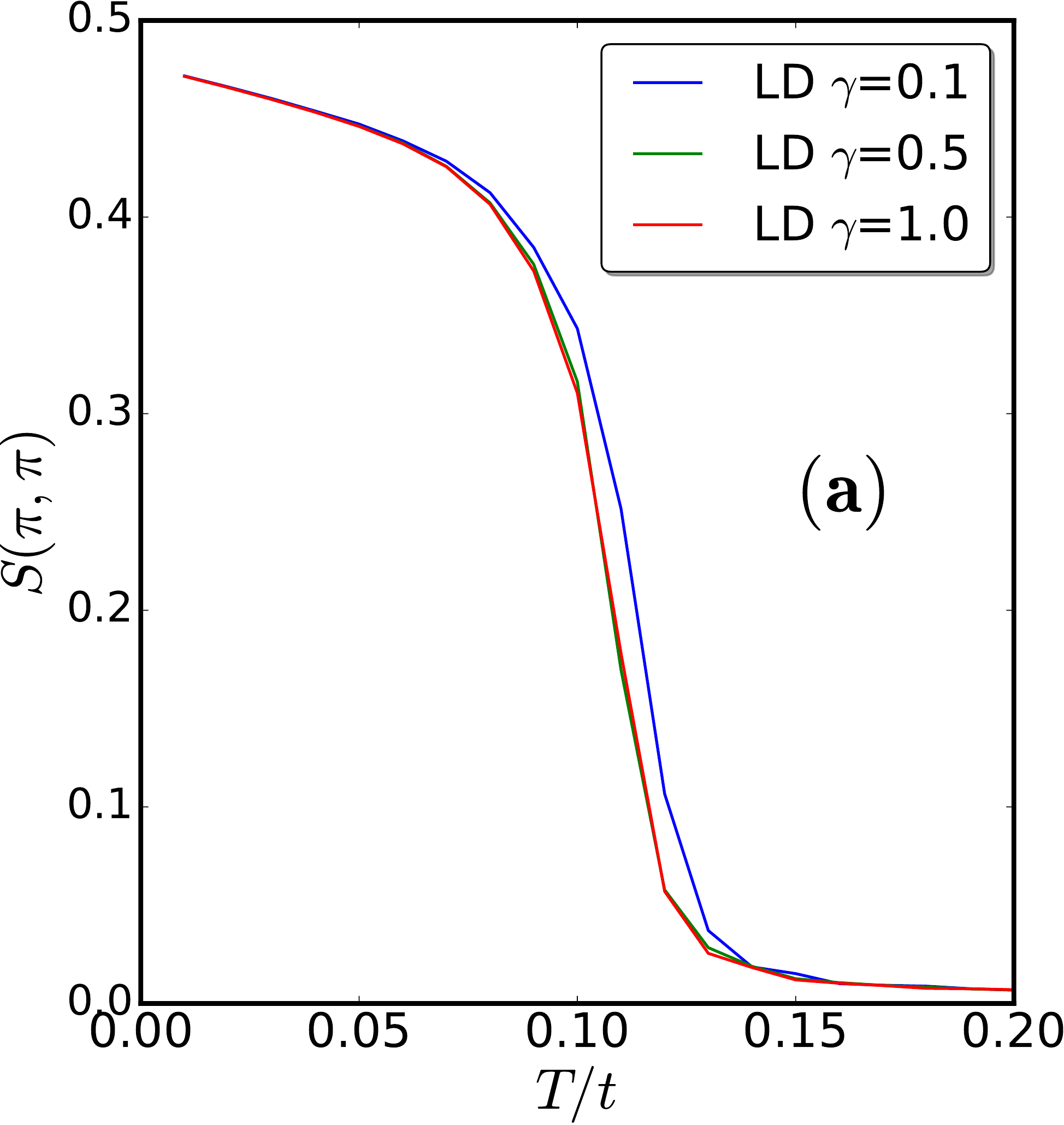}
\includegraphics[width=3.0cm,height=4cm]{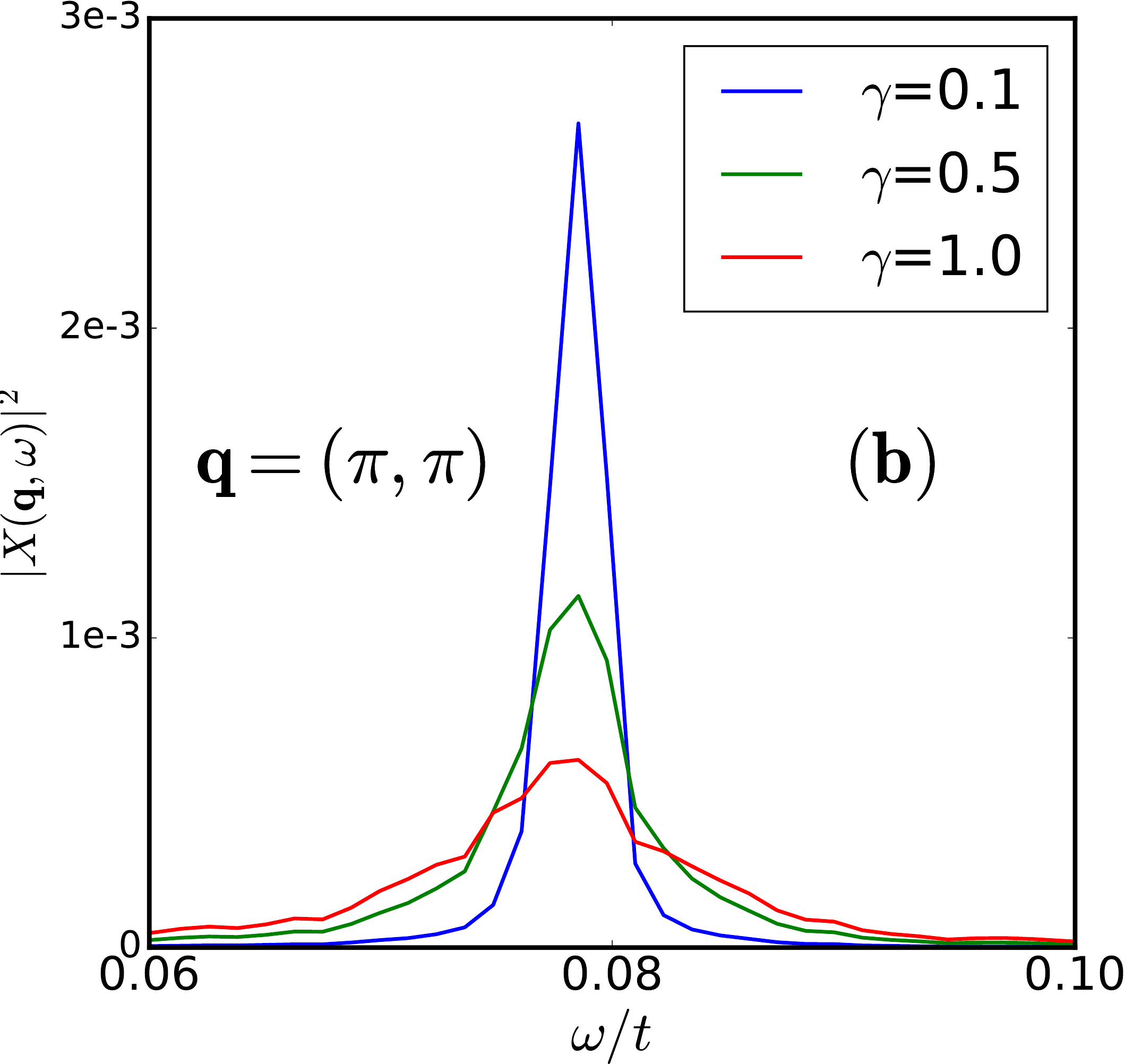}
\includegraphics[width=3.0cm,height=4cm]{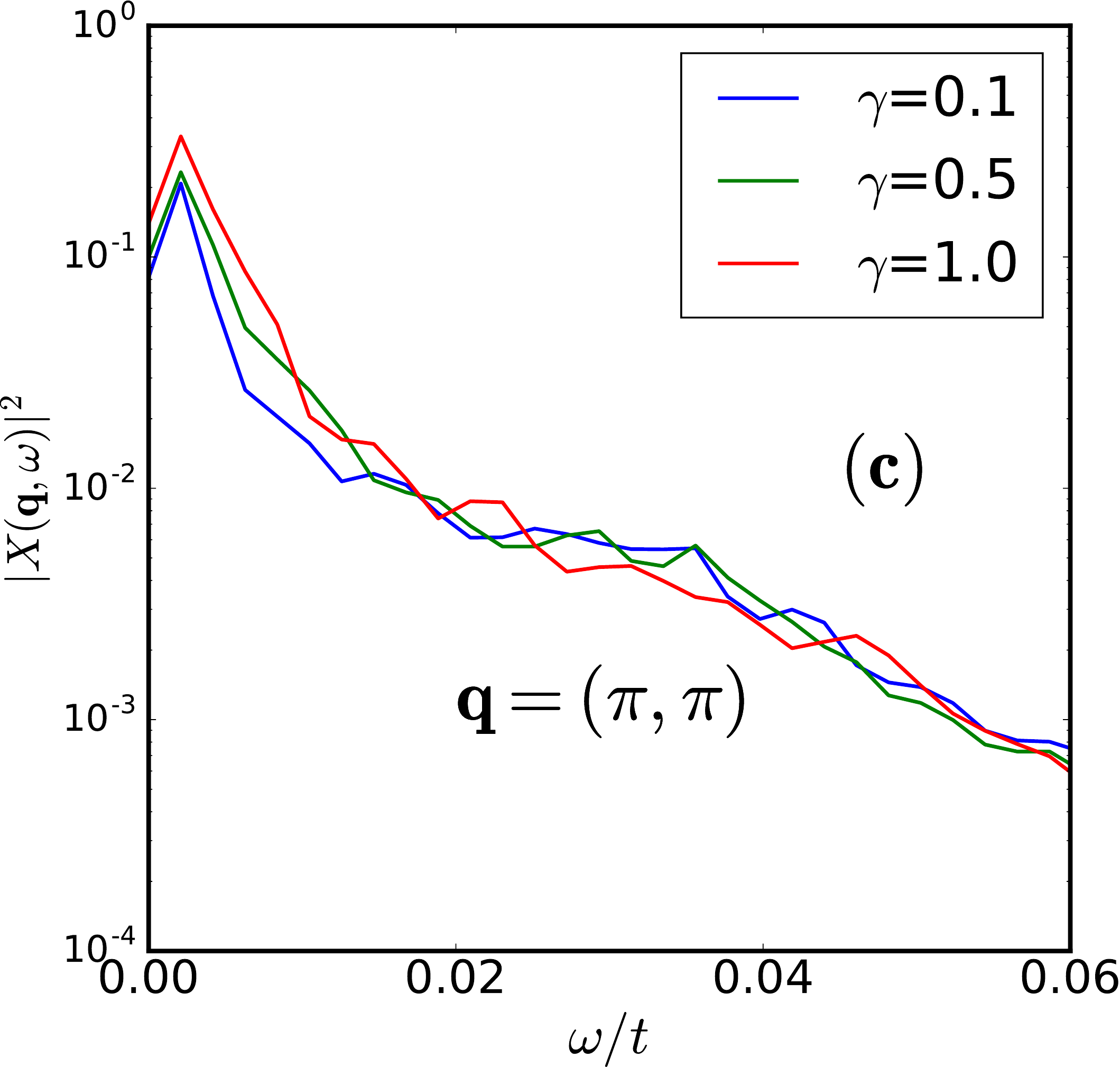}
}
\caption{
(a)~ The structure factor $(\pi,\pi)$
at $T=0.2T_c$ for three different $\gamma$ values- $(0.1,0.5,1.0)$.
The $\gamma=0.1$ result features a slightly higher $T_{c}$ compared
to the other two. There's a `saturation' at higher $\gamma$, whose
results coincide with one another.
(b)~ The ($\pi,\pi$) lineshape in the low $T$ harmonic regime.
The broadening is observed to increase proportionately with $\gamma$, as
expected from the harmonic oscillator result. (c)~ The $(\pi,\pi)$
lineshapes at $T_c$ plotted on a logarithmic scale. The $\gamma$ variation
is significantly suppressed in this critical regime.  }
\end{figure}

\begin{figure}[t]
\centerline{
\includegraphics[width=4.0cm,height=4cm]{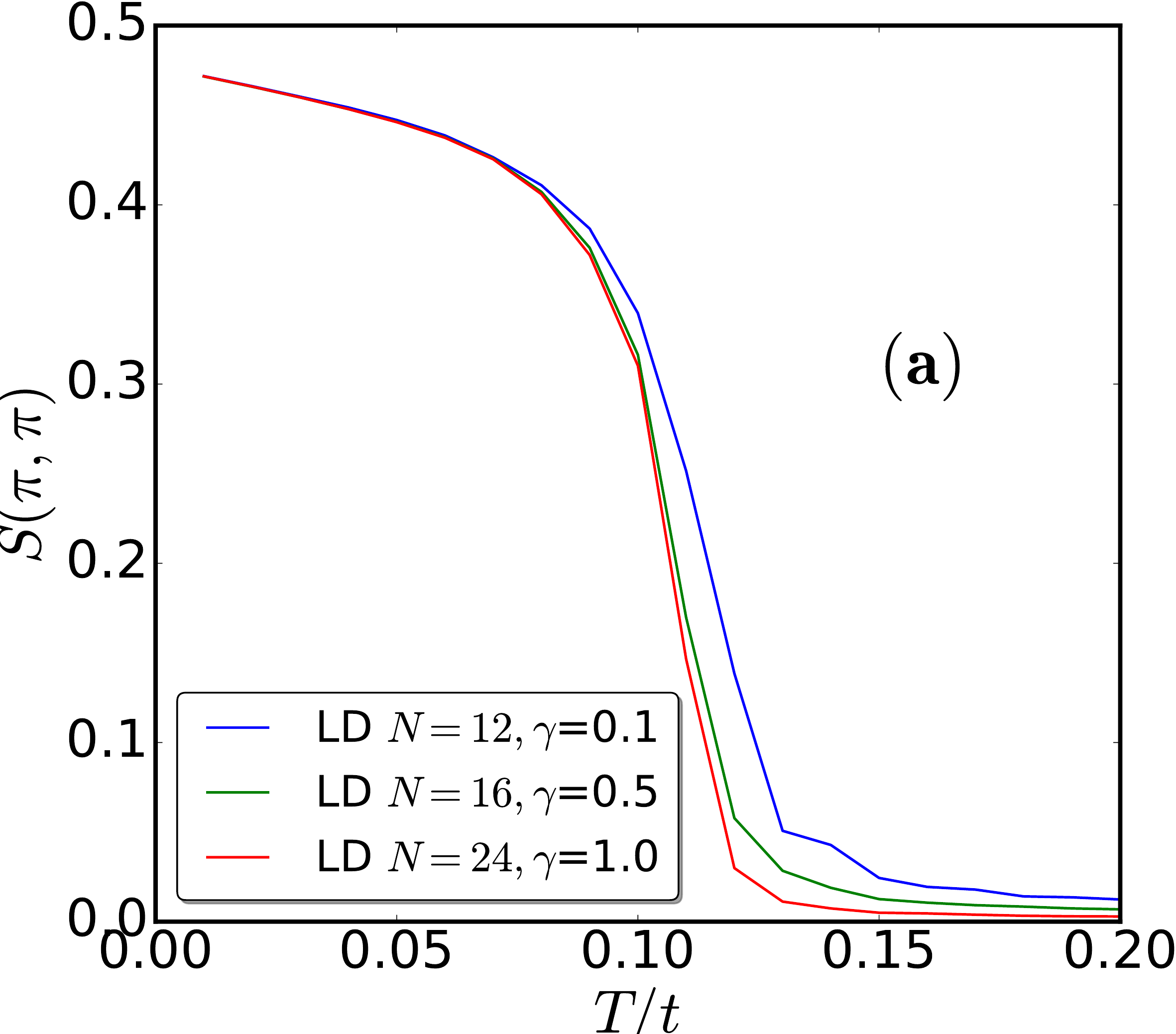}
\hspace{.2cm}
\includegraphics[width=4.0cm,height=4cm]{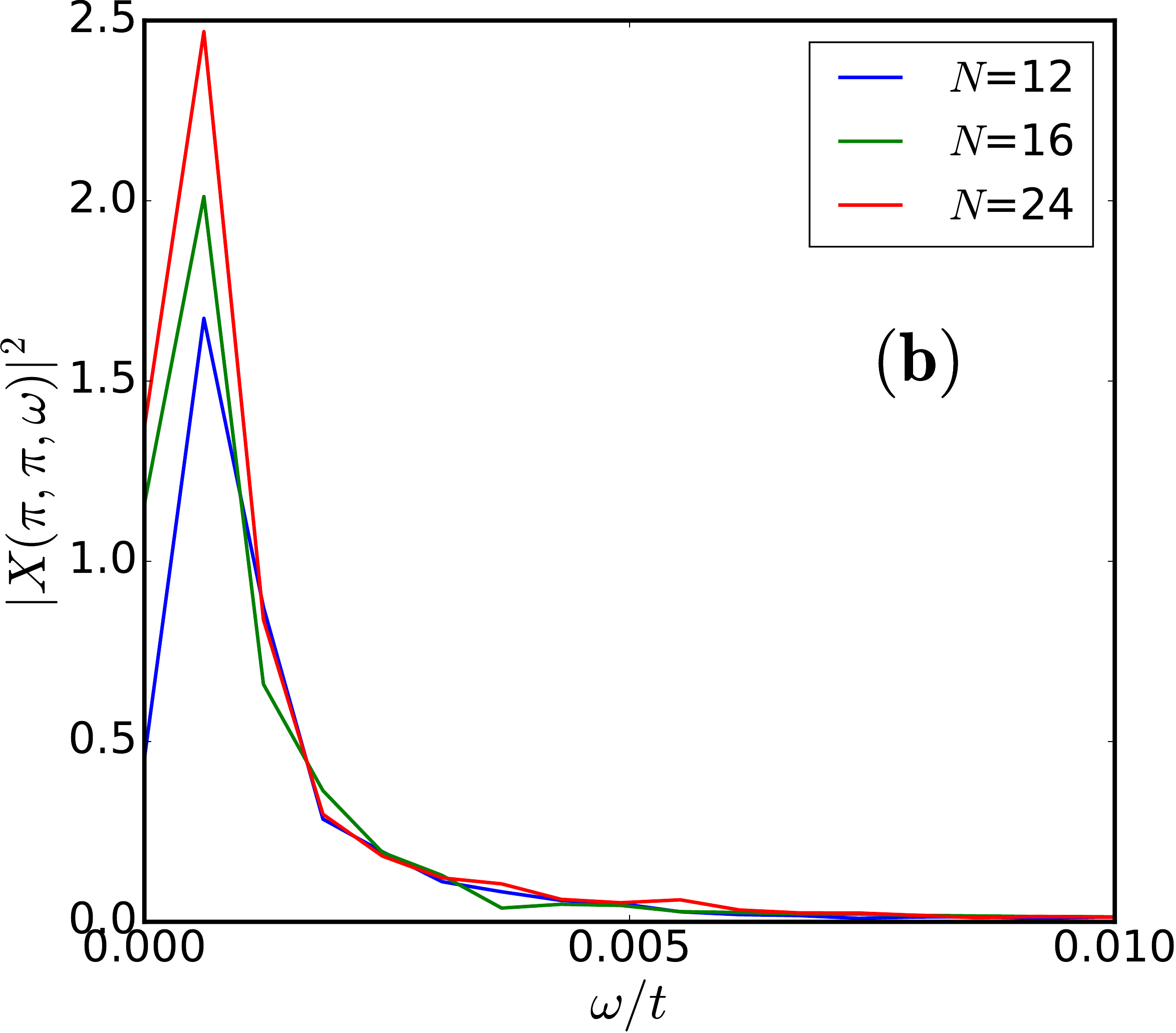}
}
\caption{
(a)~The structure factor $S(\pi,\pi)$ characterizing
order-disorder transition for three different sizes- $(12 \times 12,16
\times 16,24 \times 24)$.  The $\gamma$ value is $1.0$ for all of them.
We see a mildly sharper transition for the largest size.
(b)~Power spectra at $(\pi,\pi)$ for different sizes. The
character is similar for all the profiles, with a growing weight near
zero frequency for larger sizes.  }
\end{figure}

\section{Discussion}

\subsection{Computational checks}

\subsubsection{Dependence of results on $\gamma$ choice}

The stucture factors for different $\gamma$ values have an overall
similarity. The lower $\gamma=0.1$ gives a slightly higher $T_c$. The
$\gamma=0.5$ and $\gamma=1.0$ curves lay on top of each other, signifying
the insensitivity of the system's equal-time properties on this parameter
(shown in Fig.17(a)).
However, we comment that if one uses a much higher $\gamma$, the problem
ceases to have a correspondence with the physical Holstein model and becomes
overdamped.
To establish the dependence of power spectra on 
$\gamma$, we look at two different temperature regimes- i)
the low $T$ harmonic regime and ii) the critical regime ($T\sim T_c$).
The observations are featured in Figs.17(a) and 17(b) respectively. 
In the former, increasing $\gamma$ by a decade (0.1-1.0) has a proportionate
impact on the broadening of the $(\pi,\pi)$ lineshape. This is expected
from the analytic form of the power spectrum, as discussed in subsection A
of the results section. On the other hand, near $T_c$, the dependence
on $\gamma$ is feeble. We have superposed the $(\pi,\pi)$ lineshapes on
a logarithmic scale for three different $\gamma$ values to show this. 
The conclusion is that the critical behaviour is universal and doesn't 
depend crucially on microscopically generated dissipation scales.

\subsubsection{Size dependence}

We have checked size dependence of the $S(\pi,\pi)$, which characterises
the order-disorder transition and the $(\pi,\pi)$ lineshapes at criticality.
The results are displayed in Figs.18(a) and 18(b) respectively.
The former shows a sharper transition as we go to bigger sizes ($16\times16$ 
to $24\times24$), as expected. The low to intermediate temperature behaviour,
governed by linear phonon excitations, is very similar for all sizes. The 
latter quantity has a width that is basically resolution limited for all sizes.
The weight at `near-zero' frequency increases nominally with size. We expect
an infinitely sharp peak at zero frequency in the thermodynamic limit.

\subsection{Simple models for the different regimes}

\begin{figure}[b]
\centerline{
\includegraphics[width=5.5cm,height=4cm]{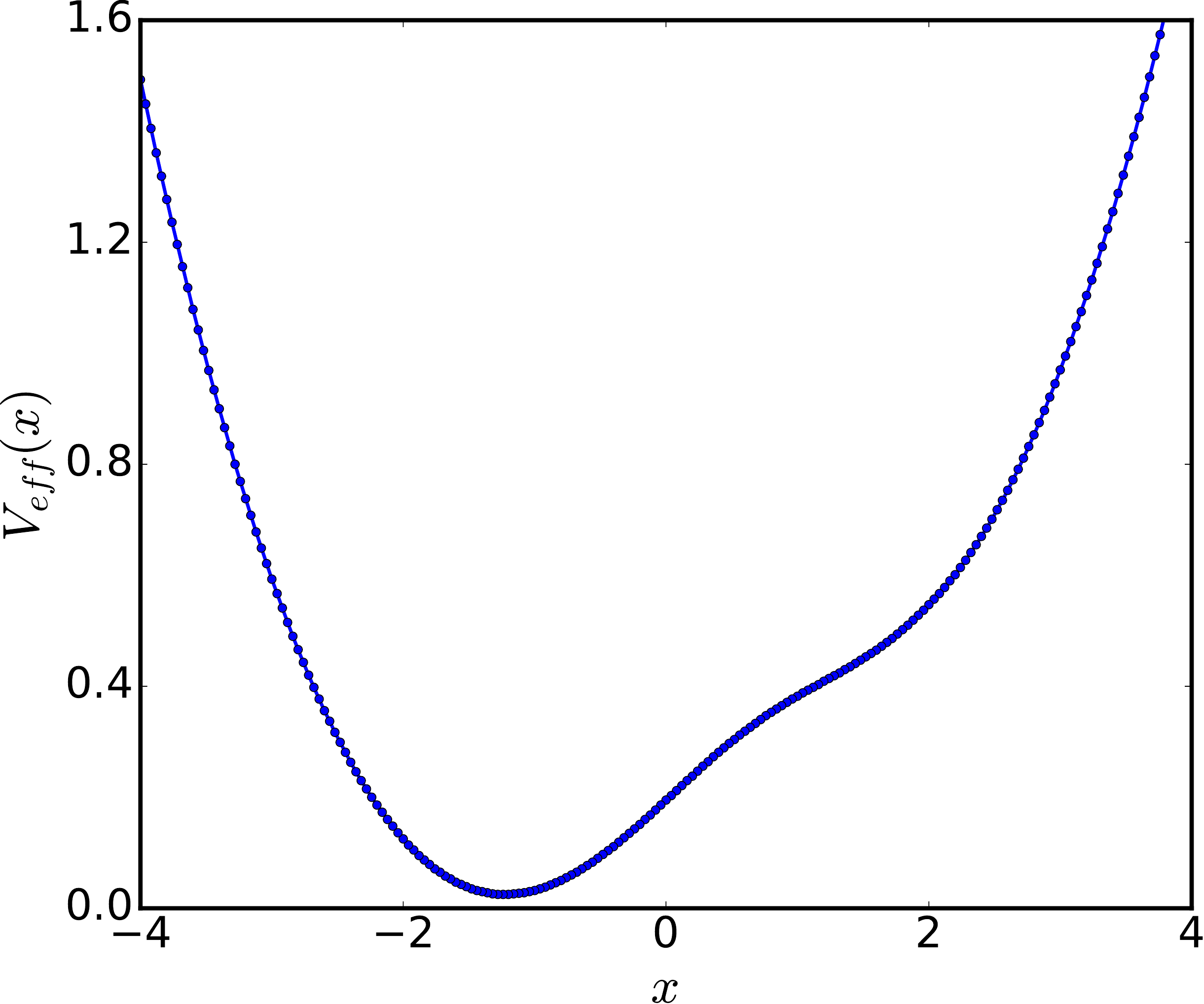}
}
\caption{
Effective potential for the `embedded two-site' problem plotted
against the difference in displacements $x=(x_1-x_2)$ of the two sites.
The background order is a perfect checkerboard with distortions $0$ and
$g/K$ respectively. The potential is obtained by keeping $x_1+x_2$ fixed,
which ensures number conservation locally. The look is that of an asymmetric
double well with an extremely shallow second minimum.
The oscillation freqeuncy at the harmonic level about the lower well is
$0.054$ and that about the overall profile (relevant at very high
temperature) is $0.069$. The potential is indeed harmonic at large range.
The splitting between the two minima is $\Delta=0.38$.
}
\end{figure}

\begin{figure}[t]
\centerline{
\includegraphics[width=4.5cm,height=4cm]{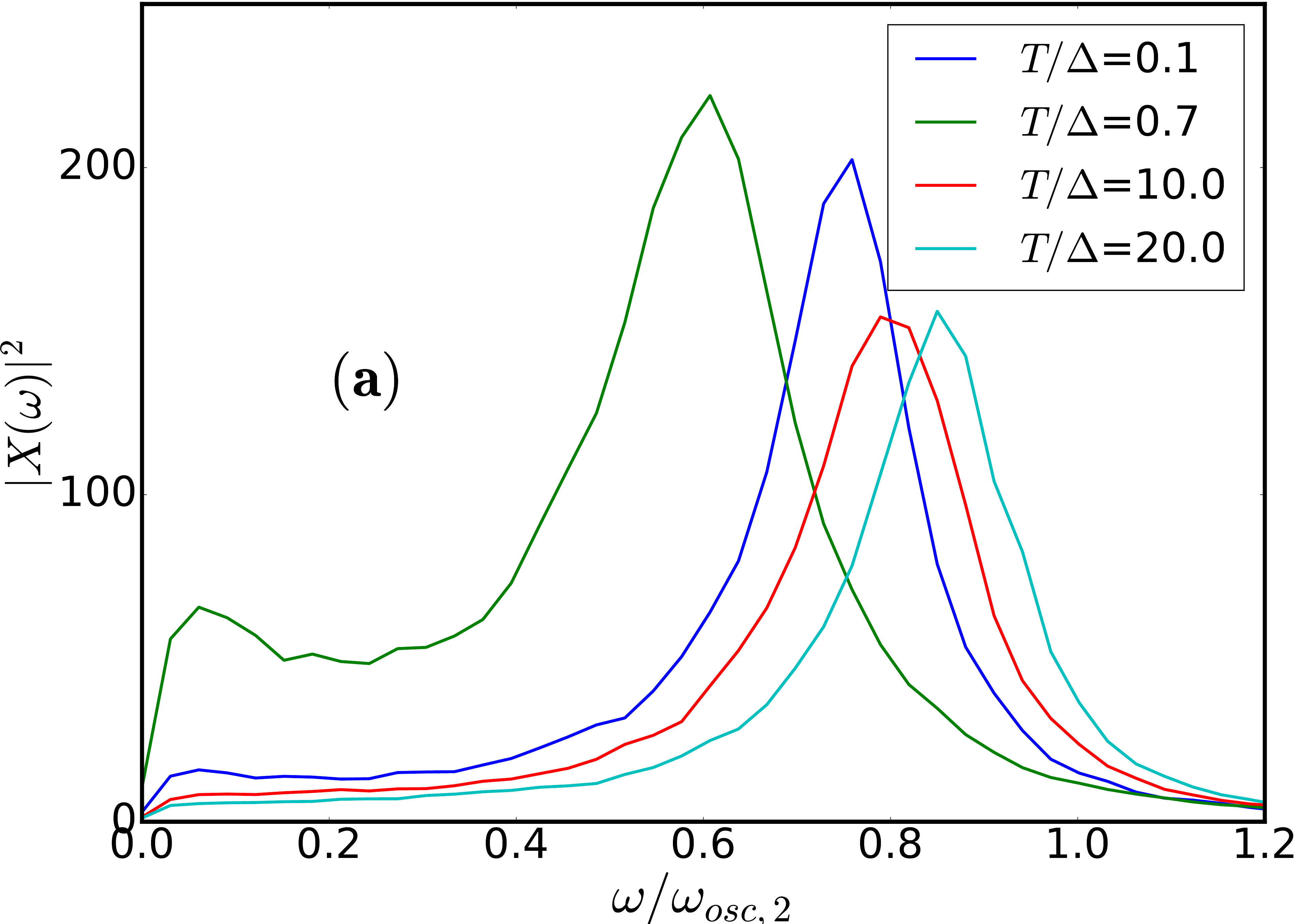}
\includegraphics[width=4.5cm,height=4cm]{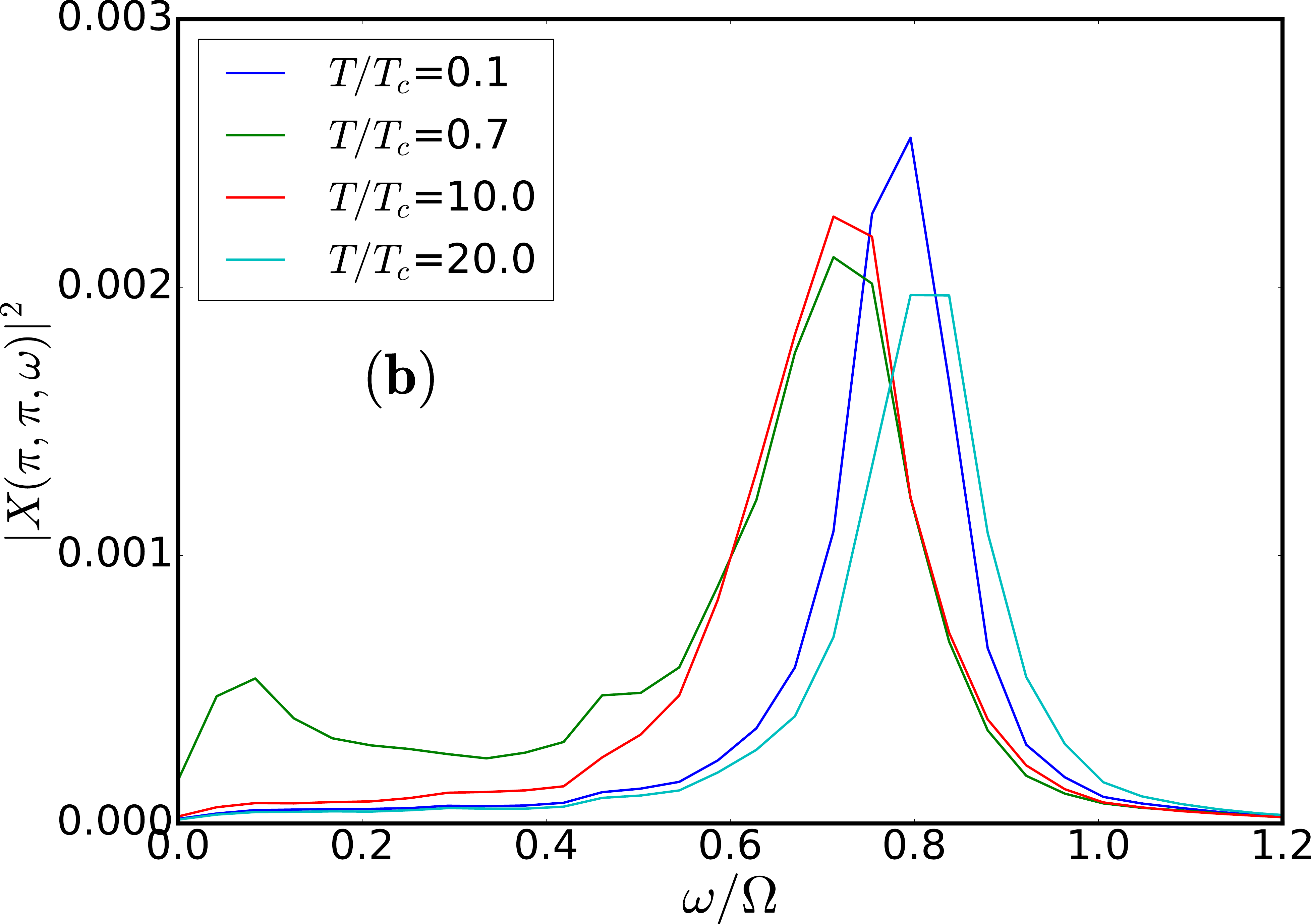}
}
\caption{
(a)~Spectra of the model problem ($|(x_{1}-x_{2})(\omega)|^2)$
compared with (b)~the
 actual power spectrum of the full problem at $(\pi,\pi)$.
A visual similarity is apparent for both the low and high $T$
regimes. Frequencies are scaled with respect to $\omega_{osc,2}$ in the
former and $\Omega$ in the latter case. Temperatures are featured in
units of $\Delta$ for the model problem and $T_c$ for the full problem.
}
\end{figure}

\subsubsection{Non critical behaviour}

We attempted an explanation of the features 
found in the power spectrum of wavevectors near the zone boundary 
through a single variable toy model. It misses the 
collective
critical dynamics but gives a fairly good account of `beyond harmonic' 
physics at low to intermediate temperatures and also 
the high $T$ regime.

The model problem constructed is a two-site Holstein dynamics
embedded in the background of the $T=0$ mean-field state, which is 
$(\pi,\pi)$ ordered. The sum of coordinates $(x_{1}+x_{2})$ is held fixed, 
to ensure particle density conservation, while $(x_{1}-x_{2})$ is varied to
first generate an effective potential numerically. Then, the dynamics
of the `collective coordinate' is studied within this potential. Although,
in the real problem, there's no density conservation locally, the
present simplification is a good enough approximation at least 
for $T<T_c$.

The look of this effective potential (shown in Fig.19) 
is that of an asymmetric double well 
at low values of $(x_{1}-x_{2})$, with a very shallow second well. 
The difference in well depths is due to correlation energies arising 
from the background order. Hence, the energy splitting $\Delta$ between
the two minima is comparable to $T_c$ in the actual problem.
At asymptotically large values of $(x_{1}-x_{2})$, 
the potential is harmonic, whose stiffness can be extracted numerically. 
Hence, the profile can't be fitted to any polynomial function over the 
full range.

The dynamics features three timescales- (i) the oscillation time
about the deeper well ($\tau_{osc,1}$), where the initial condition is 
chosen to lie, (ii) $\tau_{flip}$, for excursions to the shallower well and
(iii) the oscillation time about the overall potential profile 
($\tau_{osc,2}$). These become accessed gradually as one increases $T$.
The frequency scales corresponding to (i) and (iii) are 0.054 and 0.069
respectively.

We've analyzed the power spectrum of the difference coordinate 
$|(x_{1}-x_{2})(\omega)|^2$ for various values of the scaled temperature 
$T/\Delta$, where $\Delta$ is the energy splitting between 
two wells. The frequencies are scaled with respect to $\Omega$ (the
bare oscillation frequency) in the real problem and $\omega_{osc,2}$
in the model situation. Fig.20 (panels (a) and (b)) 
shows the result. Fig.20(a) is
for the model problem, whereas 20(b) is for the actual dynamics.
The gross behaviour is simple- at low enough $T/\Delta$, harmonic 
dynamics about the deeper well is observed, with the characteristic
frequency related to its stiffness. At slightly higher temperatures,
anharmonic effects result in well shift and quantitatively larger
damping. Then, the flips to the shallower well start showing up, with
$\tau_{flip} \gg \tau_{osc,1}$. This results in an accumulation of spectral
weight near zero frequency. For $T/\Delta \gg 1$, the oscillations
take place about the overall profile. The peak shifts towards a frequency
consistent with the corresponding stiffness and damping again becomes
$\gamma$ limited. 

\subsubsection{Critical behaviour}

We attempted to understand the critical behaviour in terms of dynamics
of an Ising model with nearest neighbour AF coupling. Since this is a discrete
variable model, the dynamics was designed in terms of the `sign' of the
evolved degree of freedom. The equation of motion (in discrete form) is-
\begin{eqnarray}
H&=&J\sum_{<ij>}S_{i}S_{j} \\ \nonumber
S_{i,t_{n+1}}&=&sgn(S_{i,t_{n}} - \epsilon*\frac{\partial H}{\partial S_{i}} + 
\sqrt{\epsilon}*\eta_{i,t_{n}})
\end{eqnarray}

The parameters for this problem were: $J=1$ and  $\epsilon=0.01$. The noise
has the properties-
\begin{eqnarray}
\langle \xi_{i}(t_{n})\rangle~~~~&=&0 \cr
\cr
\langle \xi_{i}(t_{n})\xi_{j}(t_{m}) \rangle &=&2 k_{B}T\delta_{ij}\delta_{nm}
\nonumber
\end{eqnarray}

This can be interpreted as an overdamped Langevin equation, from which the
critical dynamics can be extracted along with the thermodynamics. The power
spectrum of this model at $(\pi,\pi)$ has a Lorentzian profile about zero, 
whose width collapses as one tunes the temperature near $T_c$ ($\sim2.2J$) 
from above. At higher $T$, one observes gradual broadening, related to the 
reduction of correlation length. A direct comparison of the lineshapes 
between this model and the real problem at $(\pi,\pi)$ is shown in Figs.21(a)
and 21(b). The temperatures are normalized by the respective $T_c$ scales.
 
\begin{figure}[t]
\centerline{
\includegraphics[width=4.5cm,height=4cm]{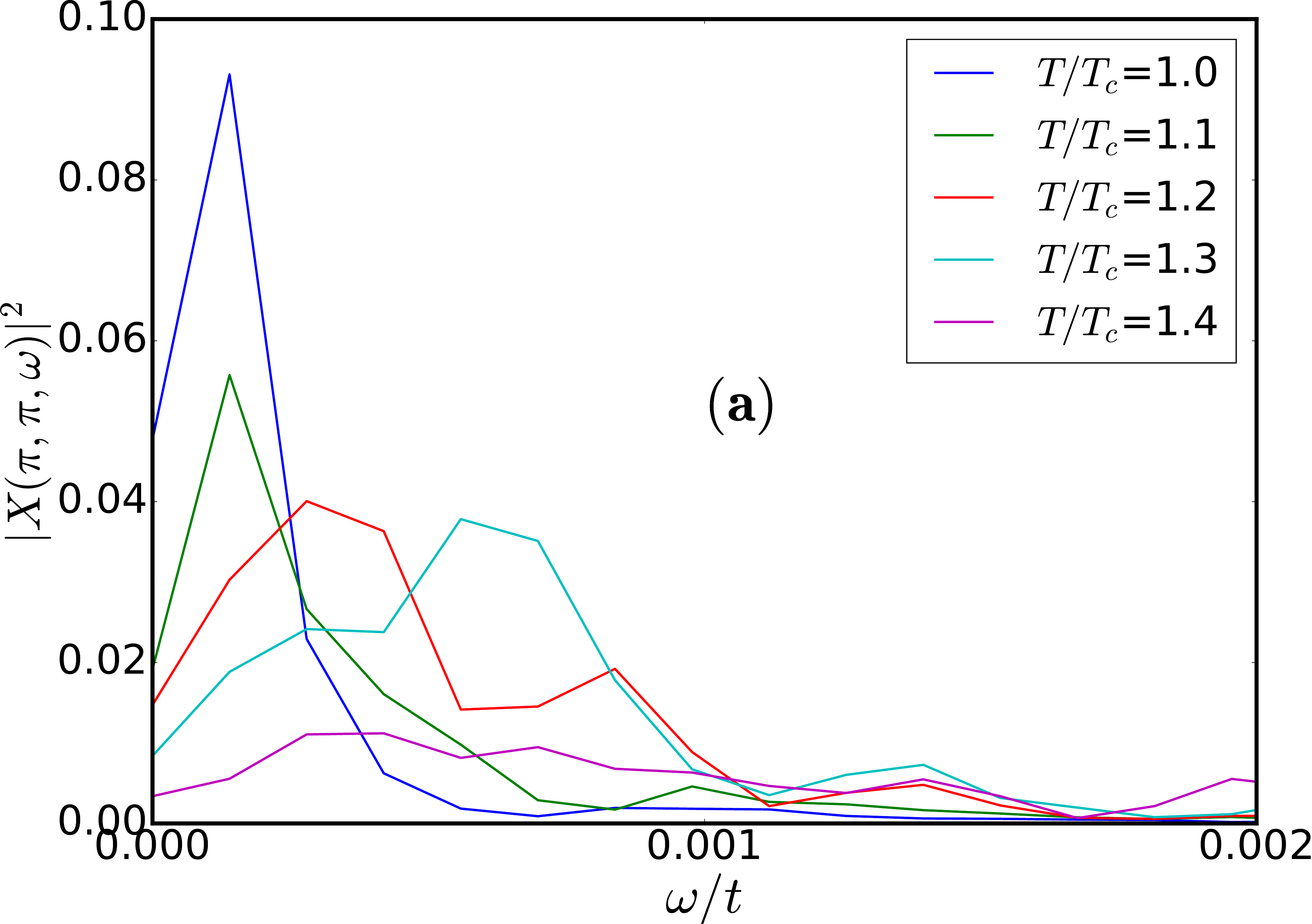}
\includegraphics[width=4.5cm,height=4cm]{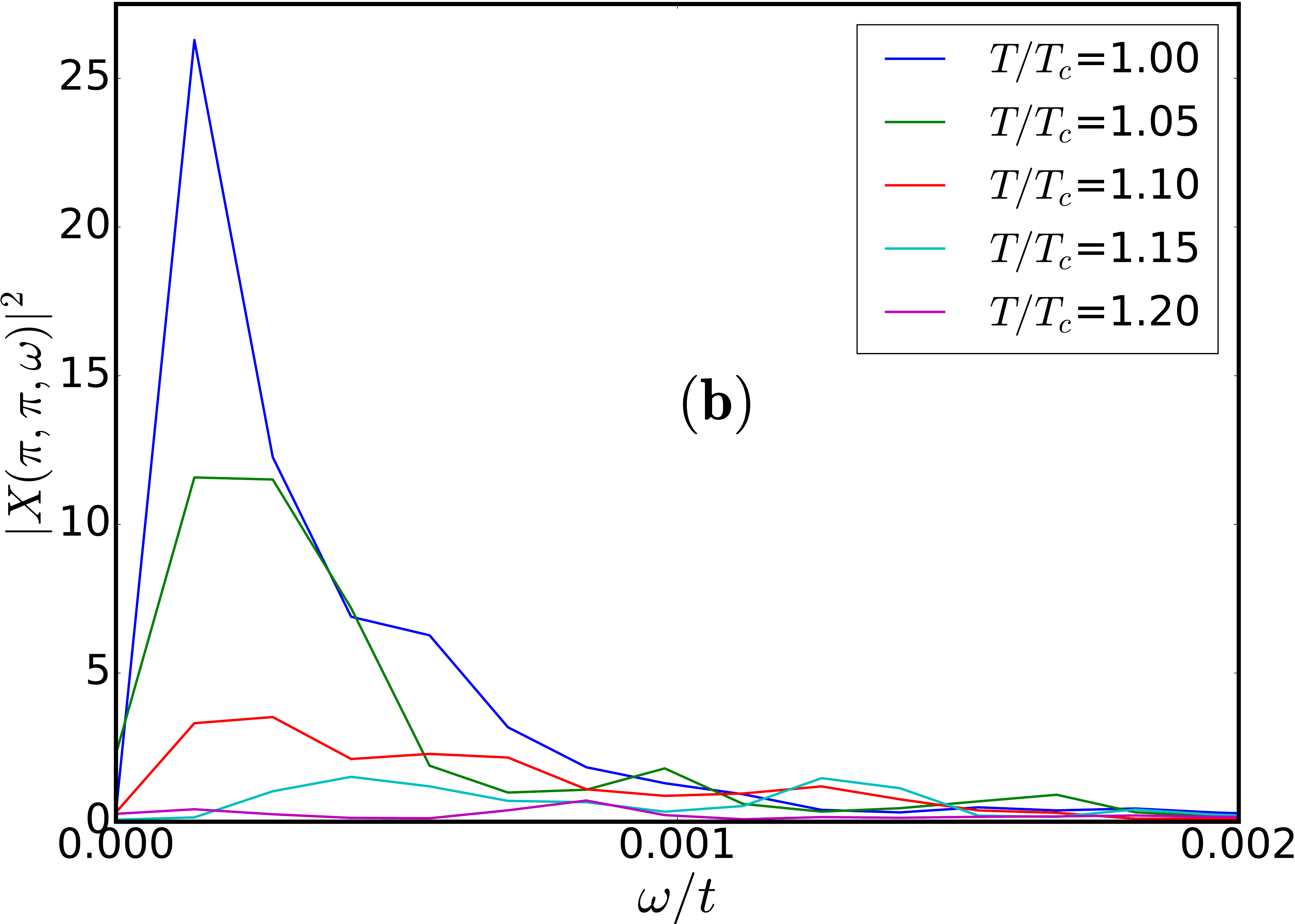}
}
\caption{
(a)~Spectra of the Ising problem compared
to the actual power spectra (b) of the full problem at $(\pi,\pi)$.
A visual similarity is apparent close to $T_c$, where other details are
irrelevant.
Frequencies are scaled with respect to $J$ in the former and $t$ in
the latter case. Temperatures are featured in units of $T_c$ for the
respective problems.
}
\end{figure}

The coarse-grained, continuum limit of the AF Ising system with a conserved
magnetization density, relevant to the present Holstein model,  
falls under Model B amongst the dynamical universality
classes, introduced by Halperin and Hohenberg\cite{hohenberg}. This model in 
the Gaussian limit ($T \gg T_{c}$) can be solved analytically to yield a 
Lorentzian power spectrum, whose width collapses on approaching $T_{c}$ from
above.

Let $S(q,\omega)$ be the coarse grained spin field. 
The power spectrum is obtained by multiplying
two factors of this field and averaging over the noise, which gives-
\begin{eqnarray}
\langle S(q,\omega)S(q^{\prime},\omega^{\prime}) \rangle  
&=& C(q,\omega)(2\pi)^{d+1}
\delta(q+q^{\prime})\delta(\omega+\omega^{\prime})\cr
\cr
C(q,\omega) &=& \frac{2k_{B}TDq^2}{\omega^{2}+(Dq^{2}(r+q^{2}))^{2}}
\nonumber
\end{eqnarray}
This profile is a Lorentzian and has width that is almost $Dq^{2}r$ for 
$T \gg T_{c}$ and almost $Dq^{4}$ for $T\sim T_c$. Hence, we have a diminishing 
width (strictly zero for $q\rightarrow0$) as one approaches $T_c$ from above.

\subsection{Experimental signature}

Several recent experiments suggest an anomalous behaviour of phonon
dispersion and damping across a thermal transition in manganites\cite{weber1,
weber2,weber3}. The inelastic neutron scattering data reveals that phonons
broaden and soften considerably as one makes a transition from a
ferromagnetically ordered, homogeneous ground state to a paramagnetic
insulator with short-range correlated small polarons. In the insulating
state, reduction of states near the Fermi level again causes a decrease in
both the softening and damping. While there are additional magnetic degrees
of freedom which are relevant in these materials, we believe our framework
can be generalized to give a theoretical explanation of these data. Part
of our ongoing work is to compute phonon properties using Langevin 
equation on the doped Holstein
model, to avoid charge ordering physics. If one couples the conduction 
electrons in this to `local moments' with a double exchange coupling, the
resulting system is a reasonable model for doped manganites. We plan
to study the phonon dynamics there. 

\subsection{Future problems}

\subsubsection{Going beyond thermal noise}

The present paper is based on numerical calculations done with a white, memory-
less noise field. While this may be a good enough approximation for 
$\omega/k_{B}T \ll 1$, to estimate the low $T$ behaviour, one has to include
memory effects encoded within $[\Pi]_{ij}^{K}(\omega)$. This is easier said
than done, but the lowest order modification will be to introduce a `thermal
correlation time' that goes to zero in the high $T$ limit. This will also
make the damping term non-local in time with the same $\tau_{th}$ timescale.
The calculation of such a characteristic time has to be done by calculating
the polarizability in the adiabatic approximation. This extension of the 
present scheme will give rise to an equation which is more complicated to
solve numerically. But, it should contain more quantum fluctuation effects
built into it which are important at low $T << \omega$. The physics of 
harmonic phonons will be better captured, as regards dispersion and
momentum dependent damping. Moreover, we hope barrier tunneling events
which restore translation symmetry for doped systems will also be accessible.

\subsubsection{Acoustic phonons, multiple atomic species}

The present study contains a single optical phonon mode coupled to conduction
electrons. To model a more realistic physical system, one has to include
additional vibrational modes like acoustic phonons. This requires a term
with intersite couplings at the non-interacting level. One may readily 
introduce such terms in the instantaeous Hamiltonian of the present scheme.
Moreover, multiple atomic species with different masses can be also brought
in within the same basic equation. This amounts to just changing the inertia 
term and has a bearing on the phonon scattering rates.

\subsubsection{Impact of disorder}

Real materials like manganites always have intrinsic disorder effects 
arising from a substitutional origin. One may introduce potential disorder
in the present scheme directly through the Hamiltonian. Moreover, one can
also put in random mass terms (taken from a binary distribution) to mimic
a disordered medium. The main effect should be on the phonon damping, which
will increase substantially even at low temperature. Disorder will also
suppress quantum tunneling effects making our strategy more justifiable
at low enough temperatures.

\section{Conclusions}

We have studied the real time dynamics of the Holstein model at
half filling using a Langevin  approach. This exploits the 
smallness of the bare phonon energy with respect to the electron
hopping to simplify the `force' acting on the phonon degrees of
freedom. Using  exact diagonalisation of the electron problem
and Newtonian evolution of the stochastic equation we establish
the phonon dynamics from the low temperature charge ordered phase,
through the critical region, into the high temperature polaron
liquid phase. This reveals non monotonic variation in both
the mean energy $\omega_{\bf q}$ and damping $\Gamma_{\bf q}$
of the phonon modes with respect to temperature. This approach,
building in large amplitude dynamical fluctuations, can help
address a wide variety of finite temperature phonon problems
including issues of thermal transport.  
Using an auxiliary field approach it can also approach the
dynamics of other correlated electron problems.

We acknowledge use of the High Performance Computing Facility at
HRI. 

\bibliographystyle{unsrt}

\begin{thebibliography}{99}

\bibitem{ziman} J. M. Ziman, {\it Electrons and phonons}, Oxford University
Press (2001).

\bibitem{bcs} 
  J. Bardeen, L.N. Cooper, and J.R.Schrieffer, Phys. Rev. {\bf 108}, 1175 
(1957). 

\bibitem{gruner}
  G. Gruner, Rev. Mod. Phys. {\bf 60}, 1129 (1988).

\bibitem{emin}
David Emin, {\it Polarons}, Cambridge University Press (2012).

\bibitem{pol} 
  A.S. Alexandrov, \textit{Polarons in Advanced Materials}, Springer (2007).

\bibitem{tokura1}
  Y. Tokura, Physics Today, {\bf 56}, 7, 50 (2003).

\bibitem{tokura2}
  Masatoshi Imada, Atsushi Fujimori, and Yoshinori Tokura, Rev. Mod. Phys., 
{\bf 70}, 1039 (1998).

\bibitem{tokura3}
  Y.Tokura, \textit{Colossal Magnetoresistive Oxides}, CRC Press (2000).

\bibitem{pol2}
 A.S. Alexandrov and N.F. Mott, \textit{Polarons and Bipolarons}, World 
Scientific, Singapore (1995).

\bibitem{millis1} 
  Stefan Blawid and Andrew J. Millis, Phys. Rev. B, {\bf 63}, 115114 (2001).

\bibitem{ins} 
  S.W. Lovesey ed., \textit{Dynamics of Solids and Liquids by Neutron 
Scattering }, Springer (1977).

\bibitem{mannella} 
 N. Mannella, W.L. Yang, X.J. Zhou, H. Zheng, J.F. Mitchell, J.Zaanen, 
T.P. Devereaux, N. Nagaosa, Z. Hussain and Z.-X. Shen, Nature, {\bf 438}, 
474 (2005).	

\bibitem{kemper} 
 Michael Sentef, Alexander F. Kemper, Brian Moritz, James K. Freericks, 
Zhi-Xun Shen, and Thomas O. Devereaux, Phys. Rev. X, {\bf 3}, 041033 (2013). 

\bibitem {phonon}
  P. Bruesch, \textit{Phonons:Theory and Experiments}, Vols. I-III, 
Springer (1982).

\bibitem{QMC} 
  R. Blankenbecler, D.J. Scalapino, and R.L. Sugar, Phys. Rev. D {\bf 24}, 
8 (1981). 

\bibitem{DMFT}
  A. Georges, G. Koliar, W. Krauth, and M.J. Rozenberg, Rev. Mod. Phys. 
{\bf 68}, 13 (1996).  

\bibitem{weber1} 
F. Weber, N. Aliouane, H. Zheng, J.F. Mitchell, D.N. Argyriou, and D.Reznik,
Nature Materials {\bf 8}, 798 (2009).  

\bibitem{weber2} 
F. Weber, S. Rosenkranz, J.-P. Castellan, R. Osborn, H. Zheng, J.F. Mitchell, 
Y. Chen, Songxue Chi, J. W. Lynn, and D. Reznik,
Phys. Rev. Lett. {\bf 107}, 207202 (2011).

\bibitem{weber3} 
M. Maschek, D. Lamago, J.-P. Castellan, A. Bosak, D. Reznik, and F. Weber, 
Phys. Rev. B {\bf 93}, 045112 (2016).

\bibitem{creff} 
  C.E. Creffield, G. Sangiovanni, and M. Capone, Eur. Phys. J. B {\bf 44}, 175 (2005).

\bibitem{hohen} 
  M Hohenadler, H Fehske, and F F Assaad, Phys. Rev. B {\bf 83}, 115105 (2011).

\bibitem{fehske}
 J. Loos, M. Hohenadler, A. Alvermann and H. Fehske, J. Phys.: Condens.
Matter {\bf 18}, 7299 (2006).

\bibitem{bulla} 
  D. Meyer, A.C. Hewson, and R. Bulla, Phys. Rev. Lett. {\bf 89}, 196401
  (2002).    

\bibitem{millis2} 
 Stefan Blawid, Andreas Deppeler, and A.J. Millis, Phys. Rev. B {\bf 67}, 
165105 (2003).
  
\bibitem{neq} 
 A. Kamenev, \textit{Field Theory of Non-Equilibrium Systems}, CUP (2011).

\bibitem{martin} 
 D. Mozyrsky, M.B. Hastings, and I. Martin, Phys. Rev. B {\bf 73}, 035104 
(2006). 

\bibitem{egger}
 Alex Zazunov and Reinhold Egger, Phys. Rev. B {\bf 81}, 014508 (2010).  

\bibitem{brandbyge} 
 Jing-Tao Lu, Mads Brandbyge, Per Hedegard, Tchavdar N. Todorov, and Daniel
Dundas, Phys. Rev. B {\bf 85}, 245444 (2012). 

\bibitem{hohenberg} 
 P.C. Hohenberg and B.I. Halperin, Rev. Mod. Phys. {\bf 49}, 3 (1977).

\bibitem{chern} 
 Gia-Wei Chern, Kipton Barros, Zhentao Wang, Hidemaro Suwa and 
Cristian D. Batista, Phys. Rev. B {\bf 97}, 035120 (2018).
 
\end{thebibliography}

\end{document}